\documentclass[twocolumn]{aastex7}

\newcommand{\Msun}{M$_{\odot}$}
\newcommand{\Mbh}{\ensuremath{M_{\rm BH}}}

\newcommand{\Lsun}{L$_\odot$}

\newcommand{\ml}{\emph{M/L}}

\newcommand{\hst}{\emph{HST}}
\newcommand{\jwst}{\emph{JWST}}
\newcommand{\kms}{km~s$^{-1}$}

\newcommand{\go}{G$_{\rm outer}$}

\newcommand{\jamcyl}{JAM$_{\rm cyl}$}
\newcommand{\jamsph}{JAM$_{\rm sph}$}
\newcommand{\vrms}{$V_{\rm rms}$}

\hypersetup{colorlinks, linkcolor=blue, citecolor=cyan, urlcolor=magenta}

\usepackage{multirow}
\usepackage{newtxtext,newtxmath}
\usepackage{amsmath}	

\graphicspath{{./}{figures/}}
\received{\today}
\submitjournal{AAS Journals}

\shorttitle{\jwst\ stellar-based SMBH measurement of M81}
\shortauthors{D.\ D.\ Nguyen, T.\ N.\ Le $\&$ M.\ Cappellari et al.}
\begin{document}

\title{The Supermassive Black Hole in the Nearby Spiral Galaxy M81: A Robust Mass from JWST/NIRSpec Stellar Dynamics}

\correspondingauthor{Dieu D.\ Nguyen} \email{dieun@umich.edu}

\author[0000-0002-5678-1008]{Dieu D.\ Nguyen}
\email{dieun@umich.edu}
\affiliation{Department of Astronomy, University of Michigan, 1085 South University Avenue, Ann Arbor, MI 48109, USA}

\author[0009-0009-0015-1208]{Tuan N.\ Le}
\affiliation{Faculty of Physics -- Engineering Physics, University of Science, Vietnam National University -- Ho Chi Minh City, Vietnam}
\email{tuan.le.nutshell@gmail.com}

\author[0000-0002-1283-8420]{Michele Cappellari}
\email{michele.cappellari@physics.ox.ac.uk}
\affiliation{Sub-Department of Astrophysics, Department of Physics, University of Oxford, Denys Wilkinson Building, Keble Road, Oxford, OX1 3RH, UK}

\author[0009-0006-5852-4538]{Hai N.\ Ngo}
\affiliation{Faculty of Physics -- Engineering Physics, University of Science, Vietnam National University -- Ho Chi Minh City, Vietnam}
\email{hai10hoalk@gmail.com}

\author[0009-0004-3689-8577]{Tinh Q.\ T.\ Le}
\email{lethongquoctinh01@gmail.com}
\affiliation{Department of Physics, International University, Vietnam National University in Ho Chi Minh City, Vietnam}

\author[0009-0005-8845-9725]{Tien H.\ T.\ Ho}
\email{htien2808@gmail.com}
\affiliation{Faculty of Physics -- Engineering Physics, University of Science, Vietnam National University -- Ho Chi Minh City, Vietnam}

\author[0009-0006-4602-1968]{Long Q.\ T.\ Nguyen}
\email{baobi123987@gmail.com}
\affiliation{Faculty of Physics -- Engineering Physics, University of Science, Vietnam National University -- Ho Chi Minh City, Vietnam}

\author[0000-0001-5802-6041]{Elena Gallo}
\email{egallo@umich.edu}
\affiliation{Department of Astronomy, University of Michigan, 1085 South University Avenue, Ann Arbor, MI 48109, USA}

\author[0000-0002-4436-6923]{Fan Zou}
\email{fanzou@umich.edu}
\affiliation{Department of Astronomy, University of Michigan, 1085 South University Avenue, Ann Arbor, MI 48109, USA}

\author[0000-0002-0362-5941]{Michele Perna}
\email{fmperna@cab.inta-csic.es}
\affiliation{Centro de Astrobiolog\'{i}a (CAB), CSIC–INTA, Departamento de Astrof\'{i}sica, Cra. de Ajalvir Km. 4, 28850 – Torrej\'{o}n de Ardoz, Madrid, Spain}

\author[0000-0002-6694-5184]{Niranjan Thatte} 
\email{niranjan.thatte@physics.ox.ac.uk}
\affiliation{Sub-Department of Astrophysics, Department of Physics, University of Oxford, Denys Wilkinson Building, Keble Road, Oxford, OX1 3RH, UK}

\author[0000-0002-4005-9619]{Miguel Pereira-Santaella}
\email{miguel.pereira@iff.csic.es}
\affiliation{Instituto de F\'isica Fundamental, CSIC, Calle Serrano 123, 28006 Madrid, Spain}

\begin{abstract}
    Despite its proximity, the mass of the supermassive black hole (SMBH) in the spiral galaxy M81 (NGC~3031) has remained uncertain, with previous dynamical measurements being unreliable. We present the first robust stellar-dynamical measurement of its mass using high-resolution, two-dimensional kinematics from \jwst/NIRSpec observations of the central $3\arcsec\times3\arcsec$. By tracing stellar motions in the near-infrared, our data penetrate the obscuring nuclear dust and allow for the separation of stellar light from the non-thermal AGN continuum. We modeled the kinematics using JAM within a Bayesian framework, exploring a comprehensive suite of models that systematically account for uncertainties in the point-spread function, orbital anisotropy, and stellar mass-to-light ratio. This ensemble modeling approach demonstrates that a central dark mass unambiguously drives the central rise in velocity dispersion. The models yield a robust SMBH mass of $M_{\rm BH} = (4.78^{+0.07}_{-0.10})\times10^7$~\Msun. This result resolves a long-standing uncertainty in the mass of M81's black hole and provides a crucial, reliable anchor point for SMBH-galaxy scaling relations.
\end{abstract} 


 
\section{Introduction}\label{sec:intro} 

The masses of supermassive black holes (SMBHs; $\Mbh\approx 10^{6-10}$~\Msun) are known to correlate tightly with the global properties of their host galaxies, establishing a fundamental link between their evolution. These scaling relations, such as the \Mbh--$\sigma$ relation, have been established and refined in numerous studies \citep[e.g.,][]{Gultekin2009, McConnell2013, Saglia2016, vandenBosch2016}, and are comprehensively reviewed in \citet{Kormendy2013}. Accurately measuring \Mbh\ is crucial for calibrating these scaling relations, but it remains a significant observational challenge. The most reliable measurements are derived from dynamical modeling of the gravitational potential, using the motions of various tracers. These include stars \citep[e.g.,][]{Verolme2002, Gebhardt2003, Cappellari2009, Walsh2016, Ahn2018, Krajnovic2018, Voggel2018, Nguyen2014, Nguyen2017, Nguyen2018, Nguyen2019, Nguyen17conf, Thater2019, Thater2022, Thater2023}, ionized gas \citep[e.g.,][]{Barth2001, Shapiro2006, Neumayer07}, and molecular gas \citep[e.g.,][]{Davis2013, Onishi2017, Davis2020, Nguyen2019conf, Nguyen2020, Nguyen2021, Nguyen2022, Ngo2025a}.

The nearby spiral galaxy M81 (NGC~3031) is a prime example of the difficulties in measuring BH masses. Throughout this paper, we adopt a distance of $D = 3.63 \pm 0.14$~Mpc \citep{Durrell2010}. This value represents the mean of five independent determinations: one based on Cepheid variables \citep{Freedman1994, Ferrarese2000_distance, McCommas2009} and four derived from the Tip of the Red Giant Branch (TRGB) method \citep{Tikhonov2005, Rizzi2007, Dalcanton2009, Durrell2010}.

Despite its proximity, a reliable measurement of the SMBH mass in M81 has remained elusive. To date, only two direct dynamical modelling determinations have been attempted, but they are both problematic (see discussion in the box at pg.~552 of the review by \citealt{Kormendy2013}). The first attempt was a preliminary stellar-dynamical measurement and BH discovery, based on a two-integral Jeans model of long-slit \hst/STIS observations \citep{Bower2000}, which is known to be an overly restrictive assumption for realistic galaxies. This preliminary value $M_{\rm BH} = 6(\pm20\%) \times 10^7$~\Msun\ was never finalized or published in a peer-reviewed paper, and the abstract in the AAS Bulletin did not include any figures. The other determination was based on ionized gas kinematics, which yielded an estimate of $M_{\rm BH} = (7 \pm 2) \times 10^7$~\Msun\ \citep{Devereux2003}. However, the underlying gas velocity field appears highly disturbed and is poorly represented by the regular rotating disk model assumed in that work (see their Figure~2), casting significant doubt on the accuracy of the result. Consequently, our work provides the first reliable and detailed measurement of the \Mbh\ in this galaxy. This large uncertainty is compounded by the presence of a low-luminosity active galactic nucleus (LLAGN; \citealt{Eracleous2010}) and complex dust structures associated with both inflows \citep{Muller2011} and outflows \citep{FangzhengShi2021}, which can significantly bias measurements made at optical wavelengths \citep{Thater2019conf}.

The advent of the \textit{James Webb Space Telescope} (\jwst) offers a transformative opportunity to overcome these challenges. By targeting the CO bandhead absorption features in the near-infrared (NIR) at $\sim2.3~\mu$m, the NIRSpec instrument can trace stellar motions directly, penetrating the obscuring dust that affects optical studies \citep[e.g.,][]{Cappellari2009}. Adopting a velocity dispersion of $\sigma = 162$~\kms\ \citep{Muller2011} the $\Mbh-\sigma$ relation by \citet{Gultekin2009} predicts a $\Mbh\approx5 \times 10^7$~\Msun. Given this mass, the SMBH's sphere of influence (SOI) is $r_{\rm SOI} \approx 0\farcs5$. This is more than three times the NIRSpec point spread function (PSF) full width at half maximum (FWHM) at this wavelength \citep{DEugenio2024, D'Eugenio2025}, ensuring the gravitational influence of the black hole is well-resolved. These capabilities are crucial for probing the demographic of intermediate-mass black holes (IMBHs, $\Mbh\approx 10^{2-5}$~\Msun) and bona-fide massive BHs \citep[$M_{\rm BH} > 1.8 \times 10^7$~\Msun;][]{Sasseville2025}, a regime that will be further explored by next-generation facilities like the 39-meter extremely large telescope \citep[ELT; e.g.,][]{Nguyen2023, Nguyen2025b, Ngo2025b, Ngo2025c, Thatte2024}.

In this work, we measure the stellar-based SMBH mass in M81 using Jeans Anisotropic Models \citep[JAM;][]{Cappellari2008, Cappellari2020} applied to archival \jwst/NIRSpec G235H/F170LP IFU data (PID: 02016, PI: Anil C. Seth), \hst/Wide Field Camera 3 (WFC3) infrared (IR) F110W-band imaging (PID: 11421, PI: H.~A.~Bushouse), and the $J$-band image from the Two Micron All Sky Survey (2MASS) \citep{Jarrett2003}. The modeling also constrains key dynamical parameters, including inclination, anisotropy, and the stellar mass-to-light ratio (\ml$_J$), and assesses their impact on the inferred \Mbh\ \citep[e.g.,][]{Nguyen2018, Nguyen2019} in M81.

It is important to note that the nucleus of M81 exhibits dust structures associated with both inflows \citep{Muller2011} and outflows \citep{FangzhengShi2021}, which can bias optical gas-dynamical measurements \citep{Thater2019conf}. The NIRSpec G235H/F170LP observations mitigate these effects by targeting the CO bandhead absorption lines \citep[e.g.,][]{Cappellari2009} in the NIR, where dust extinction is minimal and stellar motions are traced directly. Thus, our measurements not only refine the \Mbh\ estimate for M81, but also demonstrate the remarkable capability of \jwst/NIRSpec to probe SMBH demographics.

This paper is organized as follows. In Section~\ref{sec:data}, we describe the NIRSpec IFU observations, the data reduction, the correction of spectral wiggles in the central regions, and the extraction of the two-dimensional (2D) line-of-sight velocity distribution (LOSVD) from the CO bandhead features. Section~\ref{sec:hst} details the construction of the stellar photometric model from the \hst\ F110W and 2MASS $J$-band images. In Section~\ref{sec:dyna_model}, we combine the 2D LOSVD and photometric model as input for JAM models to derive \Mbh. Finally, our results and conclusions are presented in Section~\ref{sec:result}.

\begin{figure*}
    \centering\includegraphics[width=0.98\linewidth]{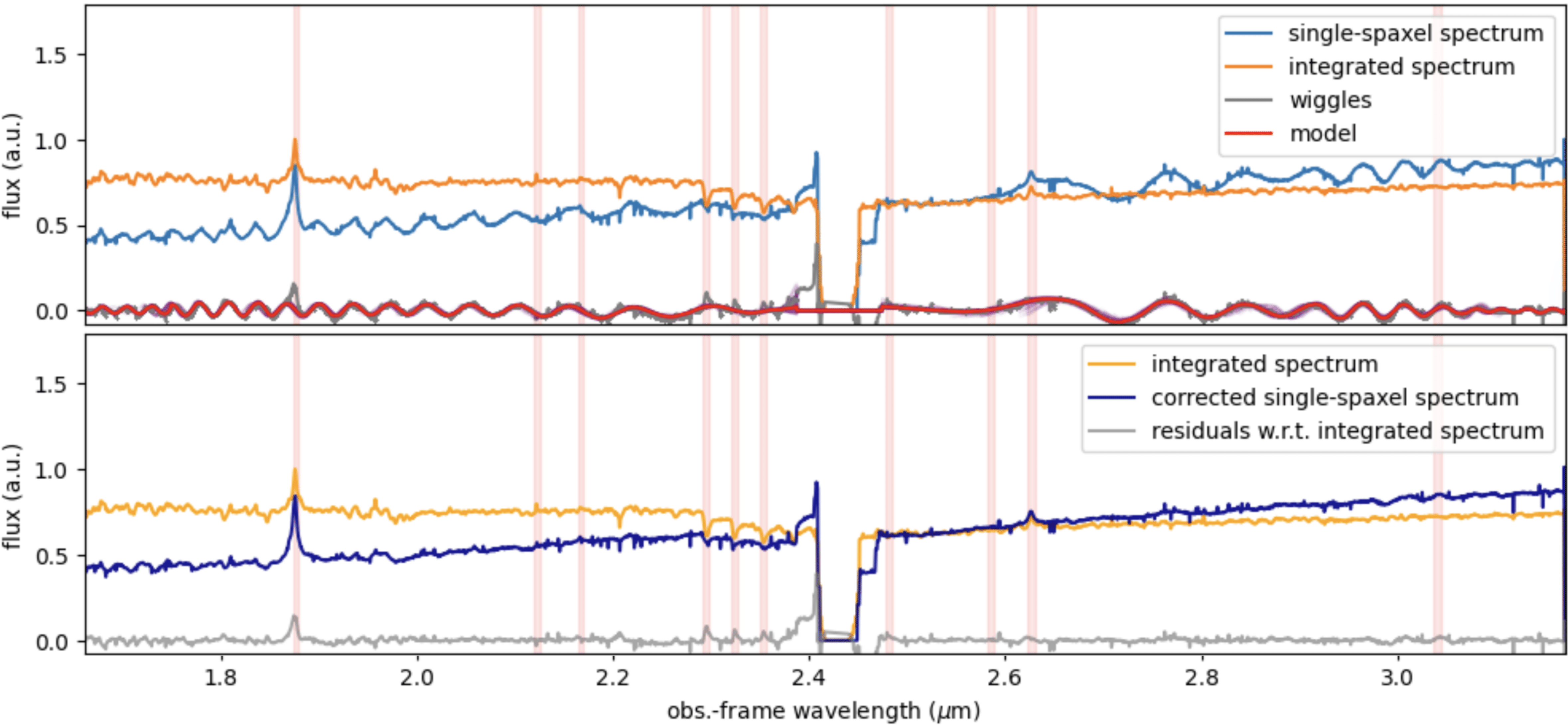}
    \caption{Modeling of wiggles in the single-spaxel NIRSpec spectrum of M81. {\it Top panel:} Integrated spectrum (orange), single-spaxel spectrum (blue), and residual wiggles (gray). The red curve shows the best-fitting wiggle model. {\it Bottom panel:} Wiggle-corrected single-spaxel spectrum (dark blue) compared to the integrated spectrum (orange); residuals after correction are shown in gray. In all panels, red shaded regions mark emission lines excluded from the fit.}
    \label{fig:Wiggles}
\end{figure*}

\section{\jwst/NIRSpec Observations and Kinematics}\label{sec:data}

\subsection{G235H/F170LP IFU Data}\label{ifu}

M81 was observed on 25 November 2022 with the \jwst/NIRSpec IFU using the high spectral resolution grating (G235H/F170LP; $R\sim2700$), covering the wavelength range between 1.66--3.17$\mu$m and a spectral sampling of 4~\AA\ per spectral pixel. This observation employed the NRSIRS2RAPID readout pattern with 14 groups per integration and one integration per exposure, with an effective integration time of 204 seconds and a total effective exposure time of 1,634 seconds. A four-point medium-cycling dither pattern was used to optimize spatial sampling. The raw data were processed using the \jwst\ calibration pipeline, specifically STScI pipeline version v1.14.0 \citep{Bushouse24_1.14.0} with the Calibration Reference Data System (CRDS) context \texttt{jwst\_1242.pmap}\footnote{\url{https://jwst-crds.stsci.edu}}. The final datacube was resampled at a pixel scale of 0\farcs1, approximately five times smaller than the SMBH's $r_{\rm SOI}$ (as discussed in Section~\ref{sec:intro}), and covers the central field of view (FoV) of $3\arcsec\times3\arcsec$. 

 \begin{figure*}
  \centering
   	\includegraphics[width=0.98\linewidth]{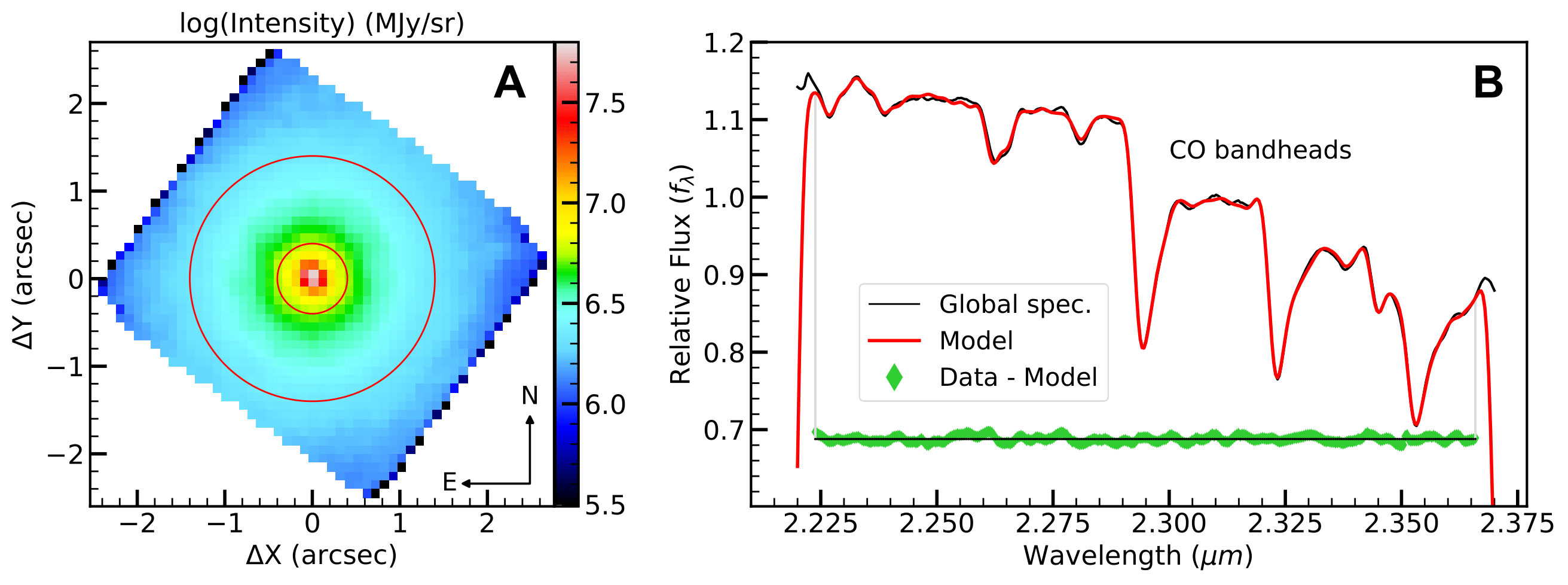} 
	\caption{{\it Panel A:} Logarithmically scaled intensity map of the NIRSpec G235H/F170LP data cube, collapsed along the spectral axis (excluding the detector gap at 2.41–2.49~\micron). The red circular annulus ($0\farcs4 < r < 1\farcs4$) indicates the region from which the global spectrum was extracted.  {\it Panel B:} The observed global spectrum (black) overlaid with the best-fit stellar template (red) for M81. The fit residuals ({\tt data - model}) are shown in green and vertically offset by +0.62 to compress the $y$-axis range and better illustrate the stellar CO absorption bandheads. This same vertical offset is applied to all subsequent figures of this type.}
	 \label{fig:Intensity_map}
\end{figure*} 

\subsection{Wiggles Corrections} \label{wiggle-correct}

Under-sampling of the NIRSpec PSF introduces sinusoidal artifacts, commonly referred to as “wiggles,” into individual spaxel spectra of bright sources. These features are largely mitigated when spectra are integrated over apertures of approximately $0\farcs2$–$0\farcs5$ \citep{Law2023}. Because the standard NIRSpec pipeline does not provide a built-in correction for these artifacts, we employed the empirical routine developed by \citet{Perna2023}\footnote{\url{https://github.com/micheleperna/JWST-NIRSpec_wiggles}}. The method is briefly summarized below. 

First, we extracted the spectrum from the brightest spaxel, where wiggles are most prominent, and masked intrinsic emission lines, absorption features, and the spectral gap between the two NIRSpec detectors. We then fit a sinusoidal model within rolling wavelength windows iteratively, allowing the fit to adapt locally to wavelength-dependent variations in wiggle characteristics as shown in Figure~\ref{fig:Wiggles}.  This model was calibrated against the spectrum extracted from a 4-pixel-radius ($0\farcs4$) circular aperture, centered on the peak of the white-light image, where wiggles are effectively smoothed out. We found that this aperture is sufficient to suppress wiggles in the NIRSpec IFS data cube of M81. Finally, the resulting wiggle frequency was adopted as a prior when subtracting the oscillations from all other affected spaxels.

All subsequent analyses were performed using the wiggle-corrected data cube.

\subsection{Defining the Galaxy Center}\label{sec:photo_cent}

Because the kinematic center is less precisely constrained due to its reliance on LOSVD extractions (Section~\ref{sec:kinematics}), we adopted the more accurate photometric center, defined as the weighted barycenter of the $\sim$150 brightest pixels, under the assumption that the black hole coincides with the surface-brightness peak. This photometric, AGN-aligned center was used as the origin in all models and figures.

\subsection{Stellar Kinematics Templates and \textsc{pPXF} Setup}\label{sec:kine}

As the nucleus of M81 is classified as a LLAGN \citep{Eracleous2010}, its nuclear spectra likely contain a substantial contribution from AGN continuum emission in several central spaxels (see Section~\ref{wiggle-correct}). This continuum emission dilutes the stellar CO bandhead absorption features used to trace stellar kinematics, resulting in underestimated velocity dispersions and reduced line strength \citep[$\gamma$; see Section 2.2 of][]{vanderMarel1993}. Additionally, a rising continuum can mimic changes in the stellar population, biasing the template mix and further reducing the reliability of the velocity-dispersion measurement.

Following \citet{Cappellari2009,Simon2024,Nguyen2025c}, we defined the observed line-strength as $\gamma={\rm stars}/({\rm stars + AGN})$, representing the fraction of flux attributed to the stellar template within the fitted spectral range. At large distances from the nucleus, $\gamma\approx1$ when continuum contamination is negligible. Toward the center, $\gamma$ decreases below unity and drops sharply in the innermost spaxels where the AGN continuum dominates.

Stellar kinematics were extracted from the \jwst/NIRSpec G235H/F170LP data cube of M81 using the Penalized PiXel-Fitting (\textsc{pPXF}) method\footnote{v8.2.1: \url{https://pypi.org/project/ppxf/}} \citep{Cappellari2004, Cappellari2017, Cappellari2023}, focusing on the commonly-used CO bandhead absorption features \citep[e.g.,][]{Cappellari2009, Nguyen2018, Krajnovic2018, Thater2023}. As demonstrated by \citet{Nguyen2025a, Nguyen2025c}, LOSVDs derived using the empirical X-shooter Spectral Library (XSL; 830 spectra from 683 stars\footnote{Data Release 3: \url{http://xsl.u-strasbg.fr}}; \citealt{Verro2022}) agree within 3\% of those obtained from the higher-resolution PHOENIX synthetic library \citep{Husser2013}. The LOSVD based on XSL is thus adopted as our primary kinematic measurements throughout our analysis. The XSL covers 3000–25,000 \AA\ at $R\sim10,000$, includes O–M and evolved stars, and has been corrected for extinction. 

\begin{figure*}
    \centering
    \includegraphics[width=0.88\linewidth]{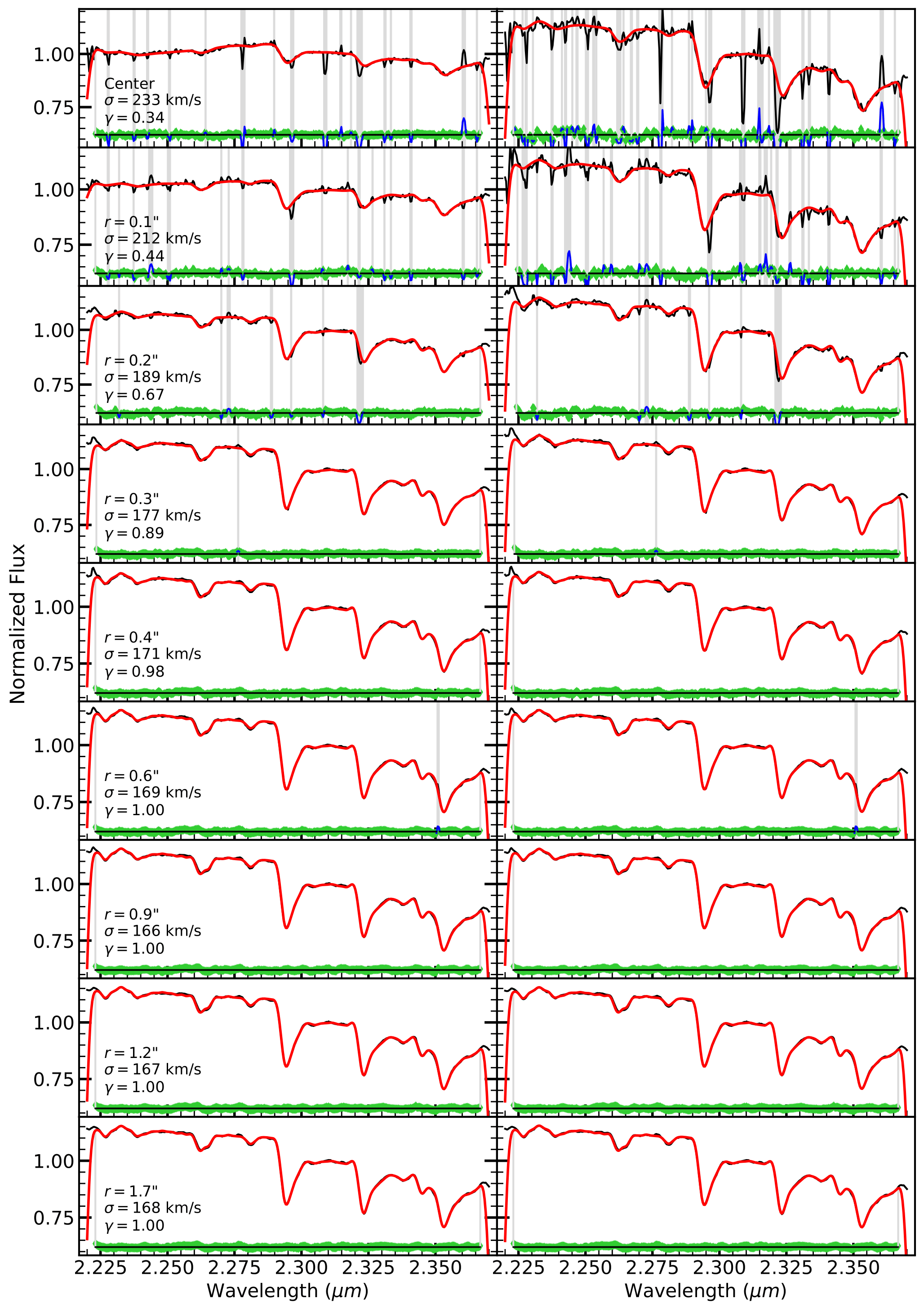}
    \caption{Radial spectral variation in M81 from the NIRSpec G235H/F170LP data cube. {\it Left column:} Observed spectra (black line) obtained by coadding spaxels within concentric annuli at radius $r$. The overplotted red line shows the \textsc{pPXF} fit, which combines the global stellar template (see left panel of Figure~\ref{fig:Intensity_map} and Section~\ref{sec:kine}), convolved with a Gaussian LOSVD and added by fourth-degree additive and multiplicative polynomials to account for the non-thermal nuclear continuum. Fit residuals ({\tt data-model}) are shown in green, with regions masked due to emission lines or artifacts indicated in gray. {\it Right column:} The convolved stellar template (red line) is compared to the observed spectrum after subtracting of the modeled non-thermal continuum (black line).} 
    \label{fig:annuli_spec}
\end{figure*}

We followed the method of \citet{Cappellari2009} to separate the stellar light from the non-thermal AGN continuum in the nucleus of M81. This process requires an optimal XSL template, derived from a large spatial region of the nucleus to avoid kinematic bias \citep{Marconi2000, Silge2003} caused by the AGN continuum. The stellar continuum was modeled using fourth-degree additive ({\tt degree = 4}) and multiplicative ({\tt mdegree = 4}) Legendre polynomials within \textsc{pPXF}, as this configuration provided the most stable LOSVD after testing various combinations of {\tt degree} and {\tt mdegree} values between 4 and 8. The resulting LOSVDs are nearly identical beyond 0\farcs2, with differences within this radius remaining below 7.5\%. These polynomials effectively capture low-order variations in line strength, residual sky-subtraction errors, spectral calibration imperfections, and AGN continuum contamination, allowing for robust measurements of the stellar velocity dispersion even in the central spaxels where AGN light dominates over the stellar CO bandheads.

\subsection{Determining the Optimal Stellar Template}\label{sec:optimal_template}

First, we constructed a global spectrum by combining all spaxels within an annulus of $0\farcs4<r<1\farcs4$ (confined by the two red rings in the left panel of Figure~\ref{fig:Intensity_map}) of the NIRSpec G235H/F170LP data cube, thereby excluding the central, AGN‐contaminated spaxels. This annular spectrum achieves a signal-to-noise ratio (S/N) of 300 per spectral pixel and was logarithmically rebinned along the spectral dimension with a constant velocity scale of  50 \kms\ per pixel calculated using Eq. 8 of \citet{Cappellari2017}. 

Second, we accounted for the instrumental broadening of the spectrum in the XSL templates by convolving them with a Gaussian whose dispersion is determined using Eq. 5 of \citet{Cappellari2017}. 

Third, we derived the fixed optimal XSL template by fitting the XSL-instrumental-broadened spectra to the global spectrum using \textsc{pPXF}, modeling the LOSVD as a simple Gaussian by setting {\tt moments = 2}. This setup returns the LOSVD parameters, including the rotational velocity $V$ (relative to the systemic velocity $V_{\rm sys}$) and the velocity dispersion $\sigma$, which are our primary kinematic quantities of interest.

The fixed optimal XSL template, returned by \textsc{pPXF}, is composed 14 giant-star spectra, which reproduce the observed spectrum over 2.22--2.37~\micron, including the CO bandhead absorption (right panel of Figure~\ref{fig:Intensity_map}). The fit is dominated by the K4III C star (HD 109871 from XSL), contributing $\approx$45\% of the flux, with additional contributions from a giant M2V C star (HIP 75423, $\approx$18\%) and an M9 D star (BMB 289, $\approx$10\%) required to match the data. From this template, \textsc{pPXF} yields $V = -31 \pm 3$ \kms\ and $\sigma = 168 \pm 4$ \kms, consistent with long-slit measurements at Calar Alto Observatory \citep{Bender1994}. The 1$\sigma$ uncertainties are estimated from the standard deviation of 200 Monte Carlo realizations \citep{Cappellari2004}.

\subsection{Separating AGN Continuum from Stellar Light}\label{sec:non-thermal}

In this section, we assess the reliability of the fixed optimal stellar template (Section~\ref{sec:kine}) for extracting stellar kinematics in the presence of non-thermal AGN emission within the NIRSpec G235H/F170LP data cube. To do so, we constructed high-S/N spectra by coadding spaxels within multiple concentric annuli spanning radii from 0\farcs1 to 1\farcs7. Each annular spectrum was then fit using the fixed optimal stellar template within the \textsc{pPXF} framework.

Figure~\ref{fig:annuli_spec} shows the radial profiles of stellar line strength ($\gamma$, left) and stellar-light fraction (right). The observed spectra in each annulus are modeled with the fixed optimal stellar template plus fourth-degree additive and multiplicative Legendre polynomials to account for AGN continuum emission. The stellar-light fraction (right panels) is obtained by subtracting the AGN component, revealing a clear non-stellar contribution from the changing spectral slope. In the nucleus, the AGN contributes 66\% of the flux, decreasing to $\approx$54\% at $r \approx 0\farcs1$, $\approx$33\% at $r \approx 0\farcs2$, and $<$10\% beyond $r = 0\farcs2$. 

\begin{figure}
    \centering\includegraphics[width=0.98\linewidth]{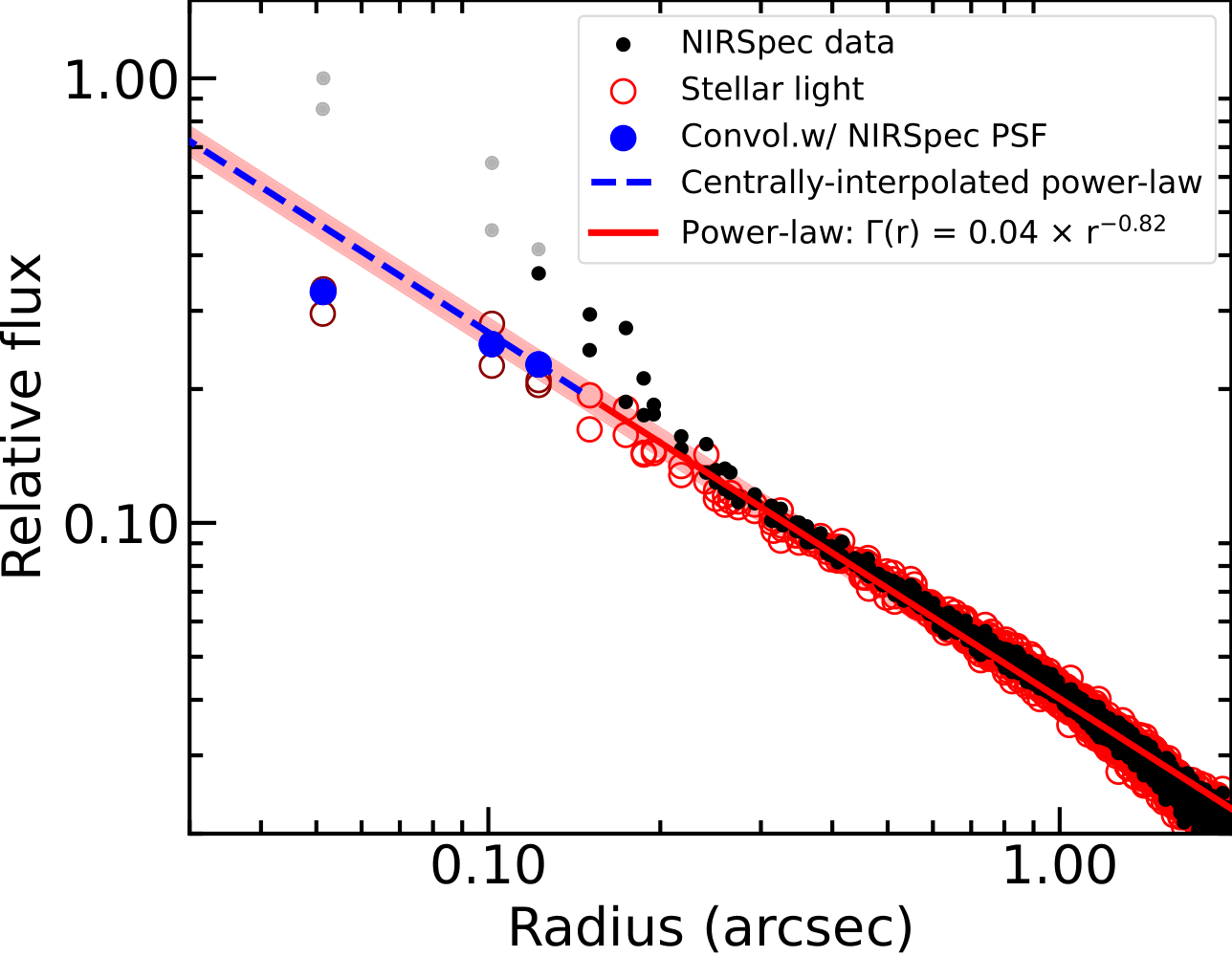}
    \caption{The radial surface brightness profile $I(r)$, derived from individual \textsc{Voronoi} bins in the NIRSpec G235H/F170LP data cube of M81 (filled black circles), is compared to the estimated stellar light profile $\Gamma(r) = I(r)\gamma(r)$ (red open circles).  The underlying stellar distribution is smooth and well approximated by a single-power-law (red line with pink region shows its $1\sigma$ uncertainty). The PSF–convolved central interpolation of this single power-law also reproduces the two innermost $\gamma$ measurements, which lie below the intrinsic (unconvolved) profile, indicating that their apparent decline is fully consistent with PSF effects. Gray points indicate pixels where $\gamma < 0.5$.} 
    \label{fig:stellar_perc}
\end{figure}

\begin{figure*}
    \centering\includegraphics[width=0.98\textwidth]{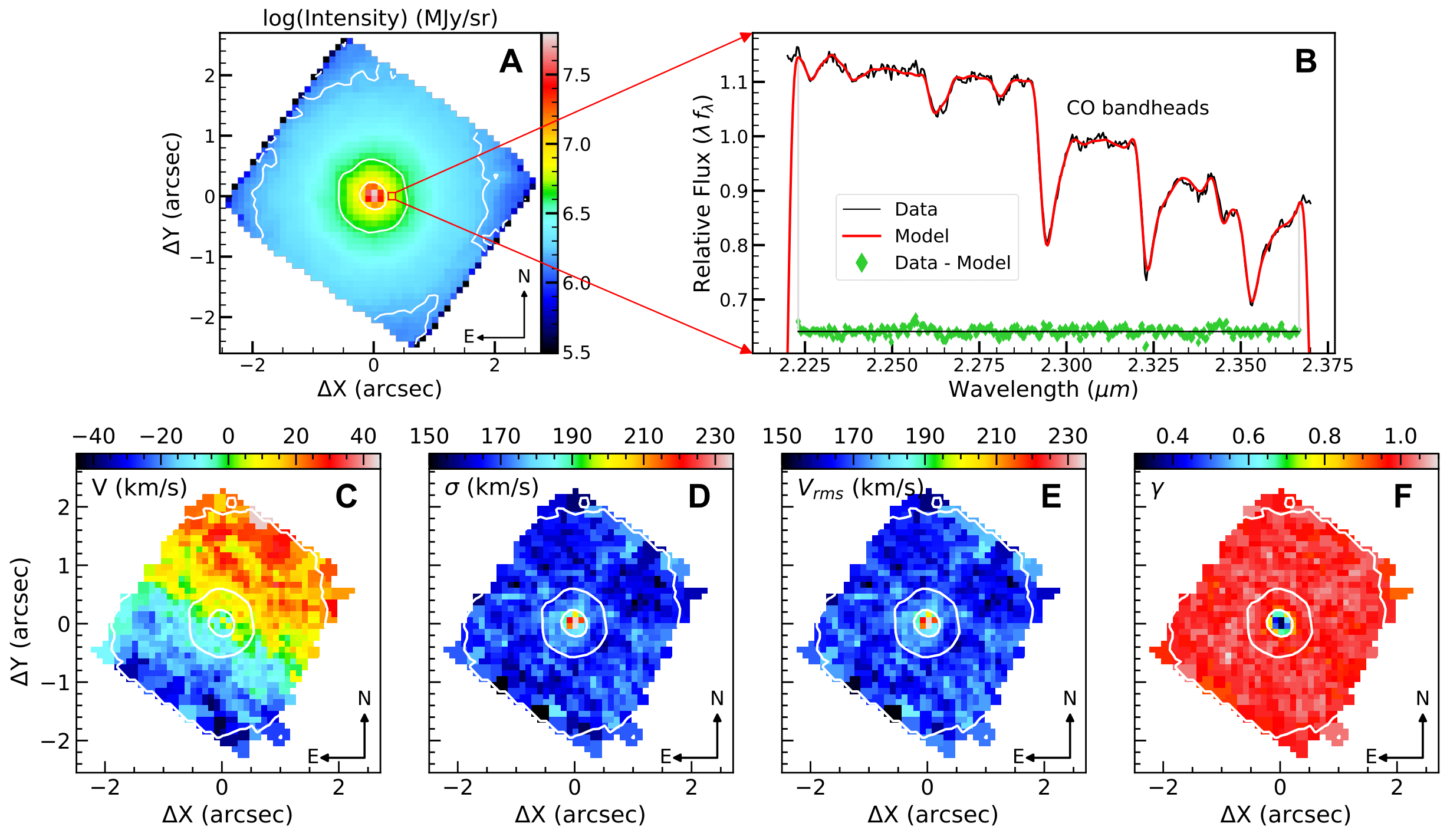}
    \caption{Stellar kinematic measurements from the NIRSpec G235H/F170LP observations of NGC~3031 are shown. {\it Panel A:} The logarithmic collapsed intensity map (cf.\ left panel of Figure~\ref{fig:Intensity_map}). {\it Panel B:} Example \textsc{pPXF} fit for a central-offset \textsc{Voronoi} bin unaffected by wiggles (location indicated in Panel A).  The observed spectrum (black) displays the CO bandhead absorptions near 2.3 $\mu$m, with the best-fit XSL template overplotted in red. Fit residuals ({\tt data-model}) are plotted in green, and the vertical gray lines mark the wavelength range used for fitting across all bins. {\it Panels C--F:} 2D maps of the of the stellar rotation velocity ($V$), velocity dispersion ($\sigma$), root-mean-square velocity ($V_{\rm rms} = \sqrt{V^2 + \sigma^2}$), and stellar light contribution fraction ($\gamma$). The central white pixels are masked due to unreliable kinematics, as the AGN continuum contributes $\approx$66\% of the total light in this region.  White contours trace the intensity, decreasing by mag arcseconds$^{-1}$ steps from the center.}
    \label{fig:kin-map}
\end{figure*}

We quantify the dilution of stellar light by the central nonthermal AGN continuum in Figure~\ref{fig:stellar_perc}, which presents the radial surface-brightness profile $I(r)$ measured from the \jwst/NIRSpec G235H/F170LP data cube of M81, together with the stellar contribution $\Gamma(r)=I(r)\gamma(r)$ derived from the \textsc{pPXF} fits in individual spatial bins. Similar analyses have been carried out for nearby systems such as M87 \citep[Figure~4 of][]{vanderMarel1994}, Centaurus~A \citep[Figure~6 of ][]{Cappellari2009}, and more recently M106  \citep[Figure~4 of][]{Nguyen2025c}. As an initial approximation in Figure~\ref{fig:stellar_perc}, we modeled the radial stellar-light distribution using a single power-law form, $\Gamma(r)=\alpha\times r^{\beta}$ (red line), with $\alpha=0.04$ and $\beta=-0.82$. This simple model overestimates the stellar-light fraction within $0\farcs1$ (blue dashed line), as it neglects the effect of the \jwst/NIRSpec PSF at 2.3~$\mu$m.

To assess the impact of PSF convolution, we performed a forward-modeling experiment. First, we generated a synthetic image with an intrinsic surface-brightness profile $\Sigma\propto r^{-0.82}$, sampled at $0\farcs025$ pixel$^{-1}$, corresponding to a factor of four oversampling relative to the native NIRSpec spatial scale \citep[PSF~1 in Table~1 of][]{Nguyen2025c}. This image was then convolved with the synthetic NIRSpec PSF at the same sampling and subsequently rebinned to the native scale of $0\farcs1$ pixel$^{-1}$ by summing $4\times4$ spaxels. We then measured the stellar fraction in the central spaxel of the rebinned image.

The resulting central stellar fraction follows the same trend defined by the three innermost measurements of $\gamma(r)$ in Figure~\ref{fig:stellar_perc}, demonstrating that the apparent break toward the nucleus is fully accounted for by PSF convolution and spatial sampling effects. This behavior does not imply a genuine change in the intrinsic stellar surface-brightness profile within the NIRSpec FoV, but rather reflects instrumental effects that are properly accounted for once the PSF is taken into consideration. The determination of $\gamma(r)$ therefore enables a robust extraction of stellar kinematics even in the unresolved nuclear region, where stellar light is strongly diluted by AGN emission.

Although stars contribute only $\sim$34\% of the total flux in the central spaxel, the resulting kinematics remain usable, albeit with larger systematic uncertainties of order $\sim$15\%, and are thus retained in our dynamical modeling (Section~\ref{sec:jam}). At radii beyond $r=0\farcs1$, the stellar contribution increases and the kinematic measurements are correspondingly more secure, with typical uncertainties below 3\%.

\subsection{Two-Dimensional Kinematic Maps}\label{sec:kinematics}

The 2D stellar LOSVD of the nucleus of M81 was derived following the fitting procedure described in Section~\ref{sec:kine}. First, the adaptive \textsc{Voronoi} binning method\footnote{v3.1.5: \url{https://pypi.org/project/vorbin/}} \citep{Cappellari2003} was applied to spatially group spaxels until a target S/N of 150 per spectral pixel was achieved, resulting in $N=700$ \textsc{Voronoi} bins. This target S/N represents a compromise: spaxels beyond $r \approx 0\farcs5 \approx r_{\rm SOI}$ were grouped, while those within this radius remained unbinned to preserve the spatial resolution necessary for precise stellar kinematic measurements inside $r_{\rm SOI}$. Within the SOI, individual unbinned spaxels exhibit S/N values ranging from 150 to 200 per spectral pixel, increasing toward the central peak. Each binned spectrum was resampled onto a logarithmic wavelength scale. The \textsc{pPXF} routine was then used to fit each binned spectrum with the fixed optimal XSL-instrumental-broadened template (Section~\ref{sec:optimal_template}), incorporating fourth-degree multiplicative and additive Legendre polynomials to derive the LOSVD (i.e., $V$ and $\sigma$).

\begin{table}[]
\centering
\caption{JWST/NIRSpec Kinematic Data of the M81 Nucleus}
\vspace{-3mm}
\begin{tabular}{cccccccc}
\hline \hline
$\Delta$R.A. & $\Delta$Decl. & $V$ & $\Delta V$ & $\sigma$ & $ \Delta\sigma$ \\
($\arcsec$)&($\arcsec$)&(\kms)&(\kms)&(\kms)&(\kms) & \\
\hline
0.012      &    0.050 &    7.66    & 14.16 & 233.77 & 14.05 \\ 
0.112      &    0.050 & $-$9.08  & 6.79 & 216.17 & 8.66 \\ 
0.012      &$-$0.150&$-$11.06 & 5.02 & 180.76 & 6.33 \\ 
0.212      &    0.050 &    10.43   & 3.76 & 179.11 & 4.72 \\ 
$-$0.188 &    0.150 &    8.23    & 3.82 & 188.36 & 4.80 \\ 
$-$0.088 &$-$0.250& $-$12.29 & 3.46 & 174.71 & 4.33 \\ 
$-$0.288 &$-$0.050& $-$9.06   & 3.29 & 172.81 & 4.12 \\ 
...            &   ...    &   ...    & ...   &  ...   & ...  \\ 
\hline
\end{tabular}
\noindent\parbox{\linewidth}{(This table is available in its entirety in machine-readable form. A portion is shown here for guidance regarding its form and content. A data copy is also available in Zenodo at \dataset[doi: 10.5281/zenodo.18345300]{https://doi.org/10.5281/zenodo.18345300})}
\label{tab:kin}
\end{table}

Figure~\ref{fig:kin-map} presents the logarithmic integrated intensity map (panel A), with a marked spaxel corresponding to the spectrum and \textsc{pPXF} fit shown in panel B. The remaining panels display the resulting maps of rotational velocity $V$ (panel C), velocity dispersion $\sigma$ (panel D), root-mean-square velocity $V_{\rm rms} = \sqrt{V^2 + \sigma^2}$ (panel E), and stellar line strength $\gamma$ (panel F). The nucleus exhibits a steeply rising rotation, with $\vert V \vert \approx 40\pm 4$~\kms\ at the edge of the NIRSpec IFU field (after subtracting the systemic heliocentric velocity of $v_{\rm sys} = -32 \pm 4$~\kms), consistent with the IFS observations by \citet{Muller2011}, as expected for a spiral galaxy. This derived $v_{\rm sys}$ from the \jwst/NIRSpec G235H/F170LP data cube is 7 \kms\ larger than thar value reported in the NASA/IPAC Extragalactic Database (NED\footnote{\url{https://ned.ipac.caltech.edu}}).

We measure a global kinematic position angle of PA$_{\rm kin} = 151.6^\circ \pm 9.5^\circ$. Both $v_{\rm sys}$ and PA$_{\rm kin}$ were derived from the velocity map using the \textsc{pafit}\footnote{v2.0.8: \url{https://pypi.org/project/pafit/}} package \citep{Krajnovic2006}. The kinematic PA$_{\rm kin}$ is consistent within the uncertainties with the photometric one PA$_{\rm phot}\approx157^\circ$ reported in the HyperLeda database\footnote{\url{https://leda.univ-lyon1.fr/}} \citep{Paturel2003}, as expected for a nearly axisymmetric disk galaxy. The $\sigma$ profile, which dominates over $V$, rises from $\sim$$165 \pm 4$~\kms\ at $r \gtrsim 1\arcsec$ ($\approx$3~\kms\ higher than the value obtained from GMOS observations by \citealt{Muller2011}), to a central peak of $\sim$$233 \pm 15$~\kms\ within $r \lesssim 0\farcs1$. This central increase in both $\sigma$ and \vrms\ provides strong evidence for the presence of a central SMBH. We presented these stellar-kinematic measurements in Table~\ref{tab:kin}.

As shown in panel~F of Figure~\ref{fig:kin-map}, the non-thermal AGN continuum exhibits only weak emission, with $\gamma \approx 1$ over most of the NIRSpec FoV, except within the innermost spaxels where $\gamma \leq 0.34$. In these central spaxels, the AGN light dominates, resulting in relatively large kinematic uncertainties of up to $\sim$15\%. In contrast, for the remaining bins—where the stellar light contributes more than 34\% of the total flux—the derived stellar kinematics are robust, with uncertainties below 3\%.

\begin{table}
\centering
\caption{ \jwst/NIRSpec AGN MGE PSF model at 2.3 $\mu$m and \hst\ {\tt Tiny Tim} WFC3/F110W MGE PSF model \label{tab:mge_psf}}
\vspace{-3mm}
\begin{tabular}{cccc}
\hline\hline
$j$&(Light fraction)$_j$&$\sigma_j~({\rm arcsec})$&FWHM$^{\rm tot}_{\rm PSF}~({\rm arcsec})$ \\
(1) & (2) & (3) & (4) \\
\hline
   &           &\underline{PSF$_{\rm AGN}$}&\\
1 & 0.770 & 0.080 & \multirow{2}{*}{0.210} \\
2 & 0.230 & 0.128 &  \\
\hline
   &           &\underline{PSF$_{\rm HST}$}& \\
1 & 0.442 & 0.028 & \multirow{4}{*}{0.131} \\
2 & 0.389 & 0.094 & \\
3 & 0.099 & 0.330 & \\
4 & 0.070 & 0.825 & \\
\hline
\end{tabular}
\noindent\parbox{\linewidth}{\textbf{Notes.} The first column gives the number of Gaussian components.  The second column reports the light fraction of each Gaussian.  The third column lists the Gaussian dispersions along the major axis. The fourth column lists the representative average FWHM of the \hst\ photometric PSF, computed from their Gaussian components.}
\end{table}

\subsection{NIRSpec AGN PSF at 2.3 $\mu$m}\label{agn_psf}

\begin{figure}
     \centering\includegraphics[width=0.9\linewidth]{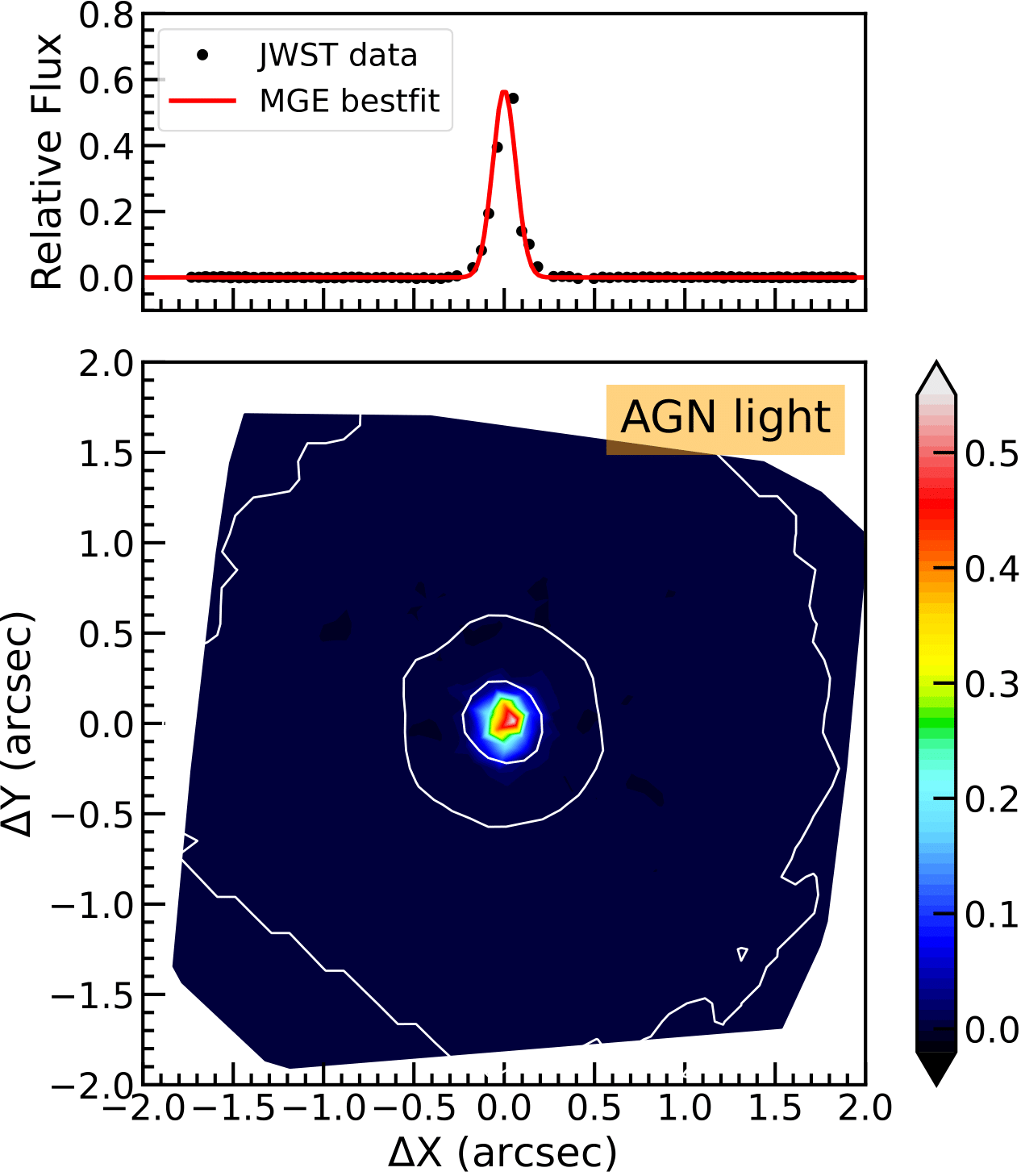}
     \caption{{\it Upper panel:} The 1D radial profiles of the spectrally extracted AGN signal from the \jwst/NIRSpec G235H/F170LP data cube, shown together with the best-fitting MGE model of the PSF. The profile is constructed by mapping each data point to its projected radius, with negative and positive values corresponding to positions on opposite sides of the nucleus. {\it Lower panel:} The 2D image of the extracted AGN light distribution displayed on a logarithmic scale.} 
     \label{fig:agn_psf}
\end{figure}

From panel F of Figure~\ref{fig:kin-map}, we derive the AGN light–fraction map as $1-\gamma$, where $\gamma$ denotes the stellar light fraction. The AGN light–distribution map, shown in the lower panel of Figure~\ref{fig:agn_psf}, is then obtained by multiplying this fraction map by the integrated-light map from the \jwst/NIRSpec G235H/F170LP data cube. We parameterize the resulting AGN light distribution using a multi-Gaussian expansion \citep[MGE;][]{Emsellem1994}, implemented with the \texttt{mge\_fit\_sector} routine from the \textsc{MgeFit} package \citep{Cappellari2002}. The 2D AGN light distribution is expressed analytically as 
\[
{\rm PSF}_{\rm AGN}(r) = \displaystyle \sum_{j=1}^n \dfrac{F_j}{2\pi \sigma_j^2} \exp\left(-\dfrac{r^2}{2\sigma_j^2}\right), 
\]
where $r$ is the radial coordinate, $\sigma_j$ is the dispersion of Gaussian component $j$, and $F_j$ is its normalized weight, with $\sum_{j=1}^{n} F_j = 1$. The best-fit MGE PSF model for the AGN light distribution comprises two Gaussians, which have an average FWHM of $0\farcs21$, as listed in Table~\ref{tab:mge_psf} and shown in the upper panel of Figure~\ref{fig:agn_psf}. This value is consistent with the STPSF synthetic \jwst/NIRSpec PSF reported in Table~1 of \citet{Nguyen2025c} and with the empirical PSF derived by \citet{Bentz2025} from observations of an isolated late-type star. We adopt this PSF model in this work to derive the black hole mass of M81 (Section~\ref{sec:jam_grid}).

\begin{figure}
     \centering\includegraphics[width=0.9\linewidth]{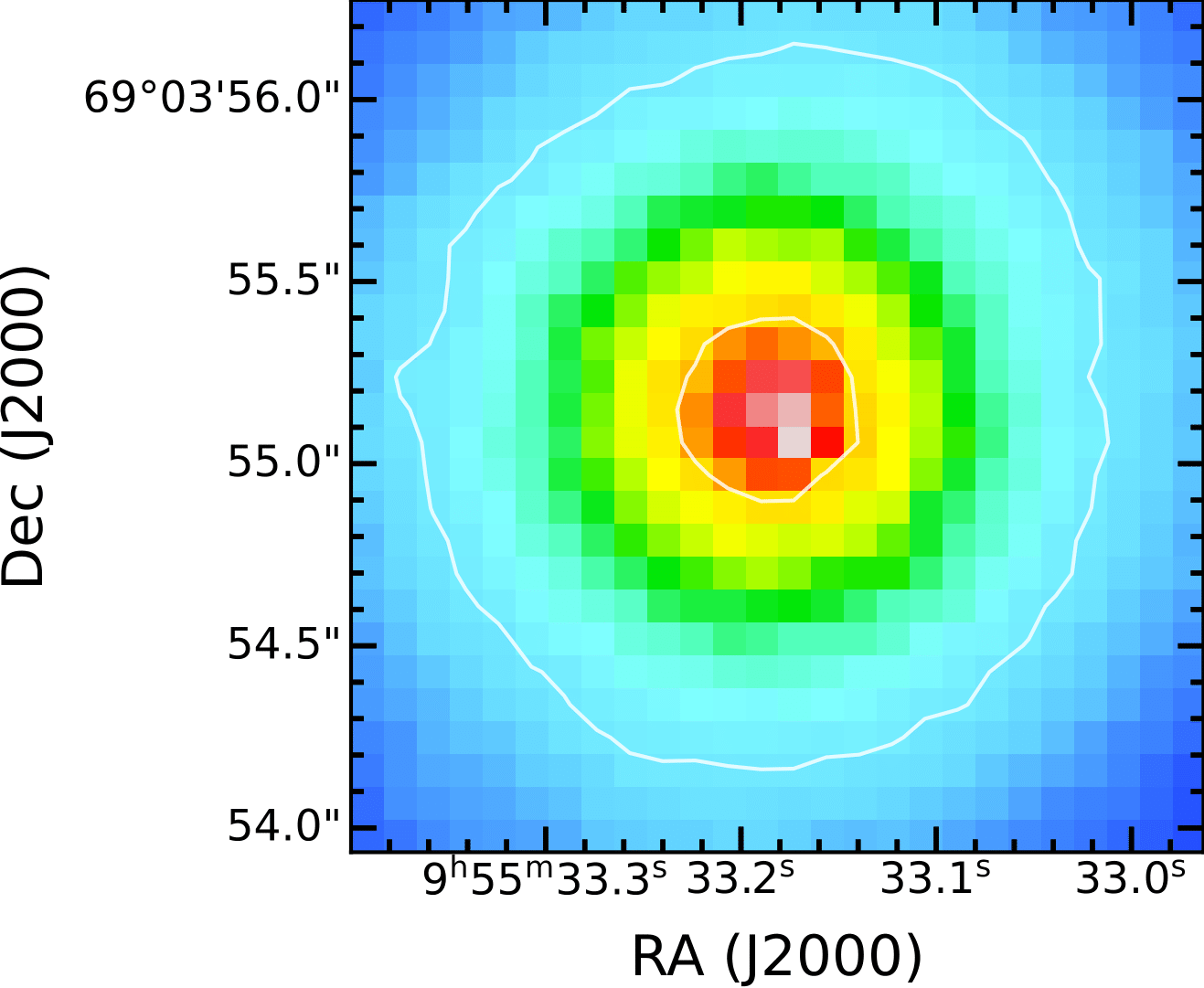}
     \caption{The central region of the \hst/WFC3 F110W image reveals a peak in surface brightness that is offset by approximately $0\farcs1$ to the southwest of the M81 nuclear centroid.} 
     \label{fig:nucleus_offset}
\end{figure}

\section{Imaging and Photometric model}\label{sec:hst}

Accurate dynamical modeling and \Mbh\ measurements require a well-constrained estimate of the gravitational potential of the galaxy. We derived the stellar potential by scaling the galaxy’s luminosity with a mass-to-light ratio (\ml). To model the stellar luminosity distribution with high fidelity, we combined high-resolution narrow-field space-based \hst\ imaging with wide-field ground-based data from the 2MASS survey.

 \begin{figure*}
    \centering
    \includegraphics[width=0.98\linewidth]{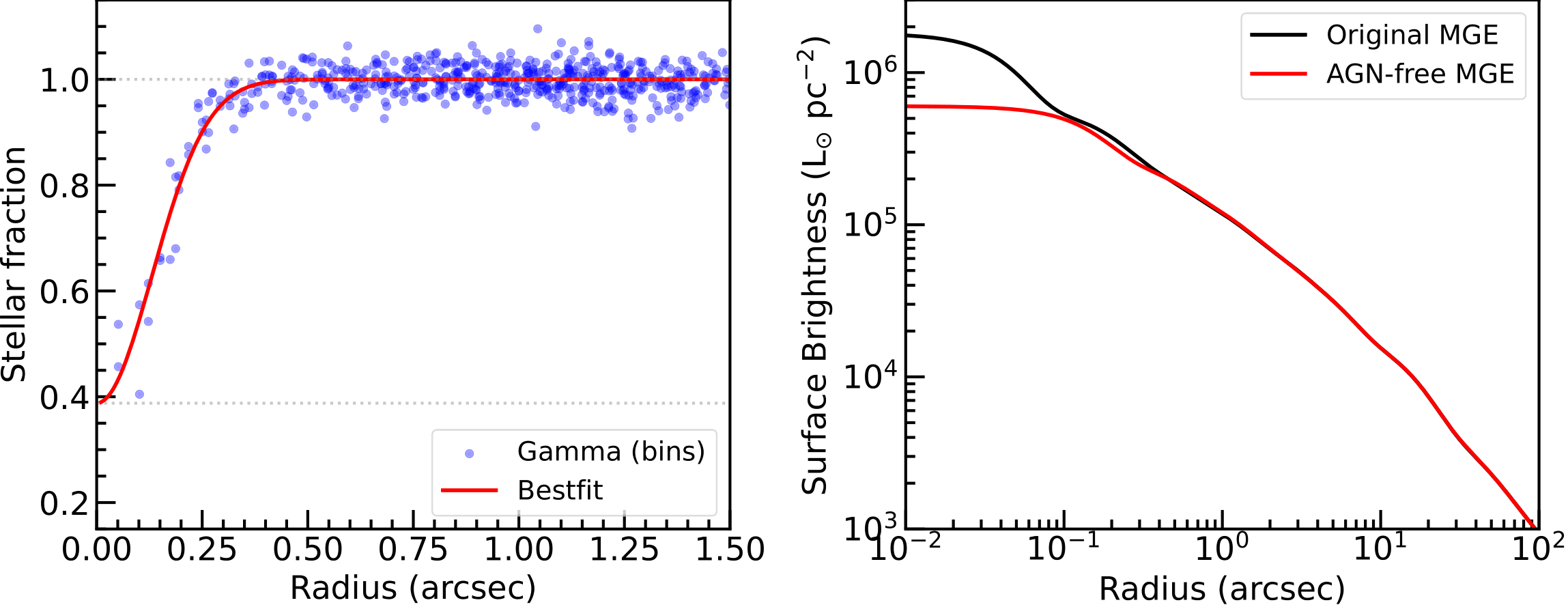}
    \caption{\textit{Left:} The stellar light fraction distribution for all kinematic bins measured in Section~\ref{sec:kinematics} and extracted from panel F of Figure~ \ref{fig:kin-map}. The red curve represents the best-fit profile, which dominates beyond 0\farcs3 but declines sharply toward the galaxy center. \textit{Right:} Comparison between the scaled \hst\ radial profile (black line) and the profile constrained by the stellar contribution (red line) measured spectroscopically with \jwst/NIRSpec G235H/F170LP.} 
    \label{fig:stars}
\end{figure*}

\begin{figure*}
    \centering
    \includegraphics[width=0.98\linewidth]{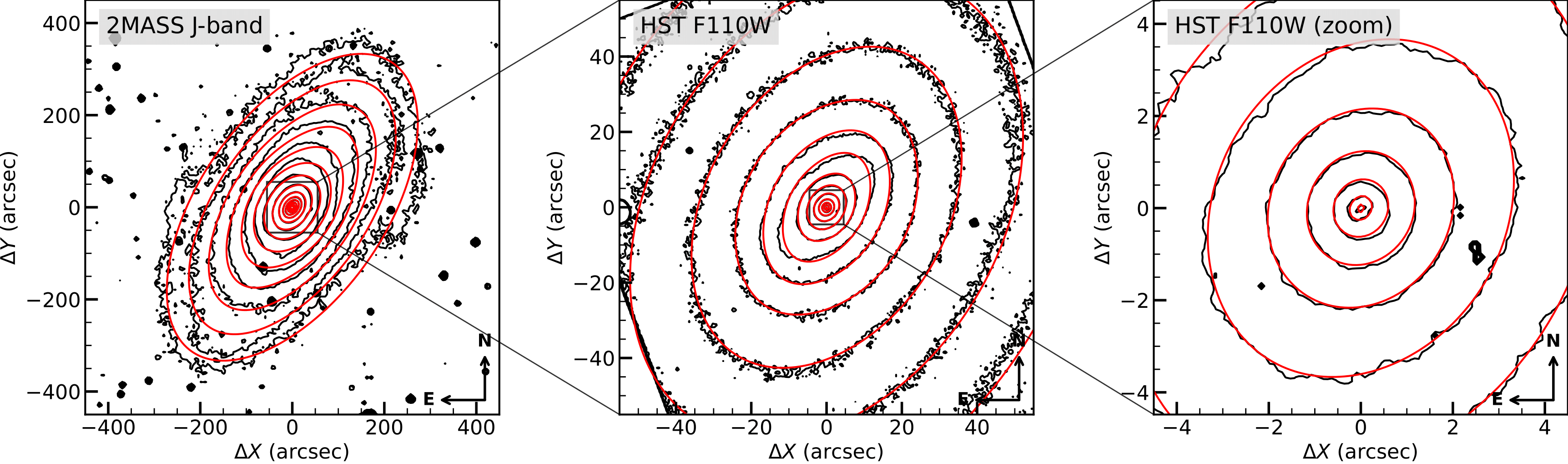}
    \caption{\textit{Left:} Isophotes of M81 from the 2MASS $J$-band image (black contours) and the corresponding MGE model (red contours) are shown over a  $1000\arcsec \times 1000\arcsec$ FoV. \textit{Middle:} The same comparison is presented for the \hst/WFC3/IR F110W image, covering a $110\arcsec \times 110\arcsec$ FoV. \textit{Right:} Zoom-in of the central $10\arcsec \times 10\arcsec$ region of the \hst\ image. The MGE model accurately reproduces the observed isophotes across all spatial scales.  Contours represent decreasing surface brightness levels, with a spacing of 1~mag arcsec$^{-2}$.}
    \label{fig:mge}
\end{figure*}

\begin{table}
\centering
\caption{Gaussian Parameters of the M81 MGE Model \label{tab:mge}}
\begin{tabular}{cccc}
\hline
$j$&lg$\Sigma_{\star,j}~(\mathrm{L_\odot~pc^{-2}})$&lg$\sigma_j$~(arcsec)&$q_j' (=b/a)$\\
(1) & (2) & (3) & (4) \\
\hline
1 & 5.511 & $-$0.940 & 0.727 \\
2 & 5.034 & $-$0.468 & 0.900 \\
3 & 4.873 & $-$0.107 & 0.900 \\
4 & 4.610 & $-$0.194 & 0.900 \\
5 & 4.172 &    0.543   & 0.681 \\
6 & 4.280 &    0.569   & 0.900 \\
7 & 4.007 &    1.020   & 0.550 \\
8 & 3.630 &    1.112   & 0.843 \\
9 & 3.234 &    1.374   & 0.621 \\
10& 3.207 &    1.516  & 0.717 \\
11& 3.070 &    1.754  & 0.766 \\
12& 2.783 &    2.078  & 0.500 \\
13& 2.353 &    2.391  & 0.500 \\
14& 1.497 &    2.468  & 0.900 \\
\hline
\end{tabular}
\noindent\parbox{\linewidth}{\textbf{Notes.} The first column gives the number of Gaussian components.  The second column reports the surface brightness of each Gaussian.  The third column lists the Gaussian dispersions along the major axis. The fourth column provides the corresponding axial ratios. }
\end{table}

\subsection{HST Observation and Its PSF Image}\label{hst_pfs} 

We modeled the stellar light distribution of M81 within a $110\arcsec\times110\arcsec$ field using \hst\ imaging obtained on 24 June 2009 with the WFC3/IR channel, which has a pixel scale of $0\farcs09$, and the broad-band F110W filter. The observation consists of a single exposure with a total on-source integration time of 143 seconds.

We required a PSF to deconvolve the \hst\ image into the intrinsic stellar light distribution (Section~\ref{sb}). We generated a synthetic PSF using the {\tt Tiny Tim} software package\footnote{\url{https://github.com/spacetelescope/tinytim}} \citep{Krist1995, Krist2011}, specifically employing the {\tt tiny1} and {\tt tiny2} routines, which simulate the PSF for a given filter, detector chip, chip position, and instrument configuration. To ensure consistency with the actual observations, the PSF was generated using the same exposure parameters as the \hst\ image, and was subsequently convolved with an appropriate charge diffusion kernel to account for the redistribution of electrons into neighboring pixels on the detector. 

\subsection{2MASS}\label{2mass} 

To characterize the stellar light distribution of M81 beyond the central $110\arcsec \times 110\arcsec$ region, we used a large-scale $J$-band image from the 2MASS survey, which covers a wavelength range of 1.12–1.36$\mu$m and has a central wavelength of 1.235~$\mu$m \citep{Jarrett2003}. The 2MASS image has a pixel scale of $1\arcsec$ per pixel.

\subsection{Center Offset in the HST F110W image?}\label{central_offset} 

We discovered that the peak brightness in the \hst\ image appears to be offset $\sim$0\farcs1 southwest from the centroid of the M81 nucleus as seen in Figure~\ref{fig:nucleus_offset}. Given that we are observing this offset from the IR F110W image, it is unlikely that this feature is caused by dust extinction. The S/N in the \hst\ image is high enough to verify that the AGN peak agrees well with the offset peak in the \hst\ image. Thus, this offset could be the morphological evidence that the SMBH is not yet settled in the center.  There are not many galaxies that are close enough for us to resolve such an offset between the SMBH's location and the photometric center, except for the well-known example of M31 nucleus \citep{Lauer2005} and the recent same discovery in the nucleus of NGC 4486B at the Virgo cluster \citep{Tahmasebzadeh2025}. 

Spatial or kinematic offsets between the location of the SMBH and the photometric or dynamical center of a galaxy have been reported in several nearby systems and interpreted in a variety of contexts. A classic example is the double nucleus of M31, in which the brighter off-center component (P1) and the true nucleus (P2) can be understood as asymmetric stellar structures in an eccentric disk around the central BH \citep{Lauer2005}, and measurements indicate a small displacement of the SMBH from the bulge center consistent with this model \citep[e.g.,][]{Lauer1998, Kormendy1999}. Similarly, early \hst\ studies suggested a projected offset of $\sim$0\farcs1 in M87 between the nuclear point source and the galaxy photocenter, though subsequent work attributed much of this apparent displacement to temporal jet variability rather than a true SMBH displacement \citep[e.g.,][]{Batcheldor2010}. Searches for recoiling or displaced SMBHs in samples of nearby early-type galaxies have also found modest ($\lesssim$10~pc) AGN-photocenter separations in a fraction of objects, which can be explained by post-merger recoil oscillations or asymmetric jet acceleration \citep[e.g.,][]{Lena2014}. Observations of velocity-offset AGN emission lines further suggest that kinematic displacements in ionized gas can arise from outflows, dual nuclei, or eccentric nuclear structures \citep[e.g.,][]{Muller-Sanchez2016}. More recent cosmological simulations indicate that off-center SMBHs, particularly in brightest cluster galaxies, may be common outcomes of hierarchical assembly, with roughly one third of systems showing significant displacements at $z=0$ due to merger dynamics \citep[e.g.,][]{Chu2022}. These studies together demonstrate that while exact SMBH–galaxy center coincidence is the norm, small offsets do occur and can provide clues to recent dynamical history beyond simple noise or measurement uncertainty.

\subsection{Separating Central AGN Emission from Stellar Light in HST image}\label{separating_agn} 

Our galaxy mass model of M81 must represent the stellar light distribution only, free from contamination by the central AGN. To achieve this, we corrected the \hst/WFC3 IR F110W image by modeling and subtracting the AGN point source before fitting the MGE. Following the approach of \citet{Simon2024}, we estimated the AGN's spatial profile from the NIRSpec stellar fraction map ($\gamma$; panel F of Figure~\ref{fig:kin-map}), assuming minor differences between the \hst/WFC3 IR F110W and \jwst/NIRSpec G325H/F170LP PSFs are negligible. The fractional contribution of the AGN at a given radius is $(1-\gamma)$. We modeled this radial profile by assuming the AGN light distribution can be approximated by a Gaussian function, which leads the purely stellar contribution to the form:
\[
\gamma(r) = 1 - \Delta\gamma\cdot\exp\left(-\frac{r^2}{2\sigma^2}\right), 
\]
where the best fit values are $\Delta\gamma = 0.612$ and $\sigma = 0\farcs131$ (left panel of Figure~\ref{fig:stars}). Using this fit, we constructed a modified (i.e., AGN-free) \hst/WFC3 IR F110W image in which the innermost sub-arcsecond region was replaced with the AGN-free stellar image. The fitted profile was then matched to the original \hst/WFC3 IR F110W surface brightness profile (i.e., AGN + stars) beyond 0\farcs3 (right panel of Figure~\ref{fig:stars}).

\subsection{Full Surface Brightness Profile and Galaxy Mass Model of M81}\label{sb} 

Before constructing the 2D stellar surface-brightness distribution of M81, we estimated and subtracted the sky background for the \hst/WFC3 IR F110W and 2MASS $J$-band images independently. For each image, the sky level was determined as the median signal measured within several $10 \times 10$ pixel$^{2}$ boxes placed in source-free regions at projected distances greater than $40\arcsec$ (for the \hst/WFC3 image) and $400\arcsec$ (for the 2MASS image) from the galaxy center. These median values were then subtracted from the corresponding images to produce sky-subtracted frames used in the subsequent analysis. Regions affected by dust obscuration and foreground stars were masked in both the \hst/WFC3 IR F110W and 2MASS $J$-band images before the fitting procedure.

We then constructed the 2D stellar surface-brightness distribution of M81 using the \texttt{mge\_fit\_sectors\_regularized} routine from the \textsc{MgeFit} package (see footnote 15), fitting the sky-subtracted \hst/WFC3 IR F110W and 2MASS $J$-band images simultaneously.   During the fitting process, we accounted for the \hst/WFC3 IR F110W PSF to recover the galaxy’s intrinsic light distribution. Specifically, the PSF image was first modeled as an MGE, whose parameters are listed in Table~\ref{tab:mge_psf}, with a FWHM of 0\farcs131, which was then convolved with the \hst/WFC3 IR F110W image during the fit. This step is essential, as the accuracy of the \Mbh\ measurement depends critically on how well the central stellar mass distribution is modeled.  

The procedure consisted of the following steps:

\begin{itemize}
	\item  We extracted multi-sector surface-brightness profiles from the sky-subtracted \hst/F110W image (describing the inner regions) and from the sky-subtracted 2MASS $J$-band image (describing the outer regions). The central region of the sky-subtracted 2MASS image overlapping with the \hst\ FoV was masked to avoid double-counting and to prevent the lower-resolution data from influencing the inner profile.
	
	\item We placed the two datasets on a consistent photometric scale. We used the \texttt{PhotometryMatch} routine to determine the relative flux offset and multiplicative flux scaling factor between the sky-subtracted \hst\ and 2MASS profiles. These parameters were derived by minimizing photometric residuals in an overlap annulus between 5\arcsec and 50\arcsec, where both datasets have a reliable signal. This step accounts for differences in pixel scale, absolute calibration, and background subtraction between the two images.
	
	\item  We combined the sky-subtracted  \hst\ radial photometry at radii $r<50\arcsec$ with the sky-subtracted 2MASS photometry at $r>50\arcsec$ into a single, continuous surface-brightness profile. 
	
	\item We then applied the MGE fit to this combined profile, with PSF convolution applied only to the \hst\ component. The resulting MGE therefore represents a joint fit to the inner high-resolution \hst\ data and the outer wide-field 2MASS data, ensuring a smooth and physically consistent stellar mass model across all radii.
\end{itemize}

Next, the flux unit of counts/second was converted to physical units of \Lsun~pc$^{-2}$ following the procedure outlined by \citet{Thatte2009}, adopting photometric zero-points and absolute magnitudes for the Sun in the Vega system \citep{Willmer2018}. For the \hst\ image, we used $m_{\rm 0, F110W} = 26.05$ mag\footnote{\url{https://www.stsci.edu/files/live/sites/www/files/home/hst/instrumentation/wfc3/documentation/instrument-science-reports-isrs/_documents/2024/WFC3-ISR-2024-13.pdf}} and $M_{\rm Vega, F110W}=3.79$ mag; for the 2MASS image, $m_{0, J} = 20.92$ mag and $M_{{\rm Vega,} J}=3.67$ mag were adopted. Galactic extinction corrections of $A_{\rm F110W} = 0.071$ mag for the \hst\ image \citep{Schlafly2011} and $A_J = 0.058$ mag\footnote{\url{https://irsa.ipac.caltech.edu/applications/DUST/}} for the 2MASS image were also applied.

Both the original and modified \hst\ images were then fitted with MGE models, which were subsequently matched to the 2MASS $J$-band MGE model at large radii. We list in Table~\ref{tab:mge} this combined and AGN-free MGE model, which consists of 14 concentric Gaussian components and has a somewhat flatter central profile (red line) compared to the AGN-contaminated one (black line), as seen in the right panel of Figure~\ref{fig:stars}. It is clear that the AGN-contaminated stellar light near the nucleus is non-negligible and, if uncorrected, would significantly bias the SMBH mass measurement for M81. Figure~\ref{fig:mge} presents this best-fitting, combined, and AGN-free MGE model overlaid on the observed surface brightness distribution of M81, based on the 2MASS $J$-band image (left) and the \hst\ WFC3/IR F110W image (middle). A zoom-in of the central $10\arcsec \times 10\arcsec$ region from the \hst\ WFC3/IR F110W image is shown in the right panel. This figure demonstrates that the combined MGE model accurately reproduces the galaxy’s isophotes across a broad range of spatial scales, confirming its suitability for subsequent dynamical modeling.

In the subsequent dynamical analysis of M81, we adopted this best-fitting, combined, and AGN-free MGE model as the fiducial stellar-mass model, while the similar best-fitting and combined but AGN-contaminated MGE model was used to assess the systematic uncertainties.

The 2D light distribution is then deprojected, assuming an axisymmetric potential and a free inclination angle ($i$), to yield a 3D axisymmetric luminosity profile. By multiplying this luminosity distribution by the \ml$_J$ ratio in the $J$ band, we obtain a model of the mass density, from which the gravitational potential can be calculated via the Poisson equation. 
 
\begin{figure*}
\centering
\includegraphics[width=0.98\textwidth]{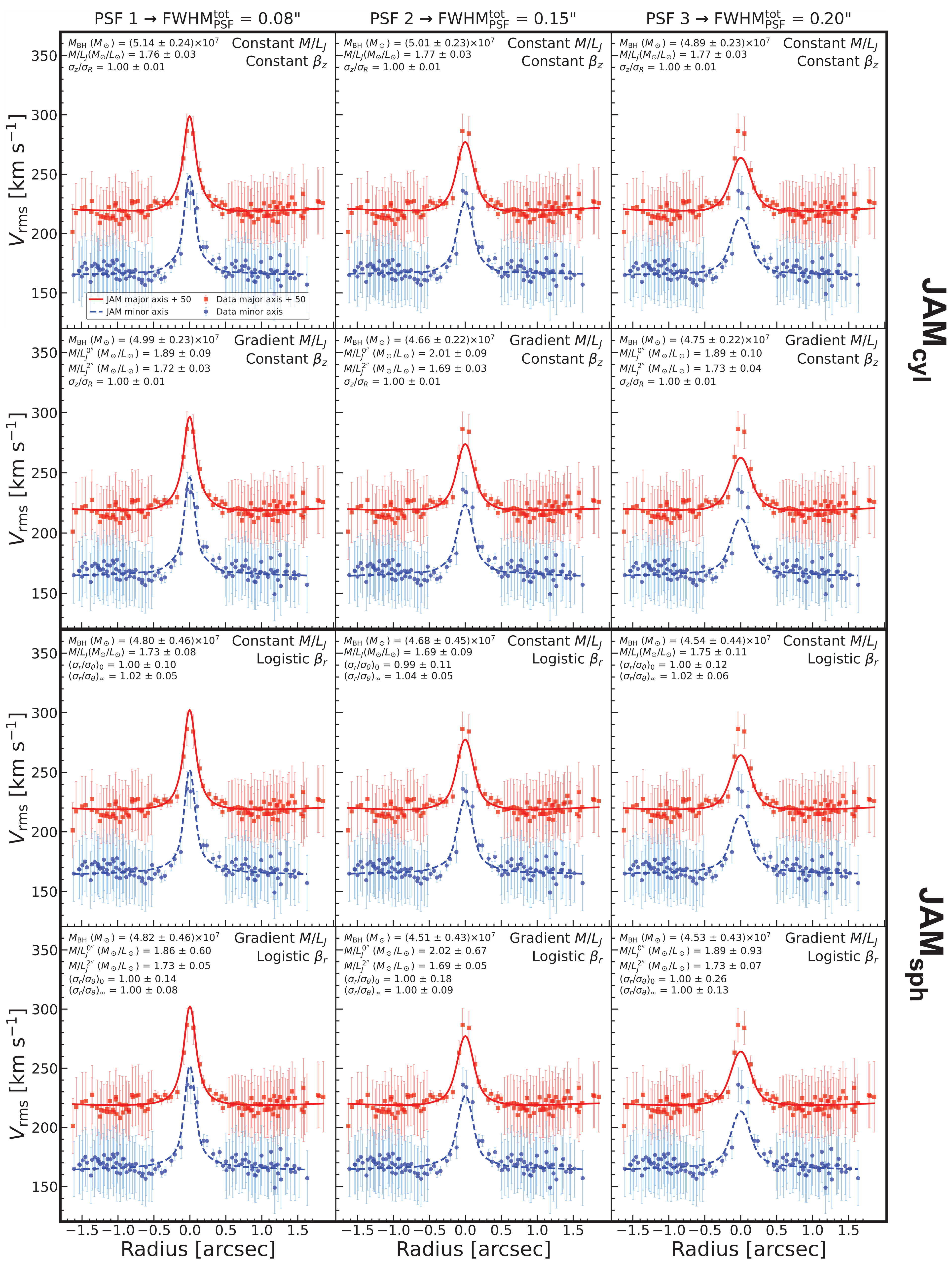}
\caption{Stellar kinematics extracted from the NIRSpec G235H/F170LP data cube of M81 are shown as filled red (major axis + 50 \kms) and blue (minor axis) points, overlaid with best-fit JAM models assuming various combinations of \ml$_J$ and orbital anisotropy. Each model includes its corresponding best-fit parameters and associated 1$\sigma$ uncertainties in the legend.}
\label{fig:1D-vrms-cyl}
\end{figure*}

\section{Dynamical Modelling}\label{sec:dyna_model}

\subsection{Jeans Anisotropic Models}\label{sec:jam}

We used both the cylindrically aligned JAM models \citep[\jamcyl;][]{Cappellari2008} and the spherically aligned version \citep[\jamsph;][]{Cappellari2020}, each assuming an axisymmetric potential, to assess how velocity-ellipsoid alignment affects the dynamical \Mbh\ determination in M81 by comparing the observed NIRSpec G235H/F170LP \vrms\ map with model predictions.
  
The \jamcyl\ model aligns the velocity ellipsoid with cylindrical coordinates ($R, \phi, z$). The anisotropy is described by $\beta_z = 1 - (\sigma_z / \sigma_R)^2$. In contrast, the \jamsph\ model aligns the velocity ellipsoid with spherical coordinates ($r, \theta, \phi$). Here, the anisotropy is quantified by $\beta_r = 1 - (\sigma_\theta / \sigma_r)^2$. In both models, all three components of the velocity ellipsoid are allowed to be different.

We constructed our models using the \texttt{jam.axi.proj} routine in \textsc{JamPy}\footnote{v7.2.0: \url{https://pypi.org/project/jampy/}} \citep{Cappellari2020}, setting {\tt align=`cyl'} for cylindrical alignment and {\tt align=`sph'} for spherical alignment to compute the LOSVD moments $\langle V_{\rm los}^2\rangle$.

\subsection{Model Grid and Parameter Space}\label{sec:jam_grid}

Both \jamcyl\ and \jamsph\ estimate \Mbh\ (sampled logarithmically) and additional parameters sampled linearly, including the mass-to-light ratio ($M/L_J$) and the orbital anisotropy. In \jamsph, the radial anisotropy ($\beta_r$) is parameterized as the ratio $\sigma_\theta/\sigma_r$, whereas in \jamcyl, the vertical anisotropy ($\beta_z$) is defined as the ratio $\sigma_z/\sigma_R$. In the \jamcyl\ models, we adopted a constant anisotropy with no imposed priors. For the physically motivated \jamsph\ models, we implemented the \citet{Simon2024} logistic anisotropy prescription with the priors described in Section~\ref{sec:logistic_anisotropy}.  For the inclination parameter ($i$, converted to $q_{\rm min}$), we fixed it to the photometrically determined value of $i \simeq 63^\circ$ ($q_{\rm min}\approx0.05$), as M81 is a spiral galaxy with a relatively low and well-constrained inclination inferred from its dust-lane morphology. Allowing for near edge-on configurations is clearly inconsistent with the observed photometry and was therefore not considered. The adopted inclination is taken from the HyperLEDA database\footnote{\url{http://atlas.obs-hp.fr/hyperleda/}}.

In these JAMs, we accounted for the effects of the NIRSpec G235H/F170LP PSF at 2.3~\micron, testing both synthetic and empirical PSFs. \citet{Nguyen2025c} showed that synthetic PSF~1 (FWHM$^{\rm tot}_{\rm PSF}=0\farcs08$) and PSF~2 (FWHM$^{\rm tot}_{\rm PSF}=0\farcs15$) provide good fits to the \jwst/NIRSpec stellar kinematics of NGC~4258, while PSF~3 (FWHM$^{\rm tot}_{\rm PSF}=0\farcs20$) underestimates the central $V_{\rm rms}$ by $\approx$20 \kms. We also tested empirical NIRSpec IFU PSFs constrained by \citet{D’Eugenio2024NaAs, D'Eugenio2025}, which are consistent with our PSF~2, whereas those reported by \citet{Bentz2025} and our AGN-based PSF derivation in Section~\ref{agn_psf} are both consistent with our synthetic PSF~3. In addition, we explored both constant and radially varying \ml$_J$ assumptions (Section~\ref{sec:default_mlvary}) to assess systematic uncertainties in the \Mbh\ measurement of M81.

In total, we performed 12 JAM model runs as listed in Table~\ref{tab:adamet} and summarized below, which together bracket the main sources of systematic uncertainty in our dynamical analysis:
\begin{enumerate}
    \item \textbf{Anisotropy alignment (2 options):}  
    Either cylindrically aligned velocity ellipsoid (\jamcyl) or spherically aligned one (\jamsph).
    \item \textbf{Point Spread Function (3 options):}  
    Three STPSF \citep{Perrin2025} synthetic JWST/NIRSpec PSFs were tested \citep[Table 1 of][]{Nguyen2025c}. 
    \item \textbf{Mass-to-light ratio (2 options):}  
    Either a constant \ml$_J$ or a radially varying \ml$_J$ within the NIRSpec FoV data cube (Section~\ref{sec:default_mlvary}).
\end{enumerate}

\subsection{Linearly Varying  $M/L_J$ Profile}\label{sec:default_mlvary}

Possible population variation within the nucleus of M81 can bias both the central \ml$_J$ and \Mbh\ measurements. We performed an additional test, in which the \ml$_J$ values beyond the NIRSpec FoV ($r > 2\arcsec$) were fixed to the best-fit value obtained from the corresponding models with a constant \ml$_J$ parameter. However, within $2\arcsec$, \ml$_J$ was allowed to vary linearly toward a central value $M/L^{0\arcsec}_{J}$. 

In JAMs, the $M/L_J(r)$ profile is implemented by associating a different $(M/L_J)_j$ to each Gaussian component with dispersion $\sigma_j$ in the MGE listed in Table~\ref{tab:mge}, as follows: 
\begin{equation}
\left(\frac{M}{L_J}\right)_j = 
\begin{cases}
    M/L_J^{0\arcsec} + \dfrac{M/L_J^{2\arcsec} - M/L_J^{0\arcsec}}{2\arcsec} \times \sigma_j, & \sigma_j < 2\arcsec\\
    M/L_J^{2\arcsec}, & \sigma_j \ge 2\arcsec
\end{cases}
\end{equation}

\subsection{Logistic Anisotropy for \jamsph\ Models}\label{sec:logistic_anisotropy}    

While formally Jeans models are subject to the mass-anisotropy degeneracy \citep{Binney1982}, our understanding of galaxy dynamics has vastly improved since then. It now allows us to place physically motivated priors on the stellar orbital structure, significantly improving the reliability of the mass measurement. Both observations and simulations of massive galaxies consistently show a characteristic anisotropy profile: orbits are nearly isotropic or mildly radially-biased at large radii, but become tangentially-biased inside the BH's SOI. This orbital structure is thought to be a natural outcome of SMBH binary scouring during galaxy mergers, which preferentially ejects stars on radial orbits \citep[e.g.,][]{Milosavljevic2001, Rantala2024}. While this scenario has been primarily studied in massive elliptical galaxies, the lack of radially-biased orbits near the BH has also been observed in lower-mass fast-rotating systems  \citep[see a review of the observations in,][Figure~10]{Cappellari2026}.

To capture this expected behavior, we explored the \jamsph\ model with a flexible, radially varying anisotropy, as done by \citet{Simon2024}. This model uses a logistic function of $\log r$ to describe the radial anisotropy profile $\beta_r(r) = 1 - \sigma_\theta^2/\sigma_r^2$:
\begin{equation}
\beta_r(r) = \beta_{r, 0} + \frac{\beta_{r,\infty} - \beta_{r,0}}{1 + (r_a/r)^\alpha}
\end{equation}
Here, $\beta_{r, 0}$ and $\beta_{r,\infty}$ are the anisotropy values at the center and at large radii, respectively, and $r_a$ is the transition radius. We re-express the anisotropy in terms of the more intuitive velocity dispersion ratio $\sigma_r/\sigma_\theta = 1/\sqrt{1-\beta_r}$.

Based on the established physics, we applied specific priors to constrain the model parameters. This approach was successfully tested on two benchmark galaxies with very accurate, independent \Mbh\ determinations: the massive elliptical galaxy M87 \citep{Simon2024} and the spiral galaxy NGC~4258 \citep{Nguyen2025c}, which has a morphology similar to M81. In both cases, the JAM models with these physically-motivated priors recovered the known BH masses, validating the reliability of this method.

Following this validated approach, we restricted the inner anisotropy to be tangentially biased or isotropic ($0.5 < (\sigma_r/\sigma_\theta)_0 < 1$) and the outer anisotropy to be isotropic or mildly radially biased ($1 < (\sigma_r/\sigma_\theta)_\infty < 1.3$). The anisotropy transition radius was fixed at $r_a = 0\farcs55$, slightly larger than the SMBH's SOI ($r_{\rm SOI}\approx0\farcs5$; see Section~\ref{sec:smbh}), to ensure the model can capture the expected transition from a tangentially-biased inner region to a more isotropic outer region. Finally, we fixed the sharpness parameter to $\alpha=2$ to reduce the dimensionality of the parameter space and minimize degeneracies.

\begin{figure*}
\centering
\includegraphics[width=0.75\textwidth]{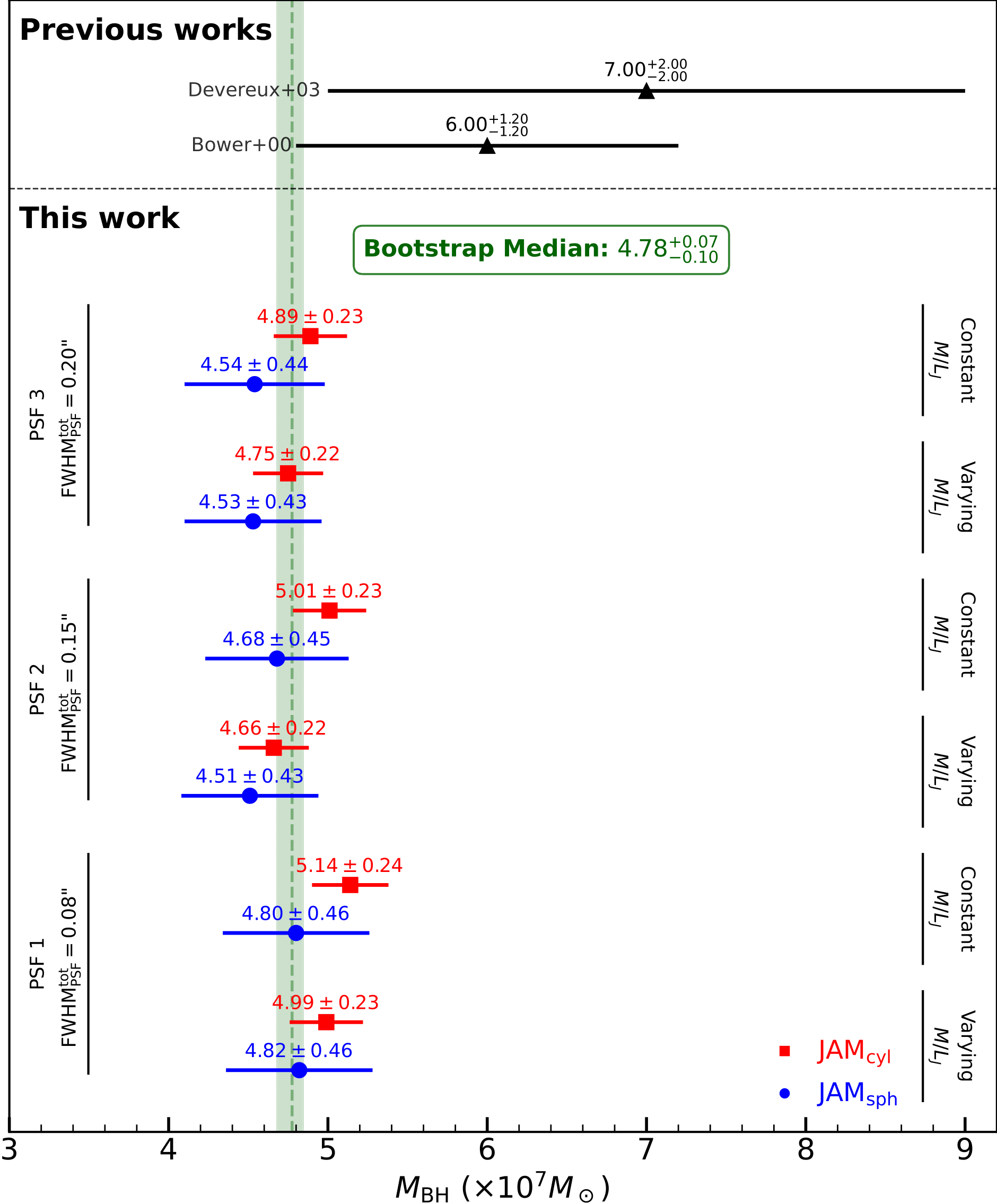}
\caption{Black hole mass measurements for M81 from 12 conventional JAM models applied to the NIRSpec G235H/F170LP data cube. Individual measurements and their 1$\sigma$ uncertainties reflect variations in PSF, orbital anisotropy, and stellar \ml$_J$ assumptions. Vertical dashed lines mark the adopted \Mbh, with shaded regions indicating the ensemble median and 68\% bootstrap confidence interval. For comparison, previous estimates from \citet{Bower2000} and \citet{Devereux2003} are shown; these earlier values relied on restrictive stellar dynamical assumptions or disturbed ionized gas kinematics. The present analysis provides the most reliable measurement of \Mbh\ in M81 to date.} 
\label{fig:summary_all_BH_ngc3031}
\end{figure*}     

\begin{figure*}
    \centering
    \includegraphics[height=0.5\textwidth]{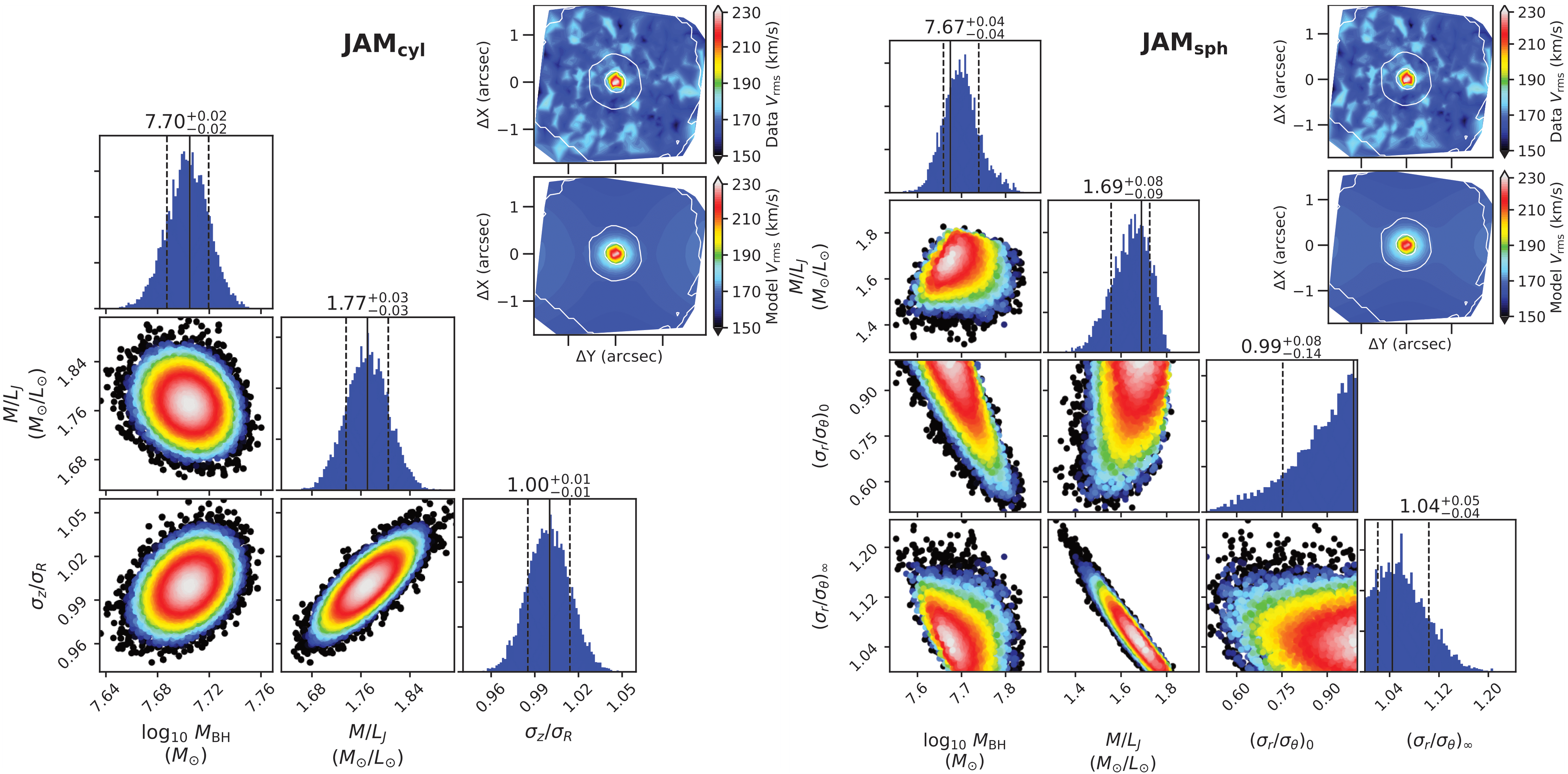}
    \caption{{\it Left:} Posterior distributions for the \jamcyl\ model with constant \ml$_J$ and anisotropy, obtained following the burn-in phase of the \textsc{adamet} MCMC optimization, are presented for the nuclear stellar kinematics of M81 using the NIRSpec G235H/F170LP data. The resulting triangle plot displays the 2D projected probability distributions of the model parameters ($M_{\rm BH}$, $M/L_J$, $\sigma_z/\sigma_R$), with 1D marginalized histograms shown along the diagonal. Solid vertical lines indicate the best-fit values, while dashed lines denote the 1$\sigma$ uncertainties. Inset panels in the upper right compare the observed and model-predicted \vrms\ on a consistent velocity scale. The observed \vrms\ map was symmetrized under the assumption of axisymmetry using the \textsc{symmetrize\_velfield} routine from the \textsc{plotbin}\footnote{v3.1.7: \url{https://pypi.org/project/plotbin/}} package. Both the symmetrized data and model prediction were visualized with \textsc{plot\_velfield}. The strong agreement across the NIRSpec FoV demonstrates the high fidelity of the model. {\it Right:} Same as the left panel, but applied the \jamsph\ models with logistic varying $\beta_r$ profile and the model parameters ($M_{\rm BH}$, $M/L_J$, $(\sigma_r/\sigma_\theta)_0$, $(\sigma_r/\sigma_\theta)_\infty$).}
    \label{fig:adamet}
\end{figure*}

\subsection{MCMC Fitting and Uncertainty Estimation}\label{sec:mcmc}

We ran the JAM models within a Markov chain Monte Carlo (MCMC) framework to explore the full parameter space. Sampling was performed using the adaptive Metropolis algorithm \citep{Haario2001}, as implemented in the \textsc{adamet}\footnote{v2.0.9: \url{https://pypi.org/project/adamet/}} package \citep{Cappellari2013a}. Each MCMC chain comprised $5 \times 10^4$ steps, with the first 20\% of samples discarded as burn-in. The remaining samples were used to construct posterior probability distributions. The most likely values for the model parameters were identified as those corresponding to the maximum likelihood, and uncertainties were estimated from the 1$\sigma$ (16th–84th percentile) and 3$\sigma$ (0.14th–99.86th percentile) confidence intervals (CLs).

Dynamical modeling of our numerous high-precision NIRSpec kinematic data points presents two challenges:
\begin{enumerate}
    \item \textbf{Underestimated Uncertainties:} The formal statistical errors on derived parameters can become unrealistically small due to the unmodelled systematic effects.
    \item \textbf{Dominance of Large-Radii Data:} The $\chi^2$ statistic can be disproportionately influenced by the large number of data points at large radii. This risks biasing the $M_{\rm BH}$ measurement, which should be dictated primarily by the kinematics within the BH's SOI.
\end{enumerate}

To mitigate the first issue, we adopt an error inflation scheme. We base our approach on the heuristic $\Delta\chi^2$-increase method of \citet{vandenBosch2009}, which, in a Bayesian framework, is equivalent to inflating the kinematic measurement errors by a factor of $(2N)^{1/4}$, where $N$ is the number of data points \citep[Section~6.1]{Mitzkus2017}. While this technique is now standard for ALMA-based $M_{\rm BH}$ measurements \citep[e.g.,][]{Nguyen2020, Smith2019, Smith2021, North2019}, a uniform inflation across all radii does not resolve the second challenge.

Therefore, we apply the inflation selectively. We preserve the formal kinematic uncertainties for all data points inside the SOI ($r \leq r_{\rm SOI} \approx 0.5\arcsec$), where the SMBH's gravitational potential dominates. For the $N_{r > r_{\rm SOI}} = 620$ kinematic bins outside this radius, we inflate their associated errors by a factor of $(2N_{r > r_{\rm SOI}})^{1/4}$. This refined strategy ensures the $M_{\rm BH}$ determination is driven by the central kinematics while still accounting for potential systematic errors at larger radii. This selective approach has been successfully employed in previous dynamical studies using integral-field data from Gemini/NIFS \citep{Drehmer2015}, VLT/MUSE \citep[Section~4.3]{Thater2022}, and recently \jwst/NIRSpec \citep{Nguyen2025c}. 

\begin{table*}
\centering
\caption{Summary of the 12 JAM models best-fitting parameters and \emph{formal} uncertainties.}
\vspace{-2mm}
\setlength{\tabcolsep}{8pt}
\begin{tabular}{l c c c c c c c c}
\hline\hline
$M/L_J$ & FWHM$_{\rm PSF}^{\rm tot}$ &
$M_{\rm BH}$ &
$M/L_J$ &
$M/L_J^{0\arcsec}$ &
$M/L_J^{2\arcsec}$ &
$\sigma_z/\sigma_R$ &
$(\sigma_r/\sigma_\theta)_0$ &
$(\sigma_r/\sigma_\theta)_\infty$ \\
            & ($\arcsec$)& ($\times 10^7$ \Msun) & (\Msun/\Lsun) & (\Msun/\Lsun) & (\Msun/\Lsun) & & & \\
(1)         & (2)        & (3)          & (4)           & (5)           & (6)           & (7) & (8) & (9) \\
\hline
\multicolumn{9}{c}{\jamcyl\ models with constant anisotropy} \\
\hline
Constant & $0.08$ & $5.14 \pm 0.24$ & $1.76 \pm 0.03$ & \dots & \dots & $1.00 \pm 0.01$ & \dots & \dots \\
Constant & $0.15$ & $5.01 \pm 0.23$ & $1.77 \pm 0.03$ & \dots & \dots & $1.00 \pm 0.01$ & \dots & \dots \\
Constant & $0.20$ & $4.89 \pm 0.23$ & $1.77 \pm 0.03$ & \dots & \dots & $1.00 \pm 0.01$ & \dots & \dots \\
Varying  & $0.08$ & $4.99 \pm 0.23$ & \dots & $1.89 \pm 0.09$ & $1.72 \pm 0.03$ & $1.00 \pm 0.01$ & \dots & \dots \\
Varying  & $0.15$ & $4.66 \pm 0.22$ & \dots & $2.01 \pm 0.09$ & $1.69 \pm 0.03$ & $1.00 \pm 0.01$ & \dots & \dots \\
Varying  & $0.20$ & $4.75 \pm 0.22$ & \dots & $1.89 \pm 0.10$ & $1.73 \pm 0.04$ & $1.00 \pm 0.01$ & \dots & \dots \\
\hline
\multicolumn{9}{c}{\jamsph\ models with radially-varying logistic anisotropy} \\
\hline
Constant & $0.08$ & $4.80 \pm 0.46$ & $1.73 \pm 0.08$ & \dots & \dots & \dots & $1.00 \pm 0.10$ & $1.02 \pm 0.05$ \\
Constant & $0.15$ & $4.68 \pm 0.45$ & $1.69 \pm 0.09$ & \dots & \dots & \dots & $0.99 \pm 0.11$ & $1.04 \pm 0.05$ \\
Constant & $0.20$ & $4.54 \pm 0.44$ & $1.75 \pm 0.11$ & \dots & \dots & \dots & $1.00 \pm 0.12$ & $1.02 \pm 0.06$ \\
Varying  & $0.08$ & $4.82 \pm 0.46$ & \dots & $1.86 \pm 0.60$ & $1.73 \pm 0.05$ & \dots & $1.00 \pm 0.14$ & $1.00 \pm 0.08$ \\
Varying  & $0.15$ & $4.51 \pm 0.43$ & \dots & $2.02 \pm 0.67$ & $1.69 \pm 0.05$ & \dots & $1.00 \pm 0.18$ & $1.00 \pm 0.09$ \\
Varying  & $0.20$ & $4.53 \pm 0.43$ & \dots & $1.89 \pm 0.93$ & $1.73 \pm 0.07$ & \dots & $1.00 \pm 0.26$ & $1.00 \pm 0.13$ \\
\hline
\end{tabular}
\label{tab:adamet}
\noindent\parbox{\linewidth}{\textit{Note:} Column (1): Assumed $M/L_J$ profile type. Column (2): Total FWHM of the synthetic PSF model. Column (3): Best-fit black hole mass. Column (4): Best-fit constant $M/L_J$. Columns (5) and (6): Best-fit central and outer $M/L_J$ for the varying profile. Column (7): Best-fit vertical anisotropy for \jamcyl\ models. Column (8): Best-fit central radial anisotropy for \jamsph\ models. Column (9): Best-fit outer radial anisotropy for \jamsph\ models.}
\end{table*}

\section{Results and Discussion}\label{sec:results}

\subsection{Dynamical Supermassive Black Hole Mass Constraints}\label{sec:smbh}

We summarize in Table~\ref{tab:adamet} the best-fit parameters and their $1\sigma$ uncertainties for the 12 conventional JAM models used to fit the NIRSpec G235H/F170LP \vrms\ data. Across all model assumptions, the inferred \Mbh\ ranges from $(4.51$--$5.14)\times 10^7$~\Msun, and the \ml$_J$ values span $(1.69$--$2.02)$~\Msun/\Lsun. While both \Mbh\ and \ml$_J$ are relatively insensitive to the choice of PSF, they are more strongly affected by the assumed velocity ellipsoid alignment. Specifically, the \jamcyl\ models yield \Mbh\ and \ml$_J$ values that are $\approx$7\% higher than those from the \jamsph\ models.

Figure~\ref{fig:1D-vrms-cyl} shows the \vrms\ profiles for all 12 best-fitting JAM models (with their 1$\sigma$ uncertainties), extracted along the major and minor axes of M81. These are directly compared to the observed \vrms\ profiles from the NIRSpec G235H/F170LP data, extracted in the same manner. Overall, all models provide a good match to both the data and to each other across the NIRSpec FoV, despite differences in the assumed \ml$_J$ profiles and orbital anisotropy. 

A systematic offset is observed between the models and the data in the innermost kinematic bins, which is likely attributable to uncertainties in the NIRSpec PSF modeling. To investigate this, we tested three synthetic PSFs from \citet{Nguyen2025c} with different widths. The models using the broadest PSF (PSF~3, FWHM$_{\rm PSF}^{\rm tot} \approx 0\farcs2$), consistently underpredict the central \vrms\ by $\approx$20~\kms\ (right column of Figure~\ref{fig:1D-vrms-cyl}). Conversely, models with the narrowest PSF (PSF~1, FWHM$_{\rm PSF}^{\rm tot} \approx 0\farcs08$) tend to overpredict the central \vrms\ by $\approx$15~\kms\ (left column). The models providing the best match to the data, including the innermost bins, are those using the intermediate PSF~2 (FWHM$_{\rm PSF}^{\rm tot} \approx 0\farcs15$; middle column). This preference for the $\approx 0\farcs15$ PSF is consistent with the findings for NGC~4258 in \citet{Nguyen2025c}, giving us confidence that our models are using a realistic representation of the instrumental resolution.

Notably, our derived \Mbh\ value is 17\% lower than that inferred from the VLT/GMOS IFU velocity dispersion and the \citet{Gultekin2009} \Mbh--$\sigma$ relation, while it is 31.7\% lower than the gas-based \Mbh\ measurement from \hst/STIS \citep{Devereux2003} and 26.5\% lower than the widely adopted value from \citet{Kormendy2013}. More specifically, the derived \Mbh\ values are insensitive to the assumed alignment of the velocity ellipsoid but are sensitive to the adopted \ml$_J$ profile in the nucleus of M81.

In contrast, the orbital anisotropy of the nuclear stellar motions in M81 is well constrained and does not depend on whether a constant or varying \ml$_J$ is assumed. The best-fit \jamcyl\ models indicate nearly isotropic kinematics, with $\sigma_z/\sigma_R \sim 1$ (i.e., $\beta_z \sim 0$). For the \jamsph\ models, we adopt a fixed transition radius $r_a=0\farcs55$ (see Section~\ref{sec:logistic_anisotropy}). The best-fit models indicate that the ratio of stellar velocity dispersions $(\sigma_r/\sigma_\theta)_0$ at the center varies from $\sim$0.55 to $\sim$1.0 depending on the adopted PSF model (i.e., $\beta_r < 0$) and suggesting a slightly tangential bias. Beyond $r_a$, the orbits transition to a radially biased structure with $(\sigma_r/\sigma_\theta)_\infty \sim 1.20$ (i.e., $\beta_r > 0$).
   
\begin{figure}
     \centering\includegraphics[width=0.98\linewidth]{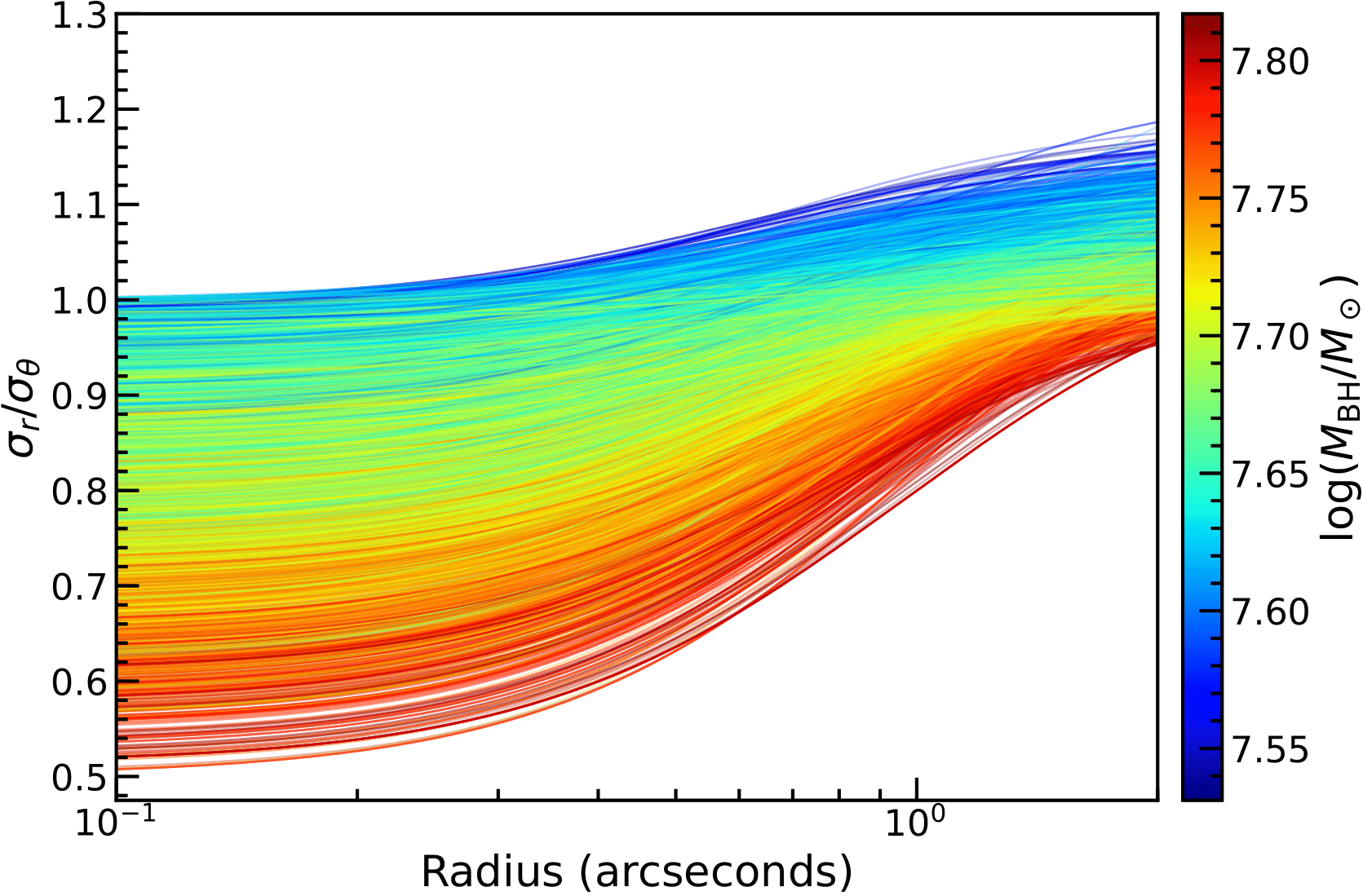}
     \caption{Plot of 10,000 anisotropy profiles randomly sampled from the MCMC chain of the \jamsph\ with logistic anisotropy Figure~\ref{fig:adamet}, color-coded by their corresponding \Mbh\ values. The profiles show clear evidence of a radially increasing anisotropy ratio and strongly exhibit the black hole mass–anisotropy degeneracy like M87 \citep{Simon2024} and NGC~4258 \citep{Nguyen2025c}. The orbital anisotropy varies with black hole mass: stars orbiting more massive black holes exhibit tangentially biased motions, whereas those around smaller black holes show radial anisotropy.}
     \label{fig:adamet_betavary}
\end{figure}

Figure~\ref{fig:summary_all_BH_ngc3031} shows our 12 \Mbh\ measurements and their $1\sigma$ uncertainties, providing a presentation-ready summary of the most precise \Mbh\ constraint for M81 based on the spatially resolved NIRSpec G235H/F170LP IFU stellar kinematics. Following the approach of \citet{Nguyen2025c}, we derive robust SMBH estimates from the 12 conventional JAM models by adopting the medians and their 68\% (1$\sigma$) bootstrap confidence intervals, yielding $M_{\rm BH} = (4.78^{+0.07}_{-0.10}) \times 10^7$~\Msun.  

Figure~\ref{fig:adamet} shows example 2D posterior distributions for the \jamcyl\ model with constant anisotropy (left) and the \jamsph\ model with logistic anisotropy (Eq.~2; right). Both models assume a constant \ml$_J$ within the NIRSpec FoV and PSF~2. Point colors indicate relative likelihood, with white corresponding to the maximum likelihood and the 1$\sigma$ confidence level (CL) region, and black marking the 3$\sigma$ CL. The accompanying 1D histograms along the diagonal are used to determine the best-fit values and 1$\sigma$ uncertainties, which incorporate the propagation of stellar kinematics and statistical errors. The figure also demonstrates the close agreement between the observed \vrms\ and the predictions from the best-fit \jamcyl\ and \jamsph\ models at the upper right corner of each triangle, assuming the major axes are aligned horizontally, with relative errors less than 5\% in each case.

In the best-fitting \jamcyl\ model, the remaining two parameters are well constrained, whereas in the best-fitting \jamsph\ model, only the \Mbh\ and \ml$_J$ parameters are well constrained. The other parameters of the logistic anisotropy are limited by their priors to enforce physically plausible stellar orbits around the SMBH and within the NIRSpec FoV.

Across all JAM models, there is no significant covariance between \Mbh\ and \ml$_J$ within the 3$\sigma$ CLs, indicating that both parameters are well constrained. In the \jamcyl\ model, \Mbh\ and \ml$_J$ each show positive covariance with the orbital anisotropy parameter ($\sigma_z/\sigma_R$), whereas in the \jamsph\ model, these covariances are negative with respect to ($\sigma_\theta/\sigma_r$). This contrast highlights the complexity of how variations in \Mbh\ or \ml$_J$ can compensate for changes in anisotropy during the fitting process.

Our stellar-based \Mbh\ measurements, derived under both cylindrical and spherical velocity ellipsoid assumptions, yield $r_{\rm SOI} \approx 0\farcs 5$ (or 9 pc), estimated using the method described in Section~\ref{sec:intro}. These values are 5 times larger than the \jwst\ spatial resolution, indicating that the \Mbh\ determinations are robust and spatially well resolved. The dynamical influence of the SMBH is clearly detected within the central $\approx$80 spaxels in area.

\begin{figure*}
	\centering
    	 \includegraphics[width=0.98\textwidth]{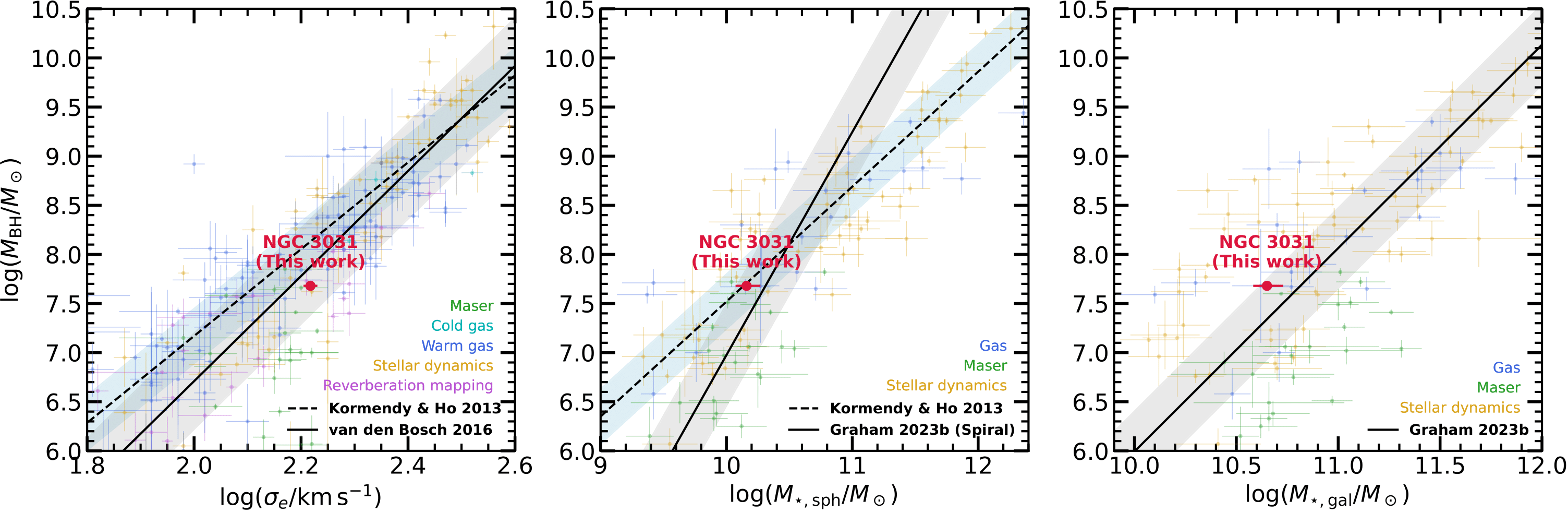} 
	 \caption{Location of M81 with respect to the \Mbh--$\sigma$ ({\it left}), \Mbh--$M_{\star, \rm sph}$ ({\it middle}),  \Mbh--$M_{\star, \rm gal}$ ({\it right}) scaling relation. Our new measurements are consistent within 1$\sigma$ with the \Mbh--$\sigma$ scaling relations for spiral galaxies.}
	 \label{bhmass_sigma}
\end{figure*}

\subsection{The Mass--Anisotropy Degeneracy}\label{sec:bh-anisotropy}

A well-known characteristic of Jeans modeling is the mass-anisotropy degeneracy, where a given kinematic profile can be fit by different combinations of mass and orbital anisotropy. This effect is clearly illustrated in Figure~\ref{fig:adamet_betavary}, which shows 10,000 anisotropy profiles randomly drawn from the posterior distribution of our logistic \jamsph\ model. The figure shows the expected covariance between \Mbh\ and the anisotropy profile: higher \Mbh\ values are required to fit the data when the stellar orbits are more tangentially biased, while lower \Mbh\ values are consistent with more radial orbits. This behavior is not a flaw in our approach but an inherent property of stellar dynamics, and it has been observed in similar analyses of other galaxies like NGC~4258 \citep[Figure~9 of][]{Nguyen2025c} and M87 \citep[Figure~15 of][]{Simon2024}.

What is crucial for a reliable mass measurement is that this degeneracy is properly sampled and accounted for in the final uncertainty budget. Our Bayesian MCMC framework is designed to do precisely this. By exploring the full parameter space, the sampler covers all degenerate solutions consistent with the data and our physically motivated priors. The final posterior probability distribution for \Mbh\ is marginalized over all other parameters, including the full range of allowed anisotropy profiles. Therefore, the reported CLs on our \Mbh\ measurement robustly incorporate the uncertainty arising from the mass-anisotropy degeneracy.

\subsection{Additional Uncertainty Budgets}\label{sec:uncer}

\subsubsection{Distances}\label{dist}

Unlike most galaxies, the distance to M81 is exceptionally well-determined. As stated in the Introduction, we adopt a distance of $D = 3.63 \pm 0.14$~Mpc, which is the mean of five high-quality determinations based on the Cepheid and Tip of the Red Giant Branch methods. This precision is corroborated by the NASA/IPAC NED, which lists 99 distance determinations, yielding a formal average of $3.675 \pm 0.049$~Mpc.

Although the distance is robust, its uncertainty remains a dominant source of systematic error in our final black hole mass. Since \Mbh\ scales linearly with distance ($M_{\rm BH} \propto D$), the $\pm 0.049$~Mpc uncertainty from the NED average and $\pm 0.14$~Mpc uncertainty from our adopted distance propagate to systematic uncertainties of $\pm 1.3\%$ and 2.9\% in our \Mbh\ measurement, respectively. This is comparable to the formal statistical errors from our dynamical models and thus represents a key component of the total error budget.

\subsubsection{AGN-Contaminated Stellar Mass Model}\label{agn}

We evaluated the impact of the \hst\ AGN contamination on the \Mbh\ and \ml$_J$ measurements by replacing our best-fitting, combined, and AGN-free MGE model (Table~\ref{tab:mge}) by the similar best-fitting and combined but AGN-contaminated MGE model (Section~\ref{sb}), which is shown in black line in the right panel of Figure~\ref{fig:stars}. The enclosed luminosity difference within $1\arcsec$ between the AGN-free and AGN-contaminated MGE models is $8.5 \times 10^6$~\Lsun. Adopting the constant \ml$_J$ from either the \jamcyl\ or \jamsph\ models, this corresponds to a stellar mass difference of $(1.4$–$1.5) \times 10^7$~\Msun, introducing a $\sim$(11–20)\% uncertainty in the \Mbh\ estimates under the fixed \ml$_J$ and anisotropy values as presented in Table~\ref{tab:adamet}. 

More realistic tests allowed both \Mbh\ and \ml$_J$ to vary in the \jamcyl\ and \jamsph\ models using the AGN-contaminated MGE model. The best-fitting \jamcyl\ model with constant \ml$_J$ yields \Mbh\ $= (4.26 \pm 0.21) \times 10^7$ \Msun\ and \ml$_J = 1.79 \pm 0.03$ (\Msun/\Lsun), while the corresponding \jamsph\ model gives \Mbh\ $= (3.81 \pm 0.36) \times 10^7$ \Msun\ and \ml$_J = 1.49 \pm 0.09$ (\Msun/\Lsun). All other parameters differ by less than 8\% from the values in Table~\ref{tab:adamet} and are discussed in Section~\ref{sec:smbh}. The close agreement with the results obtained using the AGN-free MGE model indicates that AGN light in the central region, adding to the stellar mass, introduces uncertainty in the \Mbh\ estimates.

\subsubsection{Alternative Stellar Kinematic Measurements}\label{alter_kinmaps}

We additionally derived the stellar kinematics of M81 from the NIRSpec G235H/F170LP data using an alternative setup based on the higher-resolution PHOENIX synthetic stellar library \citep[Section~\ref{sec:kine};][]{Husser2013} and adopting {\tt moments = 2} in \textsc{pPXF}. The resulting kinematics are consistent with our fiducial measurements, which employ the XSL library with {\tt moments = 2}, to within 3\%. We evaluated the impact of these alternative kinematic measurements on the inferred \Mbh\ and \ml$_J$ and find that the resulting systematic differences are negligible.

\subsection{\Mbh--Galaxy Scaling Relations}\label{sec:M-sigma_relation}

We examine the location of the central SMBH in M81 relative to the established \Mbh–$\sigma$, \Mbh–$M_{\star,\rm sph}$, and \Mbh–$M_{\star,\rm gal}$ scaling relations. We adopt a stellar velocity dispersion of $\sigma_e = (165 \pm 4)$~\kms, measured outside the SMBH's SOI, along with stellar mass estimates of $\log(M_{\star,\rm sph}) = (10.16 \pm 0.11)$~\Msun\ for the spheroidal component \citep{Davis2019} and $\log(M_{\star,\rm gal}) = (10.65 \pm 0.08)$~\Msun for the total galaxy \citep{Davis2018}. Using these values, we placed M81 on the \Mbh–$\sigma$ relations from \citet{Kormendy2013} and \citet{vandenBosch2016} (see Figure~\ref{bhmass_sigma}), as well as on the \Mbh–$M_{\star,\rm sph}$ and \Mbh–$M_{\star,\rm gal}$ relations from \citet{Kormendy2013} and \citet{Graham_Sahu2023}. The relations presented by \citet{Kormendy2013} and \citet{vandenBosch2016} combine spiral, lenticular, and elliptical galaxies, whereas \citet{Graham_Sahu2023} provides morphology-dependent fits; for the latter, we adopt the spiral-galaxy relations. The black hole masses used to define these scaling relations are predominantly derived from dynamical measurements based on gas, maser, and stellar kinematics; in the case of the \Mbh–$\sigma$ relation, they are additionally included black hole masses obtained via reverberation mapping.

Our measured \Mbh\ for M81 is consistent within $1\sigma$ of the \citet{vandenBosch2016} \Mbh–$\sigma$ relation and within $2\sigma$ of the \citet{Kormendy2013} relation. Similarly, M81 lies within $1\sigma$ of both the \citet{Kormendy2013} and \citet{Graham_Sahu2023} \Mbh–$M_{\star,\rm sph}$ relations, as well as the \citet{Graham_Sahu2023} \Mbh–$M_{\star,\rm gal}$ relation.

\section{Conclusions}\label{sec:result}

We have presented the first robust stellar-dynamical measurement of the SMBH mass in the nearby spiral galaxy M81. Previous determinations were unreliable, being based on disturbed gas kinematics or preliminary stellar-dynamical models. Our analysis leverages the power of \jwst/NIRSpec IFU observations, which provide high-resolution 2D stellar kinematics in the NIR, allowing us to penetrate the dusty nucleus and cleanly separate stellar light from the AGN continuum.    Our main results are as follows:

\begin{enumerate}

    \item The NIRSpec data clearly resolve the BH's SOI. We measured a sharp rise in the stellar velocity dispersion from $165$~\kms\ at large radii to a central peak of $233 \pm 15$~\kms, providing unambiguous kinematic evidence for a central dark mass.

    \item We performed a comprehensive dynamical analysis using an ensemble of 12 JAM models to systematically account for uncertainties in the instrumental PSF, stellar-\ml\ ratio, and orbital anisotropy. This robust approach yields a black hole mass of $M_{\rm BH} = (4.78^{+0.07}_{-0.10})\times10^7$~\Msun.

    \item Our models confirm the expected mass-anisotropy degeneracy, where more tangentially biased orbits require a more massive black hole. Our Bayesian framework fully explores this degeneracy, and its effect is incorporated into our final quoted uncertainties. 

    \item This new, reliable mass measurement for M81 resolves a long-standing uncertainty. It provides a crucial anchor point for SMBH-galaxy scaling relations, particularly for spiral galaxies, where it is consistent with the established \Mbh--$\sigma$ and \Mbh--$M_{\star, \rm gal}$ relations.

\end{enumerate}

\section*{Acknowledgements}
N.T. would like to acknowledge partial support from UKRI grant ST/X002322/1 for UK ELT Instrument Development at Oxford. M.P. acknowledges support through the grants PID2021-127718NB-I00 and RYC2023-044853-I, funded by the Spanish Ministry of Science and Innovation/State Agency of Research MCIN/AEI/10.13039/501100011033 and El Fondo Social Europeo Plus FSE+. M.P.S. acknowledges support under grants RYC2021-033094-I, CNS2023-145506, and PID2023-146667NB-I00 funded by MCIN/AEI/10.13039/501100011033 and the European Union NextGenerationEU/PRTR.

This research is based on observations made with the NASA/ESA Hubble Space Telescope obtained from the Space Telescope Science Institute, which is operated by the Association of Universities for Research in Astronomy, Inc., under NASA contract NAS 5–26555. This work is based [in part] on observations made with the NASA/ESA/CSA James Webb Space Telescope. The data were obtained from the Mikulski Archive for Space Telescopes at the Space Telescope Science Institute, which is operated by the Association of Universities for Research in Astronomy, Inc., under NASA contract NAS 5-03127 for \jwst. These observations are associated with program \#02016.

Some/all of the data presented in this article were obtained from the Mikulski Archive for Space Telescopes (MAST) at the Space Telescope Science Institute. The specific observations analyzed can be accessed via \dataset[doi:10.17909/cbpe-ft90]{https://doi.org/10.17909/cbpe-ft90}.

\facility{\jwst/NIRSpec, \hst/WFC3, and 2MASS.}

\software{{\tt Python~3.12} \citep{VanRossum2009}, 
{\tt Matplotlib~3.6} \citep{Hunter2007}, 
{\tt NumPy~1.22} \citep{Harris2020}, 
{\tt SciPy~1.3} \citep{Virtanen2020},  
{\tt photutils~0.7} \citep{bradley2024}, 
{\tt AstroPy~5.1} \citep{AstropyCollaboration2022}, 
{\tt AdaMet 2.0} \citep{Cappellari2013a}, 
{\tt JamPy~7.2} \citep{Cappellari2020}, 
{\tt pPXF~8.2} \citep{Cappellari2023}, 
{\tt vorbin~3.1} \citep{Cappellari2003}, and
{\tt MgeFit~5.0} \citep{Cappellari2002}.
}

\bibliographystyle{aasjournalv7}
\bibliography{ngc3031}

@ARTICLE{Davis2019,
       author = {{Davis}, Benjamin L. and {Graham}, Alister W. and {Cameron}, Ewan},
        title = "{Black Hole Mass Scaling Relations for Spiral Galaxies. I. M $_{BH}$-M $_{*,sph}$}",
      journal = {\apj},
         year = 2019,
        month = mar,
       volume = {873},
       number = {1},
          eid = {85},
        pages = {85},
          doi = {10.3847/1538-4357/aaf3b8},
archivePrefix = {arXiv},
       eprint = {1810.04887},
 primaryClass = {astro-ph.GA},
       adsurl = {https://ui.adsabs.harvard.edu/abs/2019ApJ...873...85D}
}

@ARTICLE{Chu2022,
       author = {{Chu}, Qingbo and {Yu}, Shenghua and {Lu}, Youjun},
        title = "{Formation and evolution of binary neutron stars: mergers and their host galaxies}",
      journal = {\mnras},
         year = 2022,
        month = jan,
       volume = {509},
       number = {2},
        pages = {1557-1586},
          doi = {10.1093/mnras/stab2882},
archivePrefix = {arXiv},
       eprint = {2110.04687},
 primaryClass = {astro-ph.GA},
       adsurl = {https://ui.adsabs.harvard.edu/abs/2022MNRAS.509.1557C}
}

@ARTICLE{Muller-Sanchez2016,
       author = {{M{\"u}ller-S{\'a}nchez}, F. and {Comerford}, J. and {Stern}, D. and {Harrison}, F.~A.},
        title = "{The Nature of Active Galactic Nuclei with Velocity Offset Emission Lines}",
      journal = {\apj},
         year = 2016,
        month = oct,
       volume = {830},
       number = {1},
          eid = {50},
        pages = {50},
          doi = {10.3847/0004-637X/830/1/50},
archivePrefix = {arXiv},
       eprint = {1606.07446},
 primaryClass = {astro-ph.GA},
       adsurl = {https://ui.adsabs.harvard.edu/abs/2016ApJ...830...50M}
}

@ARTICLE{Lena2014,
       author = {{Lena}, D. and {Robinson}, A. and {Marconi}, A. and {Axon}, D.~J. and {Capetti}, A. and {Merritt}, D. and {Batcheldor}, D.},
        title = "{Recoiling Supermassive Black Holes: A Search in the Nearby Universe}",
      journal = {\apj},
         year = 2014,
        month = nov,
       volume = {795},
       number = {2},
          eid = {146},
        pages = {146},
          doi = {10.1088/0004-637X/795/2/146},
archivePrefix = {arXiv},
       eprint = {1409.3976},
 primaryClass = {astro-ph.GA},
       adsurl = {https://ui.adsabs.harvard.edu/abs/2014ApJ...795..146L}
}

@ARTICLE{Batcheldor2010,
       author = {{Batcheldor}, D. and {Robinson}, A. and {Axon}, D.~J. and {Perlman}, E.~S. and {Merritt}, D.},
        title = "{A Displaced Supermassive Black Hole in M87}",
      journal = {\apjl},
         year = 2010,
        month = jul,
       volume = {717},
       number = {1},
        pages = {L6-L10},
          doi = {10.1088/2041-8205/717/1/L6},
archivePrefix = {arXiv},
       eprint = {1005.2173},
 primaryClass = {astro-ph.CO},
       adsurl = {https://ui.adsabs.harvard.edu/abs/2010ApJ...717L...6B}
}

@ARTICLE{Kormendy1999,
       author = {{Kormendy}, John and {Bender}, Ralf},
        title = "{The Double Nucleus and Central Black Hole of M31}",
      journal = {\apj},
         year = 1999,
        month = sep,
       volume = {522},
       number = {2},
        pages = {772-792},
          doi = {10.1086/307665},
       adsurl = {https://ui.adsabs.harvard.edu/abs/1999ApJ...522..772K}
}

@ARTICLE{Lauer1998,
       author = {{Lauer}, Tod R. and {Faber}, S.~M. and {Ajhar}, Edward A. and {Grillmair}, Carl J. and {Scowen}, Paul A.},
        title = "{M32 +/- 1}",
      journal = {\aj},
         year = 1998,
        month = nov,
       volume = {116},
       number = {5},
        pages = {2263-2286},
          doi = {10.1086/300617},
archivePrefix = {arXiv},
       eprint = {astro-ph/9806277},
 primaryClass = {astro-ph},
       adsurl = {https://ui.adsabs.harvard.edu/abs/1998AJ....116.2263L}
}

@ARTICLE{Lauer2005,
       author = {{Lauer}, Tod R. and {Faber}, S.~M. and {Gebhardt}, Karl and {Richstone}, Douglas and {Tremaine}, Scott and {Ajhar}, Edward A. and {Aller}, M.~C. and {Bender}, Ralf and {Dressler}, Alan and {Filippenko}, Alexei V. and {Green}, Richard and {Grillmair}, Carl J. and {Ho}, Luis C. and {Kormendy}, John and {Magorrian}, John and {Pinkney}, Jason and {Siopis}, Christos},
        title = "{The Centers of Early-Type Galaxies with Hubble Space Telescope. V. New WFPC2 Photometry}",
      journal = {\aj},
         year = 2005,
        month = may,
       volume = {129},
       number = {5},
        pages = {2138-2185},
          doi = {10.1086/429565},
archivePrefix = {arXiv},
       eprint = {astro-ph/0412040},
 primaryClass = {astro-ph},
       adsurl = {https://ui.adsabs.harvard.edu/abs/2005AJ....129.2138L}
}

@ARTICLE{Davis2018,
       author = {{Davis}, Benjamin L. and {Graham}, Alister W. and {Cameron}, Ewan},
        title = "{Black Hole Mass Scaling Relations for Spiral Galaxies. II. M $_{BH}$-M $_{*,tot}$ and M $_{BH}$-M $_{*,disk}$}",
      journal = {\apj},
         year = 2018,
        month = dec,
       volume = {869},
       number = {2},
          eid = {113},
        pages = {113},
          doi = {10.3847/1538-4357/aae820},
archivePrefix = {arXiv},
       eprint = {1810.04888},
 primaryClass = {astro-ph.GA},
       adsurl = {https://ui.adsabs.harvard.edu/abs/2018ApJ...869..113D}
}

@ARTICLE{Kormendy2013,
       author = {{Kormendy}, John and {Ho}, Luis C.},
        title = "{Coevolution (Or Not) of Supermassive Black Holes and Host Galaxies}",
      journal = {\araa},
         year = 2013,
        month = aug,
       volume = {51},
       number = {1},
        pages = {511-653},
          doi = {10.1146/annurev-astro-082708-101811},
archivePrefix = {arXiv},
       eprint = {1304.7762},
 primaryClass = {astro-ph.CO},
       adsurl = {https://ui.adsabs.harvard.edu/abs/2013ARA&A..51..511K}
}

@ARTICLE{Ngo2025a,
       author = {{Ngo}, Hai N. and {Nguyen}, Dieu D. and {Le}, Tinh Q.~T. and {Ho}, Khue N.~H. and {Ho}, Tien H.~T. and {Gallo}, Elena and {Nyland}, Kristina and {Imanishi}, Masatoshi and {Nakanishi}, Kouichiro and {Le}, Que T. and {Pacucci}, Fabio and {Girma}, Eden},
        title = "{Revisiting the Supermassive Black Hole Mass of NGC 7052 Using High Spatial Resolution Molecular Gas Observed with ALMA}",
      journal = {\apj},
         year = 2025,
        month = oct,
       volume = {992},
       number = {2},
          eid = {211},
        pages = {211},
          doi = {10.3847/1538-4357/ae0455},
       adsurl = {https://ui.adsabs.harvard.edu/abs/2025ApJ...992..211N}
}

@ARTICLE{Smith2019,
       author = {{Smith}, Mark D. and {Bureau}, Martin and {Davis}, Timothy A. and
         {Cappellari}, Michele and {Liu}, Lijie and {North}, Eve V. and
         {Onishi}, Kyoko and {Iguchi}, Satoru and {Sarzi}, Marc},
        title = "{WISDOM project - IV. A molecular gas dynamical measurement of the supermassive black hole mass in NGC 524}",
      journal = {\mnras},
         year = 2019,
        month = may,
       volume = {485},
       number = {3},
        pages = {4359-4374},
          doi = {10.1093/mnras/stz625},
archivePrefix = {arXiv},
       eprint = {1903.03124},
 primaryClass = {astro-ph.GA},
       adsurl = {https://ui.adsabs.harvard.edu/abs/2019MNRAS.485.4359S}
}

@ARTICLE{Smith2021,
       author = {{Smith}, Mark D. and {Bureau}, Martin and {Davis}, Timothy A. and {Cappellari}, Michele and {Liu}, Lijie and {Onishi}, Kyoko and {Iguchi}, Satoru and {North}, Eve V. and {Sarzi}, Marc and {Williams}, Thomas G.},
        title = "{WISDOM project - VII. Molecular gas measurement of the supermassive black hole mass in the elliptical galaxy NGC 7052}",
      journal = {\mnras},
         year = 2021,
        month = jun,
       volume = {503},
       number = {4},
        pages = {5984-5996},
          doi = {10.1093/mnras/stab791},
archivePrefix = {arXiv},
       eprint = {2103.08920},
 primaryClass = {astro-ph.GA},
       adsurl = {https://ui.adsabs.harvard.edu/abs/2021MNRAS.503.5984S}
}

@ARTICLE{North2019,
       author = {{North}, Eve V. and {Davis}, Timothy A. and {Bureau}, Martin and
         {Cappellari}, Michele and {Iguchi}, Satoru and {Liu}, Lijie and
         {Onishi}, Kyoko and {Sarzi}, Marc and {Smith}, Mark D. and
         {Williams}, Thomas G.},
        title = "{WISDOM project - V. Resolving molecular gas in Keplerian rotation around the supermassive black hole in NGC 0383}",
      journal = {\mnras},
         year = "2019",
        month = "Nov",
       volume = {490},
       number = {1},
        pages = {319-330},
          doi = {10.1093/mnras/stz2598},
archivePrefix = {arXiv},
       eprint = {1909.05884},
 primaryClass = {astro-ph.GA},
       adsurl = {https://ui.adsabs.harvard.edu/abs/2019MNRAS.490..319N}
}

@ARTICLE{Nguyen2025c,
       author = {{Nguyen}, Dieu D. and {Ngo}, Hai N. and {Cappellari}, Michele and {Le}, Tinh Q.~T. and {Ho}, Tien H.~T. and {Le}, Tuan N. and {Gallo}, Elena and {Thatte}, Niranjan and {Zou}, Fan and {Perna}, Michele and {Pereira-Santaella}, Miguel},
        title = "{Measuring the Central Dark Mass in NGC 4258 with JWST/NIRSpec Stellar Kinematics}",
      journal = {arXiv e-prints},
         year = 2025,
        month = sep,
          eid = {arXiv:2509.20519},
        pages = {arXiv:2509.20519},
archivePrefix = {arXiv},
       eprint = {2509.20519},
 primaryClass = {astro-ph.GA},
       adsurl = {https://ui.adsabs.harvard.edu/abs/2025arXiv250920519N}
}

@INPROCEEDINGS{Thatte2024,
       author = {{Thatte}, Niranjan A. and {Melotte}, Dave and {Neichel}, Benoit and {Le Mignant}, David and {Rees}, Phil and {Clarke}, Fraser and {Ferraro-Wood}, Vanessa and {Gonzalez}, Oscar and {Jones}, Maia and {{\'A}lvarez Urue{\~n}a}, Alonso and {Argelaguet Vilaseca}, Heribert and {Arribas Mocoroa}, Santiago and {Caballero}, Jos{\'e} Antonio and {Carracedo Carballal}, Gonzalo Jos{\'e} and {Estrada Piqueras}, Alberto and {Ferro}, Irene and {Garc{\'\i}a Garc{\'\i}a}, Miriam and {Lamperti}, Isabella and {Pereira Santaella}, Miguel and {Perna}, Michele and {Piqueras Lopez}, Javier and {Bouch{\'e}}, Nicolas and {Boudon}, Didier and {Daguise}, Eric and {Domenis}, Nicola and {Fensch}, J{\'e}r{\'e}my and {Olivier Flasseur}, Olivier and {Giroud}, R{\'e}mi and {Guibert}, Matthieu and {Jarno}, Aurelien and {Jeanneau}, Alexandre and {Krogager}, Jens-Kristian and {Langlois}, Maud and {Laurent}, Florence and {Loupias}, Magali and {Migniau}, Jean-Emmanuel and {Nguyen}, Dieu and {Piqueras}, Laure and {Remillieux}, Alban and {Richard}, Johan and {Pecontal}, Arlette and {Bardou}, Lisa and {Barr}, David and {Cetre}, Sylvain and {Dimoudi}, Sofia and {Dubbeldam}, Marc and {Dunn}, Andrew and {Gadotti}, Dimitri and {Guy}, Joss and {King}, David and {McLeod}, Anna and {Morris}, Simon and {Morris}, Tim and {O'Brien}, Kieran and {Ronson}, Emily and {Smith}, Russell and {Staykov}, Lazar and {Swinbank}, Mark and {Accardo}, Matteo and {Alvarez Mendez}, Domingo and {Fuerte Rodriguez}, Pablo Alberto and {George}, Elizabeth and {Ives}, Derek and {Mehrgan}, Leander and {Mueller}, Eric and {Reyes}, Javier and {Conzelmann}, Ralf and {Gutierrez Cheetham}, Pablo and {Alonso Sanchez}, Angel and {Battaglia}, Giuseppina and {Cagigas}, Miguel and {Castro-Almaz{\'a}n}, Julio A. and {Chulani}, Haresh and {Delgado-Garc{\'\i}a}, Graciela and {Fernandez Izquierdo}, Patricia and {Esparza-Arredondo}, Donaji and {Garc{\'\i}a-Lorenzo}, Bego{\~n}a. and {Hern{\'a}ndez Gonz{\'a}lez}, Alberto and {Hern{\'a}ndez Su{\'a}rez}, Elvio and {Licandro}, Javier and {Joven}, Enrique and {L{\'o}pez L{\'o}pez}, Roberto and {Lujan Gonzalez}, Alejandro Antonio and {Mart{\'\i}n Hernando}, Yolanda and {Mart{\'\i}n-Navarro}, Ignacio and {Mediavilla}, Evencio and {Men{\'e}ndez Mendoza}, Sa{\'u}l and {Montoya Mart{\'\i}nez}, Luz Maria and {Pe{\~n}ate Castro}, Jos{\'e} and {Murgas}, Felipe and {Pall{\'e}}, Enric and {P{\'e}rez}, {\'A}lvaro and {Rasilla}, Jose Luis and {Rebolo}, Rafael and {Rodr{\'\i}guez}, Horacio and {Rodr{\'\i}guez Ramos}, Luis Fernando and {S{\'a}nchez B{\'e}jar}, Victor and {Shahbaz}, Tariq and {Vega Moreno}, Afrodisio and {Viera}, Teodora and {Bonnefoy}, Micka{\"e}l. and {Bret}, Tony and {Carlotti}, Alexis and {Correia}, Jean-Jacques and {Curaba}, St{\'e}phane and {Delboulb{\'e}}, Alain and {Guieu}, Sylvain and {Hours}, Adrien and {Hubert}, Zoltan and {Jocou}, Laurent and {Magnard}, Yves and {Michaud}, Laurence and {Moulin}, Thibaut and {Pancher}, Fabrice and {Rabou}, Patrick and {Rochat}, Sylvain and {Stadler}, Eric and {Contini}, Thierry and {Larrieu}, Marie and {Mamessier}, S{\'e}bastien and {Boebion}, Olivier and {Fantei-Caujolle}, Yan and {Lecron}, Daniel and {Amram}, Philippe and {Blanchard}, Patrick and {Bon}, William and {Bonnefoi}, Anne and {Bozier}, Alexandre and {Ceria}, William and {Challita}, Zalpha and {Charles}, Yannick and {Choquet}, Elodie and {Costille}, Anne and {Delsanti}, Audrey and {Dohlen}, Kjetil and {Ducret}, Franck and {El Hadi}, Kacem and {Foulon}, Benjamin and {Gimenez}, Jean-Luc and {Groussin}, Olivier and {Jaquet}, Marc and {Renault}, Edgard and {Rouquette}, Paul and {Sanchez}, Patrice and {Vigan}, Arthur and {Zavagno}, Annie and {F{\'e}tick}, Romain and {Fusco}, Thierry and {H{\'e}ritier}, Cedric and {Sauvage}, Jean-Francois and {Vedrenne}, Nicolas and {Aksoy}, Demet and {Caldwell}, Martin and {Fitzpatrick}, Ann and {Geddert}, Carl and {Hiscock}, Peter and {Johnson}, Emma and {Nalagatla}, Murali and {Saraff}, Louise and {Shreeves}, Joe and {Tildesley}, Matthew and {Wells}, Mark and {Aretos}, Anastasios and {Barrett}, Lee and {Black}, Martin and {Bond}, Charlotte and {Brierley}, Saskia and {Bryson}, Ian and {Calderhead}, Amelia and {Campbell}, Kenny and {Carruthers}, James and {Chapman}, Lee and {Cochrane}, William and {Gillespie}, Rory and {Harman}, Joel and {Harvey}, Douglas and {Harvey}, Eamonn and {Johnson}, Bethany and {Louth}, Tom and {MacIntosh}, Mike and {MacIver}, Anna and {Miller}, Chris and {Montgomery}, David and {Murali}, Meenu and {Murray}, John and {O'Malley}, Norman and {Sanchez-Janssen}, Ruben and {Schwartz}, Noah and {Smith}, Patrick and {Strachan}, Jonathan and {Todd}, Stephen and {Wasley}, Dawn and {Wilson}, Sandi and {Zhou}, Junyi and {Bell}, Eric and {Gnedin}, Oleg and {Gultekin}, Kayhan and {Mateo}, Mario and {Meyer}, Michael and {Birkby}, Jayne},
        title = "{HARMONI at ELT: project status and instrument overview}",
    booktitle = {Ground-based and Airborne Instrumentation for Astronomy X},
         year = 2024,
       editor = {{Bryant}, Julia J. and {Motohara}, Kentaro and {Vernet}, Jo{\"e}l. R.~D.},
       series = {Society of Photo-Optical Instrumentation Engineers (SPIE) Conference Series},
       volume = {13096},
        month = jul,
          eid = {1309614},
        pages = {1309614},
          doi = {10.1117/12.3018520},
       adsurl = {https://ui.adsabs.harvard.edu/abs/2024SPIE13096E..14T}
}

@INPROCEEDINGS{Nguyen2019conf,
       author = {{Nguyen}, Dieu D.},
        title = "{Uncovering the Census of Black Holes in sub-Milky Way Mass Galaxies}",
    booktitle = {ALMA2019: Science Results and Cross-Facility Synergies},
         year = 2019,
        month = dec,
          eid = {106},
        pages = {106},
          doi = {10.5281/zenodo.3585410},
       adsurl = {https://ui.adsabs.harvard.edu/abs/2019asrc.confE.106N}
}

@ARTICLE{Bentz2025,
       author = {{Bentz}, Misty C.},
        title = "{The NIRSpec IFU Point Spread Function}",
      journal = {Research Notes of the American Astronomical Society},
         year = 2025,
        month = may,
       volume = {9},
       number = {5},
          eid = {128},
        pages = {128},
          doi = {10.3847/2515-5172/adddac},
       adsurl = {https://ui.adsabs.harvard.edu/abs/2025RNAAS...9..128B}
}

@ARTICLE{Tahmasebzadeh2025,
       author = {{Tahmasebzadeh}, Behzad and {Taylor}, Matthew A. and {Valluri}, Monica and {Yoshino}, Haruka and {Vasiliev}, Eugene and {Drinkwater}, Michael J. and {Thompson}, Solveig and {Dage}, Kristen and {C{\^o}t{\'e}}, Patrick and {Ferrarese}, Laura and {Akiba}, Tatsuya and {Baldassare}, Vivienne and {Bentz}, Misty C. and {Blakeslee}, John P. and {Baumgardt}, Holger and {Ko}, Youkyung and {Liu}, Chengze and {Madigan}, Ann-Marie and {Peng}, Eric W. and {Roediger}, Joel and {Wang}, Kaixiang and {Woods}, Tyrone E.},
        title = "{A JWST View of the Overmassive Black Hole in NGC 4486B}",
      journal = {\apjl},
         year = 2025,
        month = aug,
       volume = {989},
       number = {2},
          eid = {L42},
        pages = {L42},
          doi = {10.3847/2041-8213/adf728},
archivePrefix = {arXiv},
       eprint = {2505.14676},
 primaryClass = {astro-ph.GA},
       adsurl = {https://ui.adsabs.harvard.edu/abs/2025ApJ...989L..42T}
}

@ARTICLE{Ngo2025b,
       author = {{Ngo}, Hai N and {Nguyen}, Dieu D. and {Nguyen}, Truong N. and {Dang}, Trung H. and {Ho}, Tien H.~T.},
        title = "{Extending the simulations of intermediate-mass black hole mass measurements to Virgo Cluster using ELT/HARMONI high resolution integral-field stellar kinematics}",
      journal = {arXiv e-prints},
         year = 2025,
        month = sep,
          eid = {arXiv:2509.03364},
        pages = {arXiv:2509.03364},
archivePrefix = {arXiv},
       eprint = {2509.03364},
 primaryClass = {astro-ph.GA},
       adsurl = {https://ui.adsabs.harvard.edu/abs/2025arXiv250903364N}
}

@Article{Ngo2025c,
AUTHOR = {Ngo, Hai N. and Nguyen, Dieu D. and Le, Tinh T. Q. and Ho, Tien H. T. and Nguyen, Truong N. and Dang, Trung H.},
TITLE = {Detecting Intermediate-Mass Black Holes out to 20 Mpc with ELT/HARMONI: The Case of FCC 119},
JOURNAL = {Universe},
VOLUME = {11},
YEAR = {2025},
NUMBER = {11},
ARTICLE-NUMBER = {360},
URL = {https://www.mdpi.com/2218-1997/11/11/360},
ISSN = {2218-1997},
DOI = {10.3390/universe11110360}
}

@ARTICLE{D’Eugenio2024NaAs,
       author = {{D'Eugenio}, Francesco and {P{\'e}rez-Gonz{\'a}lez}, Pablo G. and {Maiolino}, Roberto and {Scholtz}, Jan and {Perna}, Michele and {Circosta}, Chiara and {{\"U}bler}, Hannah and {Arribas}, Santiago and {B{\"o}ker}, Torsten and {Bunker}, Andrew J. and {Carniani}, Stefano and {Charlot}, Stephane and {Chevallard}, Jacopo and {Cresci}, Giovanni and {Curtis-Lake}, Emma and {Jones}, Gareth C. and {Kumari}, Nimisha and {Lamperti}, Isabella and {Looser}, Tobias J. and {Parlanti}, Eleonora and {Rix}, Hans-Walter and {Robertson}, Brant and {Rodr{\'\i}guez Del Pino}, Bruno and {Tacchella}, Sandro and {Venturi}, Giacomo and {Willott}, Chris J.},
        title = "{A fast-rotator post-starburst galaxy quenched by supermassive black-hole feedback at z = 3}",
      journal = {NaAs},
         year = 2024,
        month = nov,
       volume = {8},
        pages = {1443-1456},
          doi = {10.1038/s41550-024-02345-1},
archivePrefix = {arXiv},
       eprint = {2308.06317},
 primaryClass = {astro-ph.GA},
       adsurl = {https://ui.adsabs.harvard.edu/abs/2024NatAs...8.1443D}
}

@ARTICLE{Graham_Sahu2023,
       author = {{Graham}, Alister W. and {Sahu}, Nandini},
        title = "{Reading the tea leaves in the M$_{bh}$-M$_{*,sph}$ and M$_{bh}$-R$_{e,sph}$ diagrams: dry and gaseous mergers with remnant angular momentum}",
      journal = {\mnras},
         year = 2023,
        month = apr,
       volume = {520},
       number = {2},
        pages = {1975-1996},
          doi = {10.1093/mnras/stad087},
archivePrefix = {arXiv},
       eprint = {2210.09557},
 primaryClass = {astro-ph.GA},
       adsurl = {https://ui.adsabs.harvard.edu/abs/2023MNRAS.520.1975G}
}

@ARTICLE{vandenBosch2016,
       author = {{van den Bosch}, Remco C.~E.},
        title = "{Unification of the fundamental plane and Super Massive Black Hole Masses}",
      journal = {\apj},
         year = 2016,
        month = nov,
       volume = {831},
       number = {2},
          eid = {134},
        pages = {134},
          doi = {10.3847/0004-637X/831/2/134},
archivePrefix = {arXiv},
       eprint = {1606.01246},
 primaryClass = {astro-ph.GA},
       adsurl = {https://ui.adsabs.harvard.edu/abs/2016ApJ...831..134V}
}

@ARTICLE{Cappellari2013a,
       author = {{Cappellari}, Michele and {Scott}, Nicholas and {Alatalo}, Katherine and {Blitz}, Leo and {Bois}, Maxime and {Bournaud}, Fr{\'e}d{\'e}ric and {Bureau}, M. and {Crocker}, Alison F. and {Davies}, Roger L. and {Davis}, Timothy A. and {de Zeeuw}, P.~T. and {Duc}, Pierre-Alain and {Emsellem}, Eric and {Khochfar}, Sadegh and {Krajnovi{\'c}}, Davor and {Kuntschner}, Harald and {McDermid}, Richard M. and {Morganti}, Raffaella and {Naab}, Thorsten and {Oosterloo}, Tom and {Sarzi}, Marc and {Serra}, Paolo and {Weijmans}, Anne-Marie and {Young}, Lisa M.},
        title = "{The ATLAS$^{3D}$ project - XV. Benchmark for early-type galaxies scaling relations from 260 dynamical models: mass-to-light ratio, dark matter, Fundamental Plane and Mass Plane}",
      journal = {\mnras},
         year = 2013,
        month = jul,
       volume = {432},
       number = {3},
        pages = {1709-1741},
          doi = {10.1093/mnras/stt562},
archivePrefix = {arXiv},
       eprint = {1208.3522},
 primaryClass = {astro-ph.CO},
       adsurl = {https://ui.adsabs.harvard.edu/abs/2013MNRAS.432.1709C}
}

@Book{VanRossum2009,
  author    = {Van Rossum, Guido and Drake, Fred L.},
  publisher = {CreateSpace},
  title     = {Python 3 Reference Manual},
  year      = {2009},
  address   = {Scotts Valley, CA},
  isbn      = {1441412697},
}

@Article{Hunter2007,
  author    = {Hunter, J. D.},
  journal   = {Computing In Science \& Engineering},
  title     = {Matplotlib: A 2D graphics environment},
  year      = {2007},
  number    = {3},
  pages     = {90--95},
  volume    = {9},
  doi       = {10.1109/MCSE.2007.55},
  publisher = {IEEE COMPUTER SOC},
}

@Article{Harris2020,
  author    = {Harris, Charles R. and Millman, K. Jarrod and van der Walt, St{\'{e}}fan J. and Gommers, Ralf and Virtanen, Pauli and Cournapeau, David and Wieser, Eric and Taylor, Julian and Berg, Sebastian and Smith, Nathaniel J. and Kern, Robert and Picus, Matti and Hoyer, Stephan and van Kerkwijk, Marten H. and Brett, Matthew and Haldane, Allan and del R{\'{\i}}o, Jaime Fern{\'{a}}ndez and Wiebe, Mark and Peterson, Pearu and G{\'{e}}rard-Marchant, Pierre and Sheppard, Kevin and Reddy, Tyler and Weckesser, Warren and Abbasi, Hameer and Gohlke, Christoph and Oliphant, Travis E.},
  journal   = {Nature},
  title     = {Array programming with {NumPy}},
  year      = {2020},
  month     = sep,
  number    = {7825},
  pages     = {357--362},
  volume    = {585},
  doi       = {10.1038/s41586-020-2649-2},
  publisher = {Springer Science and Business Media {LLC}},
}

@Article{Virtanen2020,
  author    = {Virtanen, Pauli and Gommers, Ralf and Oliphant, Travis E. and Haberland, Matt and Reddy, Tyler and Cournapeau, David and Burovski, Evgeni and Peterson, Pearu and Weckesser, Warren and Bright, Jonathan and van der Walt, St{\'{e}}fan J. and Brett, Matthew and Wilson, Joshua and Millman, K. Jarrod and Mayorov, Nikolay and Nelson, Andrew R. J. and Jones, Eric and Kern, Robert and Larson, Eric and Carey, C. J. and Polat, {\.{I}}lhan and Feng, Yu and Moore, Eric W. and VanderPlas, Jake and Laxalde, Denis and Perktold, Josef and Cimrman, Robert and Henriksen, Ian and Quintero, E. A. and Harris, Charles R. and Archibald, Anne M. and Ribeiro, Ant{\^{o}}nio H. and Pedregosa, Fabian and van Mulbregt, Paul and Vijaykumar, Aditya and Bardelli, Alessandro Pietro and Rothberg, Alex and Hilboll, Andreas and Kloeckner, Andreas and Scopatz, Anthony and Lee, Antony and Rokem, Ariel and Woods, C. Nathan and Fulton, Chad and Masson, Charles and Häggström, Christian and Fitzgerald, Clark and Nicholson, David A. and Hagen, David R. and Pasechnik, Dmitrii V. and Olivetti, Emanuele and Martin, Eric and Wieser, Eric and Silva, Fabrice and Lenders, Felix and Wilhelm, Florian and Young, G. and Price, Gavin A. and Ingold, Gert-Ludwig and Allen, Gregory E. and Lee, Gregory R. and Audren, Herv{\'{e}} and Probst, Irvin and Dietrich, Jörg P. and Silterra, Jacob and Webber, James T. and Slavi{\v{c}}, Janko and Nothman, Joel and Buchner, Johannes and Kulick, Johannes and Schönberger, Johannes L. and de Miranda Cardoso, Jos{\'{e}} Vin{\'{\i}}cius and Reimer, Joscha and Harrington, Joseph and Rodr{\'{\i}}guez, Juan Luis Cano and Nunez-Iglesias, Juan and Kuczynski, Justin and Tritz, Kevin and Thoma, Martin and Newville, Matthew and Kümmerer, Matthias and Bolingbroke, Maximilian and Tartre, Michael and Pak, Mikhail and Smith, Nathaniel J. and Nowaczyk, Nikolai and Shebanov, Nikolay and Pavlyk, Oleksandr and Brodtkorb, Per A. and Lee, Perry and McGibbon, Robert T. and Feldbauer, Roman and Lewis, Sam and Tygier, Sam and Sievert, Scott and Vigna, Sebastiano and Peterson, Stefan and More, Surhud and Pudlik, Tadeusz and Oshima, Takuya and Pingel, Thomas J. and Robitaille, Thomas P. and Spura, Thomas and Jones, Thouis R. and Cera, Tim and Leslie, Tim and Zito, Tiziano and Krauss, Tom and Upadhyay, Utkarsh and Halchenko, Yaroslav O. and V{\'{a}}zquez-Baeza, Yoshiki},
  journal   = {Nature Methods},
  title     = {{SciPy} 1.0: fundamental algorithms for scientific computing in Python},
  year      = {2020},
  month     = feb,
  number    = {3},
  pages     = {261--272},
  volume    = {17},
  doi       = {10.1038/s41592-019-0686-2},
  publisher = {Springer Science and Business Media {LLC}},
}

@ARTICLE{Jarrett2003,
       author = {{Jarrett}, T.~H. and {Chester}, T. and {Cutri}, R. and {Schneider}, S.~E. and {Huchra}, J.~P.},
        title = "{The 2MASS Large Galaxy Atlas}",
      journal = {\aj},
         year = 2003,
        month = feb,
       volume = {125},
       number = {2},
        pages = {525-554},
          doi = {10.1086/345794},
       adsurl = {https://ui.adsabs.harvard.edu/abs/2003AJ....125..525J}
}

@ARTICLE{Durrell2010,
       author = {{Durrell}, Patrick R. and {Sarajedini}, Ata and {Chandar}, Rupali},
        title = "{Deep HST/ACS Photometry of the M81 Halo}",
      journal = {\apj},
         year = 2010,
        month = aug,
       volume = {718},
       number = {2},
        pages = {1118-1127},
          doi = {10.1088/0004-637X/718/2/1118},
archivePrefix = {arXiv},
       eprint = {1006.2036},
 primaryClass = {astro-ph.CO},
       adsurl = {https://ui.adsabs.harvard.edu/abs/2010ApJ...718.1118D}
}

@ARTICLE{Gebhardt2003,
       author = {{Gebhardt}, Karl and {Richstone}, Douglas and {Tremaine}, Scott and {Lauer}, Tod R. and {Bender}, Ralf and {Bower}, Gary and {Dressler}, Alan and {Faber}, S.~M. and {Filippenko}, Alexei V. and {Green}, Richard and {Grillmair}, Carl and {Ho}, Luis C. and {Kormendy}, John and {Magorrian}, John and {Pinkney}, Jason},
        title = "{Axisymmetric Dynamical Models of the Central Regions of Galaxies}",
      journal = {\apj},
         year = 2003,
        month = jan,
       volume = {583},
       number = {1},
        pages = {92-115},
          doi = {10.1086/345081},
archivePrefix = {arXiv},
       eprint = {astro-ph/0209483},
 primaryClass = {astro-ph},
       adsurl = {https://ui.adsabs.harvard.edu/abs/2003ApJ...583...92G}
}

@article{Haario2001,
author = {Heikki Haario and Eero Saksman and Johanna Tamminen},
title = {{An adaptive Metropolis algorithm}},
year = 2001,
month = Apr,
volume = {7},
journal = {Bernoulli},
number = {2},
publisher = {Bernoulli Society for Mathematical Statistics and Probability},
pages = {223 - 242},
adsurl = {https://projecteuclid.org/journals/bernoulli/volume-7/issue-2/An-adaptive-Metropolis-algorithm/bj/1080222083.full},
}

@ARTICLE{Gultekin2009,
       author = {{G{\"u}ltekin}, Kayhan and {Richstone}, Douglas O. and {Gebhardt}, Karl and {Lauer}, Tod R. and {Tremaine}, Scott and {Aller}, M.~C. and {Bender}, Ralf and {Dressler}, Alan and {Faber}, S.~M. and {Filippenko}, Alexei V. and {Green}, Richard and {Ho}, Luis C. and {Kormendy}, John and {Magorrian}, John and {Pinkney}, Jason and {Siopis}, Christos},
        title = "{The M-{\ensuremath{\sigma}} and M-L Relations in Galactic Bulges, and Determinations of Their Intrinsic Scatter}",
      journal = {\apj},
         year = 2009,
        month = jun,
       volume = {698},
       number = {1},
        pages = {198-221},
          doi = {10.1088/0004-637X/698/1/198},
archivePrefix = {arXiv},
       eprint = {0903.4897},
 primaryClass = {astro-ph.GA},
       adsurl = {https://ui.adsabs.harvard.edu/abs/2009ApJ...698..198G}
}

@ARTICLE{Krajnovic2006,
   author = {{Krajnovi{\'c}}, D. and {Cappellari}, M. and {de Zeeuw}, P.~T. and 
  {Copin}, Y.},
    title = "{Kinemetry: a generalization of photometry to the higher moments of the line-of-sight velocity distribution}",
  journal = {\mnras},
   eprint = {astro-ph/0512200},
     year = 2006,
    month = mar,
   volume = 366,
    pages = {787-802},
      doi = {10.1111/j.1365-2966.2005.09902.x},
   adsurl = {http://adsabs.harvard.edu/abs/2006MNRAS.366..787K}
}

@ARTICLE{Willmer2018,
       author = {{Willmer}, Christopher N.~A.},
        title = "{The Absolute Magnitude of the Sun in Several Filters}",
      journal = {\apjs},
         year = 2018,
        month = jun,
       volume = {236},
       number = {2},
          eid = {47},
        pages = {47},
          doi = {10.3847/1538-4365/aabfdf},
archivePrefix = {arXiv},
       eprint = {1804.07788},
 primaryClass = {astro-ph.SR},
       adsurl = {https://ui.adsabs.harvard.edu/abs/2018ApJS..236...47W}
}

@ARTICLE{Schlafly2011,
       author = {{Schlafly}, Edward F. and {Finkbeiner}, Douglas P.},
        title = "{Measuring Reddening with Sloan Digital Sky Survey Stellar Spectra and Recalibrating SFD}",
      journal = {\apj},
         year = 2011,
        month = aug,
       volume = {737},
       number = {2},
          eid = {103},
        pages = {103},
          doi = {10.1088/0004-637X/737/2/103},
archivePrefix = {arXiv},
       eprint = {1012.4804},
 primaryClass = {astro-ph.GA},
       adsurl = {https://ui.adsabs.harvard.edu/abs/2011ApJ...737..103S}
}

@ARTICLE{Thater2019,
       author = {{Thater}, Sabine and {Krajnovi{\'c}}, Davor and {Cappellari}, Michele and {Davis}, Timothy A. and {de Zeeuw}, P. Tim and {McDermid}, Richard M. and {Sarzi}, Marc},
        title = "{Six new supermassive black hole mass determinations from adaptive-optics assisted SINFONI observations}",
      journal = {\aap},
         year = 2019,
        month = may,
       volume = {625},
          eid = {A62},
        pages = {A62},
          doi = {10.1051/0004-6361/201834808},
archivePrefix = {arXiv},
       eprint = {1902.10175},
 primaryClass = {astro-ph.GA},
       adsurl = {https://ui.adsabs.harvard.edu/abs/2019A&A...625A..62T}
}

@ARTICLE{Drehmer2015,
       author = {{Drehmer}, Daniel Alf and {Storchi-Bergmann}, Thaisa and {Ferrari}, Fabricio and {Cappellari}, Michele and {Riffel}, Rogemar A.},
        title = "{The benchmark black hole in NGC 4258: dynamical models from high-resolution two-dimensional stellar kinematics}",
      journal = {\mnras},
         year = 2015,
        month = jun,
       volume = {450},
       number = {1},
        pages = {128-144},
          doi = {10.1093/mnras/stv536},
archivePrefix = {arXiv},
       eprint = {1503.04540},
 primaryClass = {astro-ph.GA},
       adsurl = {https://ui.adsabs.harvard.edu/abs/2015MNRAS.450..128D}
}

@ARTICLE{vandenBosch2009,
       author = {{van den Bosch}, Remco C.~E. and {van de Ven}, Glenn},
        title = "{Recovering the intrinsic shape of early-type galaxies}",
      journal = {\mnras},
         year = "2009",
        month = "Sep",
       volume = {398},
       number = {3},
        pages = {1117-1128},
          doi = {10.1111/j.1365-2966.2009.15177.x},
archivePrefix = {arXiv},
       eprint = {0811.3474},
 primaryClass = {astro-ph},
       adsurl = {https://ui.adsabs.harvard.edu/abs/2009MNRAS.398.1117V}
}

@ARTICLE{Mitzkus2017,
   author = {{Mitzkus}, M. and {Cappellari}, M. and {Walcher}, C.~J.},
    title = "{Dominant dark matter and a counter-rotating disc: MUSE view of the low-luminosity S0 galaxy NGC 5102}",
  journal = {\mnras},
archivePrefix = "arXiv",
   eprint = {1610.04516},
     year = 2017,
    month = feb,
   volume = 464,
    pages = {4789-4806},
      doi = {10.1093/mnras/stw2677},
   adsurl = {http://adsabs.harvard.edu/abs/2017MNRAS.464.4789M}
}

@Misc{bradley2024,
  author    = {Larry Bradley and Brigitta Sip{\H o}cz and Thomas Robitaille and Erik Tollerud and Z\`e Vin{\'{\i}}cius and Christoph Deil and Kyle Barbary and Tom J Wilson and Ivo Busko and Axel Donath and Hans Moritz G{\"u}nther and Mihai Cara and P. L. Lim and Sebastian Me{\ss}linger and Simon Conseil and Zach Burnett and Azalee Bostroem and Michael Droettboom and E. M. Bray and Lars Andersen Bratholm and Adam Ginsburg and William Jamieson and Geert Barentsen and Matt Craig and Brett M. Morris and Marshall Perrin and Shivangee Rathi and Sergio Pascual and Iskren Y. Georgiev},
  month     = oct,
  title     = {astropy/photutils: 2.0.2},
  year      = {2024},
  doi       = {10.5281/zenodo.13989456},
  publisher = {Zenodo},
  url       = {https://doi.org/10.5281/zenodo.13989456},
  version   = {2.0.2},
}

@ARTICLE{Cappellari2003,
       author = {{Cappellari}, Michele and {Copin}, Yannick},
        title = "{Adaptive spatial binning of integral-field spectroscopic data using Voronoi tessellations}",
      journal = {\mnras},
         year = 2003,
        month = jun,
       volume = {342},
       number = {2},
        pages = {345-354},
          doi = {10.1046/j.1365-8711.2003.06541.x},
archivePrefix = {arXiv},
       eprint = {astro-ph/0302262},
 primaryClass = {astro-ph},
       adsurl = {https://ui.adsabs.harvard.edu/abs/2003MNRAS.342..345C}
}

@Article{Cappellari2020,
  author        = {Cappellari, Michele},
  title         = {{Efficient solution of the anisotropic spherically aligned axisymmetric Jeans equations of stellar hydrodynamics for galactic dynamics}},
  doi           = {10.1093/mnras/staa959},
  eprint        = {1907.09894},
  number        = {4},
  pages         = {4819--4837},
  volume        = {494},
  adsurl        = {https://ui.adsabs.harvard.edu/abs/2020MNRAS.494.4819C},
  archiveprefix = {arXiv},
  journal       = {\mnras},
  month         = apr,
  primaryclass  = {astro-ph.GA},
  year          = {2020},
}

@ARTICLE{AstropyCollaboration2022,
       author = {{Astropy Collaboration} and {Price-Whelan}, Adrian M. and {Lim}, Pey Lian and {Earl}, Nicholas and {Starkman}, Nathaniel and {Bradley}, Larry and {Shupe}, David L. and {Patil}, Aarya A. and {Corrales}, Lia and {Brasseur}, C.~E. and {N{\"o}the}, Maximilian and {Donath}, Axel and {Tollerud}, Erik and {Morris}, Brett M. and {Ginsburg}, Adam and {Vaher}, Eero and {Weaver}, Benjamin A. and {Tocknell}, James and {Jamieson}, William and {van Kerkwijk}, Marten H. and {Robitaille}, Thomas P. and {Merry}, Bruce and {Bachetti}, Matteo and {G{\"u}nther}, H. Moritz and {Aldcroft}, Thomas L. and {Alvarado-Montes}, Jaime A. and {Archibald}, Anne M. and {B{\'o}di}, Attila and {Bapat}, Shreyas and {Barentsen}, Geert and {Baz{\'a}n}, Juanjo and {Biswas}, Manish and {Boquien}, M{\'e}d{\'e}ric and {Burke}, D.~J. and {Cara}, Daria and {Cara}, Mihai and {Conroy}, Kyle E. and {Conseil}, Simon and {Craig}, Matthew W. and {Cross}, Robert M. and {Cruz}, Kelle L. and {D'Eugenio}, Francesco and {Dencheva}, Nadia and {Devillepoix}, Hadrien A.~R. and {Dietrich}, J{\"o}rg P. and {Eigenbrot}, Arthur Davis and {Erben}, Thomas and {Ferreira}, Leonardo and {Foreman-Mackey}, Daniel and {Fox}, Ryan and {Freij}, Nabil and {Garg}, Suyog and {Geda}, Robel and {Glattly}, Lauren and {Gondhalekar}, Yash and {Gordon}, Karl D. and {Grant}, David and {Greenfield}, Perry and {Groener}, Austen M. and {Guest}, Steve and {Gurovich}, Sebastian and {Handberg}, Rasmus and {Hart}, Akeem and {Hatfield-Dodds}, Zac and {Homeier}, Derek and {Hosseinzadeh}, Griffin and {Jenness}, Tim and {Jones}, Craig K. and {Joseph}, Prajwel and {Kalmbach}, J. Bryce and {Karamehmetoglu}, Emir and {Ka{\l}uszy{\'n}ski}, Miko{\l}aj and {Kelley}, Michael S.~P. and {Kern}, Nicholas and {Kerzendorf}, Wolfgang E. and {Koch}, Eric W. and {Kulumani}, Shankar and {Lee}, Antony and {Ly}, Chun and {Ma}, Zhiyuan and {MacBride}, Conor and {Maljaars}, Jakob M. and {Muna}, Demitri and {Murphy}, N.~A. and {Norman}, Henrik and {O'Steen}, Richard and {Oman}, Kyle A. and {Pacifici}, Camilla and {Pascual}, Sergio and {Pascual-Granado}, J. and {Patil}, Rohit R. and {Perren}, Gabriel I. and {Pickering}, Timothy E. and {Rastogi}, Tanuj and {Roulston}, Benjamin R. and {Ryan}, Daniel F. and {Rykoff}, Eli S. and {Sabater}, Jose and {Sakurikar}, Parikshit and {Salgado}, Jes{\'u}s and {Sanghi}, Aniket and {Saunders}, Nicholas and {Savchenko}, Volodymyr and {Schwardt}, Ludwig and {Seifert-Eckert}, Michael and {Shih}, Albert Y. and {Jain}, Anany Shrey and {Shukla}, Gyanendra and {Sick}, Jonathan and {Simpson}, Chris and {Singanamalla}, Sudheesh and {Singer}, Leo P. and {Singhal}, Jaladh and {Sinha}, Manodeep and {Sip{\H{o}}cz}, Brigitta M. and {Spitler}, Lee R. and {Stansby}, David and {Streicher}, Ole and {{\v{S}}umak}, Jani and {Swinbank}, John D. and {Taranu}, Dan S. and {Tewary}, Nikita and {Tremblay}, Grant R. and {de Val-Borro}, Miguel and {Van Kooten}, Samuel J. and {Vasovi{\'c}}, Zlatan and {Verma}, Shresth and {de Miranda Cardoso}, Jos{\'e} Vin{\'\i}cius and {Williams}, Peter K.~G. and {Wilson}, Tom J. and {Winkel}, Benjamin and {Wood-Vasey}, W.~M. and {Xue}, Rui and {Yoachim}, Peter and {Zhang}, Chen and {Zonca}, Andrea and {Astropy Project Contributors}},
        title = "{The Astropy Project: Sustaining and Growing a Community-oriented Open-source Project and the Latest Major Release (v5.0) of the Core Package}",
      journal = {\apj},
         year = 2022,
        month = aug,
       volume = {935},
       number = {2},
          eid = {167},
        pages = {167},
          doi = {10.3847/1538-4357/ac7c74},
archivePrefix = {arXiv},
       eprint = {2206.14220},
 primaryClass = {astro-ph.IM},
       adsurl = {https://ui.adsabs.harvard.edu/abs/2022ApJ...935..167A}
}

@ARTICLE{Nguyen17conf,
       author = {{Nguyen}, Dieu D.},
        title = "{Improved dynamical constraints on the mass of the central black hole in NGC 404}",
      journal = {arXiv e-prints},
         year = 2017,
        month = dec,
          eid = {arXiv:1712.02470},
        pages = {arXiv:1712.02470},
          doi = {10.48550/arXiv.1712.02470},
archivePrefix = {arXiv},
       eprint = {1712.02470},
 primaryClass = {astro-ph.GA},
       adsurl = {https://ui.adsabs.harvard.edu/abs/2017arXiv171202470N}
}

@ARTICLE{Davis2020,
       author = {{Davis}, Timothy A. and {Nguyen}, Dieu D. and {Seth}, Anil C. and
         {Greene}, Jenny E. and {Nyland}, Kristina and {Barth}, Aaron J. and
         {Bureau}, Martin and {Cappellari}, Michele and {den Brok}, Mark and
         {Iguchi}, Satoru and {Lelli}, Federico and {Liu}, Lijie and
         {Neumayer}, Nadine and {North}, Eve V. and {Onishi}, Kyoko and
         {Sarzi}, Marc and {Smith}, Mark D. and {Williams}, Thomas G.},
        title = "{Revealing the intermediate-mass black hole at the heart of the dwarf galaxy NGC 404 with sub-parsec resolution ALMA observations}",
      journal = {\mnras},
         year = 2020,
        month = jul,
       volume = {496},
       number = {4},
        pages = {4061-4078},
          doi = {10.1093/mnras/staa1567},
archivePrefix = {arXiv},
       eprint = {2007.05536},
 primaryClass = {astro-ph.GA},
       adsurl = {https://ui.adsabs.harvard.edu/abs/2020MNRAS.496.4061D}
}

@ARTICLE{Nguyen2019,
       author = {{Nguyen}, Dieu D. and {Seth}, Anil C. and {Neumayer}, Nadine and {Iguchi}, Satoru and {Cappellari}, Michelle and {Strader}, Jay and {Chomiuk}, Laura and {Tremou}, Evangelia and {Pacucci}, Fabio and {Nakanishi}, Kouichiro and {Bahramian}, Arash and {Nguyen}, Phuong M. and {den Brok}, Mark and {Ahn}, Christopher C. and {Voggel}, Karina T. and {Kacharov}, Nikolay and {Tsukui}, Takafumi and {Ly}, Cuc K. and {Dumont}, Antoine and {Pechetti}, Renuka},
        title = "{Improved Dynamical Constraints on the Masses of the Central Black Holes in Nearby Low-mass Early-type Galactic Nuclei and the First Black Hole Determination for NGC 205}",
      journal = {\apj},
         year = 2019,
        month = feb,
       volume = {872},
       number = {1},
          eid = {104},
        pages = {104},
          doi = {10.3847/1538-4357/aafe7a},
archivePrefix = {arXiv},
       eprint = {1901.05496},
 primaryClass = {astro-ph.GA},
       adsurl = {https://ui.adsabs.harvard.edu/abs/2019ApJ...872..104N}
}

@ARTICLE{Saglia2016,
       author = {{Saglia}, R.~P. and {Opitsch}, M. and {Erwin}, P. and {Thomas}, J. and {Beifiori}, A. and {Fabricius}, M. and {Mazzalay}, X. and {Nowak}, N. and {Rusli}, S.~P. and {Bender}, R.},
        title = "{The SINFONI Black Hole Survey: The Black Hole Fundamental Plane Revisited and the Paths of (Co)evolution of Supermassive Black Holes and Bulges}",
      journal = {\apj},
         year = 2016,
        month = feb,
       volume = {818},
       number = {1},
          eid = {47},
        pages = {47},
          doi = {10.3847/0004-637X/818/1/47},
archivePrefix = {arXiv},
       eprint = {1601.00974},
 primaryClass = {astro-ph.GA},
       adsurl = {https://ui.adsabs.harvard.edu/abs/2016ApJ...818...47S}
}

@ARTICLE{Ahn2018,
       author = {{Ahn}, Christopher P. and {Seth}, Anil C. and {Cappellari}, Michele and {Krajnovi{\'c}}, Davor and {Strader}, Jay and {Voggel}, Karina T. and {Walsh}, Jonelle L. and {Bahramian}, Arash and {Baumgardt}, Holger and {Brodie}, Jean and {Chilingarian}, Igor and {Chomiuk}, Laura and {den Brok}, Mark and {Frank}, Matthias and {Hilker}, Michael and {McDermid}, Richard M. and {Mieske}, Steffen and {Neumayer}, Nadine and {Nguyen}, Dieu D. and {Pechetti}, Renuka and {Romanowsky}, Aaron J. and {Spitler}, Lee},
        title = "{The Black Hole in the Most Massive Ultracompact Dwarf Galaxy M59-UCD3}",
      journal = {\apj},
         year = 2018,
        month = may,
       volume = {858},
       number = {2},
          eid = {102},
        pages = {102},
          doi = {10.3847/1538-4357/aabc57},
archivePrefix = {arXiv},
       eprint = {1804.02399},
 primaryClass = {astro-ph.GA},
       adsurl = {https://ui.adsabs.harvard.edu/abs/2018ApJ...858..102A}
}

@ARTICLE{Nguyen2018,
       author = {{Nguyen}, Dieu D. and {Seth}, Anil C. and {Neumayer}, Nadine and {Kamann}, Sebastian and {Voggel}, Karina T. and {Cappellari}, Michele and {Picotti}, Arianna and {Nguyen}, Phuong M. and {B{\"o}ker}, Torsten and {Debattista}, Victor and {Caldwell}, Nelson and {McDermid}, Richard and {Bastian}, Nathan and {Ahn}, Christopher C. and {Pechetti}, Renuka},
        title = "{Nearby Early-type Galactic Nuclei at High Resolution: Dynamical Black Hole and Nuclear Star Cluster Mass Measurements}",
      journal = {\apj},
         year = 2018,
        month = may,
       volume = {858},
       number = {2},
          eid = {118},
        pages = {118},
          doi = {10.3847/1538-4357/aabe28},
archivePrefix = {arXiv},
       eprint = {1711.04314},
 primaryClass = {astro-ph.GA},
       adsurl = {https://ui.adsabs.harvard.edu/abs/2018ApJ...858..118N}
}

@ARTICLE{Thater2023,
       author = {{Thater}, Sabine and {Lyubenova}, Mariya and {Fahrion}, Katja and {Mart{\'\i}n-Navarro}, Ignacio and {Jethwa}, Prashin and {Nguyen}, Dieu D. and {van de Ven}, Glenn},
        title = "{Effect of the initial mass function on the dynamical SMBH mass estimate in the nucleated early-type galaxy FCC 47}",
      journal = {\aap},
         year = 2023,
        month = jul,
       volume = {675},
          eid = {A18},
        pages = {A18},
          doi = {10.1051/0004-6361/202245362},
archivePrefix = {arXiv},
       eprint = {2304.13310},
 primaryClass = {astro-ph.GA},
       adsurl = {https://ui.adsabs.harvard.edu/abs/2023A&A...675A..18T}
}

@ARTICLE{Thater2022,
       author = {{Thater}, Sabine and {Krajnovi{\'c}}, Davor and {Weilbacher}, Peter M. and {Nguyen}, Dieu D. and {Bureau}, Martin and {Cappellari}, Michele and {Davis}, Timothy A. and {Iguchi}, Satoru and {McDermid}, Richard and {Onishi}, Kyoko and {Sarzi}, Marc and {van de Ven}, Glenn},
        title = "{Cross-checking SMBH mass estimates in NGC 6958 - I. Stellar dynamics from adaptive optics-assisted MUSE observations}",
      journal = {\mnras},
         year = 2022,
        month = feb,
       volume = {509},
       number = {4},
        pages = {5416-5436},
          doi = {10.1093/mnras/stab3210},
archivePrefix = {arXiv},
       eprint = {2111.01620},
 primaryClass = {astro-ph.GA},
       adsurl = {https://ui.adsabs.harvard.edu/abs/2022MNRAS.509.5416T}
}

@ARTICLE{Nguyen2022,
       author = {{Nguyen}, Dieu D. and {Bureau}, Martin and {Thater}, Sabine and {Nyland}, Kristina and {den Brok}, Mark and {Cappellari}, Michele and {Davis}, Timothy A. and {Greene}, Jenny E. and {Neumayer}, Nadine and {Imanishi}, Masatoshi and {Izumi}, Takuma and {Kawamuro}, Taiki and {Baba}, Shunsuke and {Nguyen}, Phuong M. and {Iguchi}, Satoru and {Tsukui}, Takafumi and {Lam}, T.~N. and {Ho}, Than},
        title = "{The MBHBM$^{{\ensuremath{\star}}}$ Project - II. Molecular gas kinematics in the lenticular galaxy NGC 3593 reveal a supermassive black hole}",
      journal = {\mnras},
         year = 2022,
        month = jan,
       volume = {509},
       number = {2},
        pages = {2920-2939},
          doi = {10.1093/mnras/stab3016},
archivePrefix = {arXiv},
       eprint = {2110.08476},
 primaryClass = {astro-ph.GA},
       adsurl = {https://ui.adsabs.harvard.edu/abs/2022MNRAS.509.2920N}
}

@ARTICLE{Nguyen2021,
       author = {{Nguyen}, Dieu D. and {Izumi}, Takuma and {Thater}, Sabine and {Imanishi}, Masatoshi and {Kawamuro}, Taiki and {Baba}, Shunsuke and {Nakano}, Suzuka and {Turner}, Jean L. and {Kohno}, Kotaro and {Matsushita}, Satoki and {Mart{\'\i}n}, Sergio and {Meier}, David S. and {Nguyen}, Phuong M. and {Nguyen}, Lam T.},
        title = "{Black hole mass measurement using ALMA observations of [CI] and CO emissions in the Seyfert 1 galaxy NGC 7469}",
      journal = {\mnras},
         year = 2021,
        month = jul,
       volume = {504},
       number = {3},
        pages = {4123-4142},
          doi = {10.1093/mnras/stab1002},
archivePrefix = {arXiv},
       eprint = {2104.03539},
 primaryClass = {astro-ph.GA},
       adsurl = {https://ui.adsabs.harvard.edu/abs/2021MNRAS.504.4123N}
}

@ARTICLE{Nguyen2020,
       author = {{Nguyen}, Dieu D. and {den Brok}, Mark and {Seth}, Anil C. and
         {Davis}, Timothy A. and {Greene}, Jenny E. and {Cappellari}, Michelle and
         {Jensen}, Joseph B. and {Thater}, Sabine and {Iguchi}, Satoru and
         {Imanishi}, Masatoshi and {Izumi}, Takuma and {Nyland}, Kristina and
         {Neumayer}, Nadine and {Nakanishi}, Kouichiro and {Nguyen}, Phuong M. and
         {Tsukui}, Takafumi and {Bureau}, Martin and {Onishi}, Kyoko and
         {Nguyen}, Quang L. and {Le}, Ngan M.},
        title = "{The MBHBM$_{{\ensuremath{\star}}}$ Project. I. Measurement of the Central Black Hole Mass in Spiral Galaxy NGC 3504 Using Molecular Gas Kinematics}",
      journal = {\apj},
         year = 2020,
        month = mar,
       volume = {892},
       number = {1},
          eid = {68},
        pages = {68},
          doi = {10.3847/1538-4357/ab77aa},
archivePrefix = {arXiv},
       eprint = {1902.03813},
 primaryClass = {astro-ph.GA},
       adsurl = {https://ui.adsabs.harvard.edu/abs/2020ApJ...892...68N}
}

@ARTICLE{Voggel2018,
       author = {{Voggel}, Karina T. and {Seth}, Anil C. and {Neumayer}, Nadine and {Mieske}, Steffen and {Chilingarian}, Igor and {Ahn}, Christopher and {Baumgardt}, Holger and {Hilker}, Michael and {Nguyen}, Dieu D. and {Romanowsky}, Aaron J. and {Walsh}, Jonelle L. and {den Brok}, Mark and {Strader}, Jay},
        title = "{Upper Limits on the Presence of Central Massive Black Holes in Two Ultra-compact Dwarf Galaxies in Centaurus A}",
      journal = {\apj},
         year = 2018,
        month = may,
       volume = {858},
       number = {1},
          eid = {20},
        pages = {20},
          doi = {10.3847/1538-4357/aabae5},
archivePrefix = {arXiv},
       eprint = {1803.09750},
 primaryClass = {astro-ph.GA},
       adsurl = {https://ui.adsabs.harvard.edu/abs/2018ApJ...858...20V}
}

@ARTICLE{Nguyen2017,
       author = {{Nguyen}, Dieu D. and {Seth}, Anil C. and {den Brok}, Mark and {Neumayer}, Nadine and {Cappellari}, Michele and {Barth}, Aaron J. and {Caldwell}, Nelson and {Williams}, Benjamin F. and {Binder}, Breanna},
        title = "{Improved Dynamical Constraints on the Mass of the Central Black Hole in NGC 404}",
      journal = {\apj},
         year = 2017,
        month = feb,
       volume = {836},
       number = {2},
          eid = {237},
        pages = {237},
          doi = {10.3847/1538-4357/aa5cb4},
archivePrefix = {arXiv},
       eprint = {1610.09385},
 primaryClass = {astro-ph.GA},
       adsurl = {https://ui.adsabs.harvard.edu/abs/2017ApJ...836..237N}
}

@ARTICLE{Nguyen2023,
       author = {{Nguyen}, Dieu D. and {Cappellari}, Michele and {Pereira-Santaella}, Miguel},
        title = "{Simulating supermassive black hole mass measurements for a sample of ultramassive galaxies using ELT/HARMONI high-spatial-resolution integral-field stellar kinematics}",
      journal = {\mnras},
         year = 2023,
        month = dec,
       volume = {526},
       number = {3},
        pages = {3548-3569},
          doi = {10.1093/mnras/stad2860},
archivePrefix = {arXiv},
       eprint = {2302.10012},
 primaryClass = {astro-ph.GA},
       adsurl = {https://ui.adsabs.harvard.edu/abs/2023MNRAS.526.3548N}
}

@ARTICLE{Nguyen2025b,
       author = {{Nguyen}, Dieu D. and {Cappellari}, Michele and {Ngo}, Hai N. and {Le}, Tinh Q.~T. and {Le}, Tuan N. and {Ho}, Khue N.~H. and {Nguyen}, An K. and {On}, Phong T. and {Tong}, Huy G. and {Thatte}, Niranjan and {Pereira-Santaella}, Miguel},
        title = "{Simulating Intermediate Black Hole Mass Measurements for a Sample of Galaxies with Nuclear Star Clusters Using ELT/HARMONI High Spatial Resolution Integral-field Stellar Kinematics}",
      journal = {\aj},
         year = 2025,
        month = aug,
       volume = {170},
       number = {2},
          eid = {124},
        pages = {124},
          doi = {10.3847/1538-3881/ade9ba},
       adsurl = {https://ui.adsabs.harvard.edu/abs/2025AJ....170..124N}
}

@article{McConnell2013,
  author = {McConnell, Nicholas J. and Ma, Chung-Pei},
  title = {Revisiting the scaling relations of black hole masses and host galaxy properties},
  journal = {The Astrophysical Journal},
  year = {2013},
  volume = {764},
  number = {2},
  pages = {184},
  doi = {10.1088/0004-637X/764/2/184}
}

@ARTICLE{Devereux2003,
       author = {{Devereux}, Nick and {Ford}, Holland and {Tsvetanov}, Zlatan and {Jacoby}, George},
        title = "{STIS Spectroscopy of the Central 10 Parsecs of M81: Evidence for a Massive Black Hole}",
      journal = {\aj},
         year = 2003,
        month = mar,
       volume = {125},
       number = {3},
        pages = {1226-1235},
          doi = {10.1086/367595},
       adsurl = {https://ui.adsabs.harvard.edu/abs/2003AJ....125.1226D}
}

@INPROCEEDINGS{Thater2019conf,
       author = {{Thater}, Sabine},
        title = "{Testing the robustness of massive black hole mass measurements using ALMA and MUSE}",
    booktitle = {ALMA2019: Science Results and Cross-Facility Synergies},
         year = 2019,
        month = dec,
          eid = {129},
        pages = {129},
          doi = {10.5281/zenodo.3585459},
archivePrefix = {arXiv},
       eprint = {1911.11491},
 primaryClass = {astro-ph.GA},
       adsurl = {https://ui.adsabs.harvard.edu/abs/2019asrc.confE.129T}
}

@ARTICLE{Perna2023,
       author = {{Perna}, M. and {Arribas}, S. and {Marshall}, M. and {D'Eugenio}, F. and {{\"U}bler}, H. and {Bunker}, A. and {Charlot}, S. and {Carniani}, S. and {Jakobsen}, P. and {Maiolino}, R. and {Rodr{\'\i}guez Del Pino}, B. and {Willott}, C.~J. and {B{\"o}ker}, T. and {Circosta}, C. and {Cresci}, G. and {Curti}, M. and {Husemann}, B. and {Kumari}, N. and {Lamperti}, I. and {P{\'e}rez-Gonz{\'a}lez}, P.~G. and {Scholtz}, J.},
        title = "{GA-NIFS: The ultra-dense, interacting environment of a dual AGN at z {\ensuremath{\sim}} 3.3 revealed by JWST/NIRSpec IFS}",
      journal = {\aap},
         year = 2023,
        month = nov,
       volume = {679},
          eid = {A89},
        pages = {A89},
          doi = {10.1051/0004-6361/202346649},
archivePrefix = {arXiv},
       eprint = {2304.06756},
 primaryClass = {astro-ph.GA},
       adsurl = {https://ui.adsabs.harvard.edu/abs/2023A&A...679A..89P}
}

@ARTICLE{vanderMarel1993,
       author = {{van der Marel}, Roeland P. and {Franx}, Marijn},
        title = "{A New Method for the Identification of Non-Gaussian Line Profiles in Elliptical Galaxies}",
      journal = {\apj},
         year = 1993,
        month = apr,
       volume = {407},
        pages = {525},
          doi = {10.1086/172534},
       adsurl = {https://ui.adsabs.harvard.edu/abs/1993ApJ...407..525V}
}

@ARTICLE{Cappellari2009,
       author = {{Cappellari}, Michele and {Neumayer}, N. and {Reunanen}, J. and {van der Werf}, P.~P. and {de Zeeuw}, P.~T. and {Rix}, H. -W.},
        title = "{The mass of the black hole in Centaurus A from SINFONI AO-assisted integral-field observations of stellar kinematics}",
      journal = {\mnras},
         year = 2009,
        month = apr,
       volume = {394},
       number = {2},
        pages = {660-674},
          doi = {10.1111/j.1365-2966.2008.14377.x},
archivePrefix = {arXiv},
       eprint = {0812.1000},
 primaryClass = {astro-ph},
       adsurl = {https://ui.adsabs.harvard.edu/abs/2009MNRAS.394..660C}
}

@ARTICLE{Law2023,
       author = {{Law}, David R. and {E. Morrison}, Jane and {Argyriou}, Ioannis and {Patapis}, Polychronis and {{\'A}lvarez-M{\'a}rquez}, J. and {Labiano}, Alvaro and {Vandenbussche}, Bart},
        title = "{A 3D Drizzle Algorithm for JWST and Practical Application to the MIRI Medium Resolution Spectrometer}",
      journal = {\aj},
         year = 2023,
        month = aug,
       volume = {166},
       number = {2},
          eid = {45},
        pages = {45},
          doi = {10.3847/1538-3881/acdddc},
archivePrefix = {arXiv},
       eprint = {2306.05520},
 primaryClass = {astro-ph.IM},
       adsurl = {https://ui.adsabs.harvard.edu/abs/2023AJ....166...45L}
}

@ARTICLE{Bender1994,
       author = {{Bender}, R. and {Saglia}, R.~P. and {Gerhard}, O.~E.},
        title = "{Line-of-sight velocity distributions of elliptical galaxies.}",
      journal = {\mnras},
         year = 1994,
        month = aug,
       volume = {269},
        pages = {785-813},
          doi = {10.1093/mnras/269.3.785},
       adsurl = {https://ui.adsabs.harvard.edu/abs/1994MNRAS.269..785B}
}

@INPROCEEDINGS{Krist1995,
       author = {{Krist}, J.},
        title = "{Simulation of HST PSFs using Tiny Tim}",
    booktitle = {Astronomical Data Analysis Software and Systems IV},
         year = 1995,
       editor = {{Shaw}, R.~A. and {Payne}, H.~E. and {Hayes}, J.~J.~E.},
       series = {Astronomical Society of the Pacific Conference Series},
       volume = {77},
        month = jan,
        pages = {349},
       adsurl = {https://ui.adsabs.harvard.edu/abs/1995ASPC...77..349K}
}

@ARTICLE{vanderMarel1994,
       author = {{van der Marel}, R.~P.},
        title = "{Velocity profiles of galaxies with claimed black holes - III. Observations and models for M 87.}",
      journal = {\mnras},
         year = 1994,
        month = sep,
       volume = {270},
        pages = {271-297},
          doi = {10.1093/mnras/270.2.271},
       adsurl = {https://ui.adsabs.harvard.edu/abs/1994MNRAS.270..271V}
}

@ARTICLE{Simon2024,
       author = {{Simon}, David A. and {Cappellari}, Michele and {Hartke}, Johanna},
        title = "{Supermassive black hole mass in the massive elliptical galaxy M87 from integral-field stellar dynamics using OASIS and MUSE with adaptive optics: assessing systematic uncertainties}",
      journal = {\mnras},
         year = 2024,
        month = jan,
       volume = {527},
       number = {2},
        pages = {2341-2361},
          doi = {10.1093/mnras/stad3309},
archivePrefix = {arXiv},
       eprint = {2303.18229},
 primaryClass = {astro-ph.GA},
       adsurl = {https://ui.adsabs.harvard.edu/abs/2024MNRAS.527.2341S}
}

@ARTICLE{Silge2003,
       author = {{Silge}, Julia D. and {Gebhardt}, Karl},
        title = "{Dust and the Infrared Kinematic Properties of Early-Type Galaxies}",
      journal = {\aj},
         year = 2003,
        month = jun,
       volume = {125},
       number = {6},
        pages = {2809-2823},
          doi = {10.1086/375324},
archivePrefix = {arXiv},
       eprint = {astro-ph/0303590},
 primaryClass = {astro-ph},
       adsurl = {https://ui.adsabs.harvard.edu/abs/2003AJ....125.2809S}
}

@ARTICLE{Marconi2000,
       author = {{Marconi}, Alessandro and {Schreier}, Ethan J. and {Koekemoer}, Anton and {Capetti}, Alessandro and {Axon}, David and {Macchetto}, Duccio and {Caon}, Nicola},
        title = "{Unveiling the Active Nucleus of Centaurus A}",
      journal = {\apj},
         year = 2000,
        month = jan,
       volume = {528},
       number = {1},
        pages = {276-291},
          doi = {10.1086/308168},
archivePrefix = {arXiv},
       eprint = {astro-ph/9907378},
 primaryClass = {astro-ph},
       adsurl = {https://ui.adsabs.harvard.edu/abs/2000ApJ...528..276M}
}

@ARTICLE{Cappellari2004,
       author = {{Cappellari}, Michele and {Emsellem}, Eric},
        title = "{Parametric Recovery of Line-of-Sight Velocity Distributions from Absorption-Line Spectra of Galaxies via Penalized Likelihood}",
      journal = {\pasp},
         year = 2004,
        month = feb,
       volume = {116},
       number = {816},
        pages = {138-147},
          doi = {10.1086/381875},
archivePrefix = {arXiv},
       eprint = {astro-ph/0312201},
 primaryClass = {astro-ph},
       adsurl = {https://ui.adsabs.harvard.edu/abs/2004PASP..116..138C}
}

@Article{Cappellari2008,
  author        = {Cappellari, M.},
  title         = {Measuring the inclination and mass-to-light ratio of axisymmetric galaxies via anisotropic Jeans models of stellar kinematics},
  doi           = {10.1111/j.1365-2966.2008.13754.x},
  eprint        = {0806.0042},
  pages         = {71--86},
  volume        = {390},
  adsurl        = {https://ui.adsabs.harvard.edu/abs/2008MNRAS.390...71C},
  archiveprefix = {arXiv},
  journal       = {\mnras},
  month         = oct,
  year          = {2008},
}

@ARTICLE{Nguyen2014,
   author = {{Nguyen}, D.~D. and {Seth}, A.~C. and {Reines}, A.~E. and {den Brok}, M. and 
  {Sand}, D. and {McLeod}, B.},
    title = "{Extended Structure and Fate of the Nucleus in Henize 2-10}",
  journal = {\apj},
archivePrefix = "arXiv",
   eprint = {1408.4446},
     year = 2014,
    month = oct,
   volume = 794,
      eid = {34},
    pages = {34},
      doi = {10.1088/0004-637X/794/1/34},
   adsurl = {http://adsabs.harvard.edu/abs/2014ApJ...794...34N}
}

@ARTICLE{Muller2011,
       author = {{Schnorr M{\"u}ller}, Allan and {Storchi-Bergmann}, Thaisa and {Riffel}, Rogemar A. and {Ferrari}, Fabricio and {Steiner}, J.~E. and {Axon}, David J. and {Robinson}, Andrew},
        title = "{Gas streaming motions towards the nucleus of M81}",
      journal = {\mnras},
         year = 2011,
        month = may,
       volume = {413},
       number = {1},
        pages = {149-161},
          doi = {10.1111/j.1365-2966.2010.18116.x},
archivePrefix = {arXiv},
       eprint = {1012.3015},
 primaryClass = {astro-ph.CO},
       adsurl = {https://ui.adsabs.harvard.edu/abs/2011MNRAS.413..149S}
}

@article{Sasseville2025,
doi = {10.3847/1538-4357/ad93d4},
url = {https://dx.doi.org/10.3847/1538-4357/ad93d4},
year = {2024},
month = {dec},
publisher = {The American Astronomical Society},
volume = {978},
number = {1},
pages = {48},
author = {Sasseville, Gabriel and Hlavacek-Larrondo, Julie and Berek, Samantha C. and Eadie, Gwendolyn M. and Rhea, Carter Lee and Springford, Aaron and Mezcua, Mar and Haggard, Daryl},
title = {A Novel Approach to Understanding the Link between Supermassive Black Holes and Host Galaxies},
journal = {The Astrophysical Journal}
}

@ARTICLE{Nguyen2025a,
       author = {{Nguyen}, Dieu D. and {Ngo}, Hai N. and {Le}, Tinh Q.~T. and {Graham}, Alister W. and {Soria}, Roberto and {Chilingarian}, Igor V. and {Thatte}, Niranjan and {Phuong}, N.~T. and {Hoang}, Thiem and {Pereira-Santaella}, Miguel and {Durre}, Mark and {Pham}, Diep N. and {Ngoc Tram}, Le and {Ngoc}, Nguyen B. and {L{\^e}}, Ng{\^a}n},
        title = "{Supermassive black hole mass measurement in the spiral galaxy NGC 4736 using JWST/NIRSpec stellar kinematics}",
      journal = {\aap},
         year = 2025,
        month = jun,
       volume = {698},
          eid = {L9},
        pages = {L9},
          doi = {10.1051/0004-6361/202554672},
archivePrefix = {arXiv},
       eprint = {2505.09941},
 primaryClass = {astro-ph.GA},
       adsurl = {https://ui.adsabs.harvard.edu/abs/2025A&A...698L...9N}
}

@misc{FangzhengShi2021,
      title={An Energetic Hot Wind from the Low-luminosity Active Galactic Nucleus M81*}, 
      author={Fangzheng Shi and Zhiyuan Li and Feng Yuan and Bocheng Zhu},
      year={2021},
      eprint={2106.04041},
      archivePrefix={arXiv},
      primaryClass={astro-ph.HE},
      url={https://arxiv.org/abs/2106.04041}, 
}

@misc{DEugenio2024,
      title={A fast-rotator post-starburst galaxy quenched by supermassive black-hole feedback at z=3}, 
      author={Francesco D'Eugenio and Pablo Perez-Gonzalez and Roberto Maiolino and Jan Scholtz and Michele Perna and Chiara Circosta and Hannah Uebler and Santiago Arribas and Torsten Boeker and Andrew Bunker and Stefano Carniani and Stephane Charlot and Jacopo Chevallard and Giovanni Cresci and Emma Curtis-Lake and Gareth Jones and Nimisha Kumari and Isabella Lamperti and Tobias Looser and Eleonora Parlanti and Hans-Walter Rix and Brant Robertson and Bruno Rodriguez Del Pino and Sandro Tacchella and Giacomo Venturi and Chris Willott},
      year={2023},
      eprint={2308.06317},
      archivePrefix={arXiv},
      primaryClass={astro-ph.GA},
      url={https://arxiv.org/abs/2308.06317}, 
}

@software{Bushouse24_1.14.0,
       author = {{Bushouse}, Howard and {Eisenhamer}, Jonathan and {Dencheva}, Nadia and {Davies}, James and {Greenfield}, Perry and {Morrison}, Jane and {Hodge}, Phil and {Simon}, Bernie and {Grumm}, David and {Droettboom}, Michael and {Slavich}, Edward and {Sosey}, Megan and {Pauly}, Tyler and {Miller}, Todd and {Jedrzejewski}, Robert and {Hack}, Warren and {Davis}, David and {Crawford}, Steven and {Law}, David and {Gordon}, Karl and {Regan}, Michael and {Cara}, Mihai and {MacDonald}, Ken and {Bradley}, Larry and {Shanahan}, Clare and {Jamieson}, William and {Teodoro}, Mairan and {Williams}, Thomas and {Pena-Guerrero}, Maria},
        title = "{JWST Calibration Pipeline}",
         year = 2024,
        month = mar,
          eid = {10.5281/zenodo.10870758},
          doi = {10.5281/zenodo.10870758},
      version = {1.14.0},
    publisher = {Zenodo},
       adsurl = {https://ui.adsabs.harvard.edu/abs/2024zndo..10870758B}
}

@ARTICLE{Cappellari2017,
       author = {{Cappellari}, Michele},
        title = "{Improving the full spectrum fitting method: accurate convolution with Gauss-Hermite functions}",
      journal = {\mnras},
         year = 2017,
        month = apr,
       volume = {466},
       number = {1},
        pages = {798-811},
          doi = {10.1093/mnras/stw3020},
archivePrefix = {arXiv},
       eprint = {1607.08538},
 primaryClass = {astro-ph.GA},
       adsurl = {https://ui.adsabs.harvard.edu/abs/2017MNRAS.466..798C}
}

@ARTICLE{Cappellari2023,
       author = {{Cappellari}, Michele},
        title = "{Full spectrum fitting with photometry in PPXF: stellar population versus dynamical masses, non-parametric star formation history and metallicity for 3200 LEGA-C galaxies at redshift z {\ensuremath{\approx}} 0.8}",
      journal = {\mnras},
         year = 2023,
        month = dec,
       volume = {526},
       number = {3},
        pages = {3273-3300},
          doi = {10.1093/mnras/stad2597},
archivePrefix = {arXiv},
       eprint = {2208.14974},
 primaryClass = {astro-ph.GA},
       adsurl = {https://ui.adsabs.harvard.edu/abs/2023MNRAS.526.3273C}
}

@ARTICLE{Cappellari2002,
       author = {{Cappellari}, Michele},
        title = "{Efficient multi-Gaussian expansion of galaxies}",
      journal = {\mnras},
         year = 2002,
        month = jun,
       volume = {333},
       number = {2},
        pages = {400-410},
          doi = {10.1046/j.1365-8711.2002.05412.x},
archivePrefix = {arXiv},
       eprint = {astro-ph/0201430},
 primaryClass = {astro-ph},
       adsurl = {https://ui.adsabs.harvard.edu/abs/2002MNRAS.333..400C}
}

@inproceedings{Krist2011,
author = {John E. Krist and Richard N. Hook and Felix Stoehr},
title = {{20 years of Hubble Space Telescope optical modeling using Tiny Tim}},
volume = {8127},
booktitle = {Optical Modeling and Performance Predictions V},
editor = {Mark A. Kahan},
organization = {International Society for Optics and Photonics},
publisher = {SPIE},
pages = {81270J},
year = {2011},
doi = {10.1117/12.892762},
URL = {https://doi.org/10.1117/12.892762}
}

@ARTICLE{Emsellem1994,
       author = {{Emsellem}, E. and {Monnet}, G. and {Bacon}, R.},
        title = "{The multi-gaussian expansion method: a tool for building realistic photometric and kinematical models of stellar systems I. The formalism}",
      journal = {\aap},
         year = 1994,
        month = may,
       volume = {285},
        pages = {723-738},
       adsurl = {https://ui.adsabs.harvard.edu/abs/1994A&A...285..723E}
}

@ARTICLE{Verro2022,
       author = {{Verro}, K. and {Trager}, S.~C. and {Peletier}, R.~F. and {Lan{\c{c}}on}, A. and {Gonneau}, A. and {Vazdekis}, A. and {Prugniel}, P. and {Chen}, Y. -P. and {Coelho}, P.~R.~T. and {S{\'a}nchez-Bl{\'a}zquez}, P. and {Martins}, L. and {Arentsen}, A. and {Lyubenova}, M. and {Falc{\'o}n-Barroso}, J. and {Dries}, M.},
        title = "{The X-shooter Spectral Library (XSL): Data Release 3}",
      journal = {\aap},
         year = 2022,
        month = apr,
       volume = {660},
          eid = {A34},
        pages = {A34},
          doi = {10.1051/0004-6361/202142388},
archivePrefix = {arXiv},
       eprint = {2110.10188},
 primaryClass = {astro-ph.SR},
       adsurl = {https://ui.adsabs.harvard.edu/abs/2022A&A...660A..34V}
}

@article{Husser2013,
   title={A new extensive library of PHOENIX stellar atmospheres and synthetic spectra},
   volume={553},
   ISSN={1432-0746},
   url={http://dx.doi.org/10.1051/0004-6361/201219058},
   DOI={10.1051/0004-6361/201219058},
   journal={\aap},
   publisher={EDP Sciences},
   author={Husser, T.-O. and Wende-von Berg, S. and Dreizler, S. and Homeier, D. and Reiners, A. and Barman, T. and Hauschildt, P. H.},
   year={2013},
   month=apr, pages={A6} 
}

@ARTICLE{Eracleous2010,
       author = {{Eracleous}, Michael and {Hwang}, Jason A. and {Flohic}, H{\'e}l{\`e}ne M.~L.~G.},
        title = "{Spectral Energy Distributions of Weak Active Galactic Nuclei Associated with Low-Ionization Nuclear Emission Regions}",
      journal = {\apjs},
         year = 2010,
        month = mar,
       volume = {187},
       number = {1},
        pages = {135-148},
          doi = {10.1088/0067-0049/187/1/135},
archivePrefix = {arXiv},
       eprint = {1001.2924},
 primaryClass = {astro-ph.GA},
       adsurl = {https://ui.adsabs.harvard.edu/abs/2010ApJS..187..135E}
}

@ARTICLE{Krajnovic2018,
  author       = {Krajnovi{\'c}, Davor and Cappellari, M. and McDermid, Richard M. and Thater, Sabine and Nyland, Kristina and de Zeeuw, P. T. and Falc{\'o}n-Barroso, Jes{\'u}s and Khochfar, Sadegh and Kuntschner, Harald and Sarzi, Marc},
  year         = 2018,
  journaltitle = {\mnras},
  title        = {{A quartet of black holes and a missing duo: probing the low end of the \(M_{BH}-\sigma\) relation with the adaptive optics assisted integral-field spectroscopy}},
  doi          = {10.1093/mnras/sty778},
  eprint       = {1803.08055},
  eprintclass  = {astro-ph.GA},
  eprinttype   = {arXiv},
  number       = {3},
  pages        = {3030--3064},
  volume       = {477},
  adsurl       = {https://ui.adsabs.harvard.edu/abs/2018MNRAS.477.3030K},
}

@ARTICLE{Walsh2016,
  author       = {Walsh, J. L. and van den Bosch, R. C. E. and Gebhardt, K. and Y{\i}ld{\i}r{\i}m, A. and Richstone, D. O. and G{\"u}ltekin, K. and Husemann, B.},
  year         = 2016,
  journaltitle = {\apj},
  title        = {{A 5 x 10$^{9}$ Msun Black Hole in NGC 1277 from Adaptive Optics Spectroscopy}},
  doi          = {10.3847/0004-637X/817/1/2},
  eid          = {2},
  eprint       = {1511.04455},
  eprinttype   = {arXiv},
  pages        = {2},
  volume       = {817},
  adsurl       = {https://ui.adsabs.harvard.edu/abs/2016ApJ...817....2W},
}

@Article{Barth2001,
  author       = {Barth, Aaron J. and Sarzi, Marc and Rix, Hans-Walter and Ho, Luis C. and Filippenko, Alexei V. and Sargent, Wallace L.~W.},
  journaltitle = {\apj},
  title        = {Evidence for a Supermassive Black Hole in the S0 Galaxy NGC 3245},
  doi          = {10.1086/321523},
  eprint       = {astro-ph/0012213},
  eprintclass  = {astro-ph},
  eprinttype   = {arXiv},
  number       = {2},
  pages        = {685--708},
  url          = {https://ui.adsabs.harvard.edu/abs/2001ApJ...555..685B},
  volume       = {555},
  year         = {2001},
}

@Article{Shapiro2006,
  author  = {Shapiro, K. L. and Cappellari, M. and de Zeeuw, T. and McDermid, R. M. and Gebhardt, K. and van den Bosch, R. C. E. and Statler, T. S.},
  title   = {The black hole in NGC 3379: a comparison of gas and stellar dynamical mass measurements with HST and integral-field data},
  doi     = {10.1111/j.1365-2966.2006.10537.x},
  eprint  = {arXiv:astro-ph/0605479},
  pages   = {559--579},
  volume  = {370},
  adsurl  = {https://ui.adsabs.harvard.edu/abs/2006MNRAS.370..559S},
  journal = {\mnras},
  month   = aug,
  year    = {2006},
}

@Article{Neumayer07,
  author   = {Neumayer, N. and Cappellari, M. and Reunanen, J. and Rix, H.-W. and van der Werf, P. P. and de Zeeuw, P. T. and Davies, R. I.},
  title    = {The Central Parsecs of Centaurus A: High-excitation Gas, a Molecular Disk, and the Mass of the Black Hole},
  doi      = {10.1086/523039},
  eprint   = {arXiv:0709.1877},
  pages    = {1329--1344},
  volume   = {671},
  adsurl   = {https://ui.adsabs.harvard.edu/abs/2007ApJ...671.1329N},
  journal  = {\apj},
  month    = dec,
  year     = {2007},
}

@Article{McCommas2009,
  author       = {{McCommas}, L. P. and {Macri}, L. M. and {Rejkuba}, M. and {Freedman}, W. L. and {Madore}, B. F. and {Stetson}, P. B. and {Ngeow}, C. -C. and {Kanbur}, S. M. and {Andreuzzi}, G.},
  year         = {2009},
  journaltitle = {\apjs},
  title        = {{The Cepheid Period-Luminosity Relation in the Near-Infrared and the Distance to M81}},
  doi          = {10.1088/0067-0049/185/1/1},
  eprint       = {0908.2121},
  eprintclass  = {astro-ph.CO},
  eprinttype   = {arXiv},
  number       = {1},
  pages        = {1-16},
  volume       = {185},
  adsurl       = {https://ui.adsabs.harvard.edu/abs/2009ApJS..185....1M},
}

@Article{Tikhonov2005,
  author       = {{Tikhonov}, N. A. and {Galazutdinova}, O. A.},
  journaltitle = {Astronomy Letters},
  title        = {{The Tip of the Red Giant Branch and the Distance to the Galaxy M81}},
  doi          = {10.1134/1.1894510},
  eprint       = {astro-ph/0501328},
  eprintclass  = {astro-ph},
  eprinttype   = {arXiv},
  number       = {4},
  pages        = {228-233},
  volume       = {31},
  adsurl       = {https://ui.adsabs.harvard.edu/abs/2005AstL...31..228T},
  year         = {2005},
}

@Article{Freedman1994,
  author       = {{Freedman}, Wendy L. and {Hughes}, Shaun M. and {Madore}, Barry F. and {Mould}, Jeremy R. and {Lee}, Myung Gyoon and {Stetson}, Peter and {Kennicutt}, Robert C. and {Turner}, Anne and {Sakai}, Shoko and {Ferrarese}, Laura and {Ford}, Holland and {Graham}, John A. and {Hill}, Robert and {Hoessel}, John G. and {Huchra}, John and {Illingworth}, Garth D.},
  journaltitle = {\apj},
  title        = {{Distance to M81 with the Hubble Space Telescope}},
  doi          = {10.1086/174172},
  pages        = {628},
  volume       = {427},
  adsurl       = {https://ui.adsabs.harvard.edu/abs/1994ApJ...427..628F},
  year         = {1994},
}

@Article{Dalcanton2009,
  author       = {{Dalcanton}, Julianne J. and {Williams}, Benjamin F. and {Seth}, Anil C. and {Dolphin}, Andrew and {Holtzman}, Jon and {Rosema}, Keith and {Skillman}, Evan D. and {Cole}, Andrew and {Girardi}, L{\'e}o and {Gogarten}, Stephanie M. and {Karachentsev}, Igor D. and {Olsen}, Knut and {Weisz}, Daniel and {Christensen}, Charlotte and {Freeman}, Ken and {Gilbert}, Karoline and {Gallart}, Carme and {Harris}, Jason and {Hodge}, Paul and {de Jong}, Roelof S. and {Karachentseva}, Valentina and {Mateo}, Mario and {Stetson}, Peter B. and {Tavarez}, Marla and {Zaritsky}, Dennis and {Governato}, Fabio and {Quinn}, Tom},
  year         = {2009},
  journaltitle = {\apjs},
  title        = {{The ACS Nearby Galaxy Survey Treasury}},
  doi          = {10.1088/0067-0049/183/1/67},
  eprint       = {0905.3737},
  eprintclass  = {astro-ph.CO},
  eprinttype   = {arXiv},
  number       = {1},
  pages        = {67-108},
  volume       = {183},
  adsurl       = {https://ui.adsabs.harvard.edu/abs/2009ApJS..183...67D},
}

@Article{Rizzi2007,
  author       = {{Rizzi}, Luca and {Tully}, R. Brent and {Makarov}, Dmitry and {Makarova}, Lidia and {Dolphin}, Andrew E. and {Sakai}, Shoko and {Shaya}, Edward J.},
  year         = {2007},
  journaltitle = {\apj},
  title        = {{The ACS Nearby Galaxy Survey Treasury. II. The Star Formation History of the M81 Group}},
  doi          = {10.1086/516829},
  eprint       = {astro-ph/0701693},
  eprintclass  = {astro-ph},
  eprinttype   = {arXiv},
  number       = {2},
  pages        = {815-828},
  volume       = {661},
  adsurl       = {https://ui.adsabs.harvard.edu/abs/2007ApJ...661..815R},
}

@Article{Ferrarese2000_distance,
  author       = {Ferrarese, Laura and Mould, Jeremy R. and Kennicutt, Jr., Robert C. and Huchra, John and Ford, Holland C. and Freedman, Wendy L. and Stetson, Peter B. and Madore, Barry F. and Sakai, Shoko and Gibson, Brad K. and Graham, John A. and Hughes, Shaun M. and Illingworth, Garth D. and Kelson, Daniel D. and Macri, Lucas and Sebo, Kim and Silbermann, N.~A.},
  journaltitle = {\apj},
  title        = {The Hubble Space Telescope Key Project on the Extragalactic Distance Scale. XXVI. The Calibration of Population II Secondary Distance Indicators and the Value of the Hubble Constant},
  doi          = {10.1086/308309},
  eprint       = {astro-ph/9908192},
  eprintclass  = {astro-ph},
  eprinttype   = {arXiv},
  number       = {2},
  pages        = {745-767},
  url          = {https://ui.adsabs.harvard.edu/abs/2000ApJ...529..745F},
  volume       = {529},
  year         = {2000},
}

@InProceedings{Bower2000,
  author    = {Bower, G.~A. and Wilson, A.~S. and Heckman, T.~M. and Magorrian, J. and Gebhardt, K. and Richstone, D.~O. and Peterson, B.~M. and Green, R.~F.},
  booktitle = {American Astronomical Society Meeting Abstracts},
  title     = {The Stellar Dynamics in the Centers of the LINER Galaxies M81 and NGC 3998},
  year      = {2000},
  pages     = {92.03},
  series    = {American Astronomical Society Meeting Abstracts},
  volume    = {197},
  eid       = {92.03},
  url       = {https://ui.adsabs.harvard.edu/abs/2000AAS...197.9203B},
}

@Article{Onishi2017,
  author        = {{Onishi}, Kyoko and {Iguchi}, Satoru and {Davis}, Timothy A. and {Bureau}, Martin and {Cappellari}, Michele and {Sarzi}, Marc and {Blitz}, Leo},
  journal       = {\mnras},
  title         = {{WISDOM project - I. Black hole mass measurement using molecular gas kinematics in NGC 3665}},
  year          = {2017},
  month         = jul,
  number        = {4},
  pages         = {4663-4674},
  volume        = {468},
  adsurl        = {https://ui.adsabs.harvard.edu/abs/2017MNRAS.468.4663O},
  archiveprefix = {arXiv},
  doi           = {10.1093/mnras/stx631},
  eprint        = {1703.05247},
  primaryclass  = {astro-ph.GA},
}

@Article{Davis2013,
  author        = {Davis, Timothy A. and Bureau, Martin and Cappellari, Michele and Sarzi, Marc and Blitz, Leo},
  journal       = {\nat},
  title         = {{A black-hole mass measurement from molecular gas kinematics in NGC4526}},
  year          = {2013},
  month         = feb,
  number        = {7437},
  pages         = {328--330},
  volume        = {494},
  adsurl        = {https://ui.adsabs.harvard.edu/abs/2013Natur.494..328D},
  archiveprefix = {arXiv},
  doi           = {10.1038/nature11819},
  eprint        = {1301.7184},
  primaryclass  = {astro-ph.CO},
}

@Article{Verolme2002,
  author  = {Verolme, E. K. and Cappellari, M. and Copin, Y. and van der Marel, R. P. and Bacon, R. and Bureau, M. and Davies, R. L. and Miller, B. M. and de Zeeuw, P. T.},
  journal = {\mnras},
  title   = {A SAURON study of M32: measuring the intrinsic flattening and the central black hole mass},
  year    = {2002},
  month   = sep,
  pages   = {517--525},
  volume  = {335},
  adsurl  = {https://ui.adsabs.harvard.edu/abs/2002MNRAS.335..517V},
  doi     = {10.1046/j.1365-8711.2002.05664.x},
  eprint  = {arXiv:astro-ph/0201086},
}

@Article{Paturel2003,
  author  = {Paturel, G. and Petit, C. and Prugniel, P. and Theureau, G. and Rousseau, J. and Brouty, M. and Dubois, P. and Cambr{\'e}sy, L.},
  journal = {\aap},
  title   = {HYPERLEDA. I. Identification and designation of galaxies},
  year    = {2003},
  month   = dec,
  pages   = {45--55},
  volume  = {412},
  adsurl  = {https://ui.adsabs.harvard.edu/abs/2003A%26A...412...45P},
  doi     = {10.1051/0004-6361:20031411},
}

@Misc{Perrin2025,
  author    = {Perrin, Marshall and Long, Joseph and Osborne, Shannon and Geda, Robel and Sappington, Bradley and Meléndez, Marcio and Lajoie, Charles-Philippe and Leisenring, Jarron and Zimmerman, Neil and Brooks, Keira and Otor, O. Justin and Kulp, Trey and Chambers, Lauren and Jurling, Alden},
  title     = {STPSF},
  year      = {2025},
  copyright = {BSD 3-Clause "New" or "Revised" License},
  doi       = {10.5281/ZENODO.15747364},
  publisher = {Zenodo},
}

@Manual{Thatte2009,
  title        = {NICMOS Data Handbook, Version 8.0},
  address      = {Baltimore, MD},
  author       = {Thatte, Deepashri and Dahlen, Tomas and Barker, Elizabeth and de Jong, Roelof and Koekemoer, Anton},
  month        = {May},
  note         = {Version 8.0 (May 2009)},
  organization = {Space Telescope Science Institute},
  year         = {2009},
  adsurl       = {https://ui.adsabs.harvard.edu/abs/2009nicm.book.....T},
  url          = {http://www.stsci.edu/hst/nicmos},
}

@Article{Rantala2024,
  author        = {Rantala, Antti and Rawlings, Alexander and Naab, Thorsten and Thomas, Jens and Johansson, Peter H.},
  journal       = {\mnras},
  title         = {The supermassive black hole merger-driven evolution of high-redshift red nuggets into present-day cored early-type galaxies},
  year          = {2024},
  month         = nov,
  number        = {1},
  pages         = {1202--1227},
  volume        = {535},
  adsurl        = {https://ui.adsabs.harvard.edu/abs/2024MNRAS.535.1202R},
  archiveprefix = {arXiv},
  doi           = {10.1093/mnras/stae2424},
  eprint        = {2407.18303},
  primaryclass  = {astro-ph.GA},
}

@Article{Milosavljevic2001,
  author  = {Milosavljevi{\'c}, M. and Merritt, D.},
  journal = {\apj},
  title   = {Formation of Galactic Nuclei},
  year    = {2001},
  month   = dec,
  pages   = {34--62},
  volume  = {563},
  adsurl  = {https://ui.adsabs.harvard.edu/abs/2001ApJ...563...34M},
  doi     = {10.1086/323830},
  eprint  = {arXiv:astro-ph/0103350},
}

@InCollection{Cappellari2026,
  author        = {Cappellari, Michele},
  booktitle     = {Encyclopedia of Astrophysics, Volume 4},
  publisher     = {Elsevier},
  title         = {Early-type galaxies: Elliptical and S0 galaxies, or fast and slow rotators},
  year          = {2026},
  address       = {Amsterdam, The Netherlands},
  editor        = {Mandel, Ilya},
  isbn          = {9780443214394},
  note          = {Preprint available at arXiv:2503.02746},
  pages         = {122--152},
  volume        = {4},
  adsurl        = {https://ui.adsabs.harvard.edu/abs/2026enap....4..122C},
  archiveprefix = {arXiv},
  doi           = {10.1016/B978-0-443-21439-4.00109-7},
  eprint        = {2503.02746},
  primaryclass  = {astro-ph.GA},
}

@Article{Binney1982,
  author   = {Binney, J. and Mamon, G.~A.},
  journal  = {\mnras},
  title    = {M/L and velocity anisotropy from observations of spherical galaxies, or must M87 have a massive black hole ?},
  year     = {1982},
  month    = jul,
  pages    = {361--375},
  volume   = {200},
  adsurl   = {https://ui.adsabs.harvard.edu/abs/1982MNRAS.200..361B},
  doi      = {10.1093/mnras/200.2.361},
}

\label{lastpage}
\end{document}